\documentclass[fleqn,usenatbib]{mnras}

\usepackage[T1]{fontenc}
\usepackage{ae,aecompl}

\usepackage{graphicx}	% Including figure files
\usepackage{amsmath}	% Advanced maths commands
\usepackage{amssymb}	% Extra maths symbols
\usepackage{bm}
\usepackage{multirow}
\usepackage{orcidlink}

\usepackage[normalem]{ulem}

\usepackage{newtxtext,newtxmath}
\newcommand{\fb}{FIREbox}
\newcommand{\pf}{FIREbox}

\newcommand{\pfdm}{FIREbox$^{\rm DM}$}
\title[FIREbox]{FIREbox: Simulating galaxies at high dynamic range in a cosmological volume}

\author[R. Feldmann et al.]{Robert Feldmann$^{1}$\thanks{E-mail: robert.feldmann@uzh.ch}\orcidlink{0000-0002-1109-1919},
Eliot Quataert$^{2}$\orcidlink{0000-0001-9185-5044},
Claude-André Faucher-Giguère$^{3}$\orcidlink{0000-0002-4900-6628},
Philip F. Hopkins$^{4}$\orcidlink{0000-0003-3729-1684},
\newauthor
Onur Çatmabacak$^{1}$\orcidlink{0000-0003-4067-1434},
Dušan Kereš$^{5}$\orcidlink{0000-0002-1666-7067}\thanks{Subsequent authors listed in alphabetical order},
Luigi Bassini$^{1}$\orcidlink{0000-0002-6864-7762},
Mauro Bernardini$^{1}$\orcidlink{0000-0002-2930-9509},
James S. Bullock$^{6}$\orcidlink{0000-0003-4298-5082},
\newauthor
Elia Cenci$^{1}$\orcidlink{0000-0002-0766-1704},
Jindra Gensior$^{1}$\orcidlink{0000-0001-6119-9883},
Lichen Liang$^{7}$\orcidlink{0000-0001-9422-0095},
Jorge Moreno$^{8}$\orcidlink{0000-0002-3430-3232},
Andrew Wetzel$^{9}$\orcidlink{0000-0003-0603-8942}
\\
$^{1}$Institute for Computational Science, University of Zurich, Zurich CH-8057, Switzerland\\
$^{2}$Department of Astrophysical Sciences, Princeton University, Princeton, NJ 08544, USA\\
$^{3}$CIERA and Department of Physics and Astronomy, Northwestern University, 1800 Sherman Ave, Evanston, IL 60201, USA\\
$^{4}$California Institute of Technology, TAPIR, Mailcode 350-17, Pasadena, CA 91125, USA\\
$^{5}$Center for Astrophysics and Space Sciences, University of California San Diego, San Diego, CA 92093, USA\\
$^{6}$Department of Physics and Astronomy, University of California, Irvine, CA 92697, USA\\
$^{7}$Canadian Institute for Theoretical Astrophysics, University of Toronto, Toronto, ON. M5S 3H8, Canada\\
$^{8}$Department of Physics and Astronomy, Pomona College, Claremont, CA 91711, USA\\
$^{9}$Department of Physics and Astronomy, University of California, Davis, CA 95616, USA
}

\date{Accepted XXX. Received YYY; in original form ZZZ}

\pubyear{2020}

\begin{document}
\label{firstpage}
\pagerange{\pageref{firstpage}--\pageref{lastpage}}
\maketitle

% Abstract: It should be a single paragraph not more than 250 words (200 words for Letters)
\begin{abstract}
We introduce a suite of cosmological volume simulations to study the evolution of galaxies as part of the \emph{Feedback in Realistic Environments} project. \pf{}, the principal simulation of the present suite, provides a representative sample of galaxies ($\sim{}1000$ galaxies with $M_{\rm star}>10^8\,M_\odot$ at $z=0$) at a resolution ($\Delta{}x\sim{}20\,{\rm pc}$, $m_{\rm b}\sim{}6\times{}10^4\,M_\odot$) comparable to state-of-the-art galaxy zoom-in simulations. \pf{} captures the multiphase nature of the interstellar medium in a fully cosmological setting ($L=22.1$ Mpc) thanks to its exceptionally high dynamic range ($\gtrsim{}10^6$) and the inclusion of multi-channel stellar feedback. 
Here, we focus on validating the simulation predictions by comparing to observational data. We find that simulated galaxies with $M_{\rm star}<10^{10.5-11}\,M_\odot$ have star formation rates, gas masses, and metallicities in broad agreement with observations. These galaxy scaling relations extend to low masses ($M_{\rm star}\sim{}10^7\,M_\odot$) and follow a (broken) power-law relationship. Also reproduced are the evolution of the cosmic ${\rm H_I}$ density and the ${\rm H_I}$ column density distribution at $z\sim{}0-5$. 
At low $z$, \pf{} predicts a peak in the stellar-mass--halo-mass relation, but also a higher abundance of massive galaxies and a higher cosmic star formation rate density than observed, showing that stellar feedback alone is insufficient to reproduce the properties of massive galaxies at late times. Given its high resolution and sample size, \pf{} offers a baseline prediction of galaxy formation theory in a $\Lambda{\rm CDM}$ Universe while also highlighting modeling challenges to be addressed in next-generation galaxy simulations.
\end{abstract}


%%%%%%%%%%%%%%%%%%%%%%%%%%%%%%%%%%%%%%%%%%%%%%%%%%

%%%%%%%%%%%%%%%%% BODY OF PAPER %%%%%%%%%%%%%%%%%%

\section{Introduction}

High-resolution galaxy surveys, e.g., with MUSE \citep{Bacon2010, Emsellem2021}, ALMA \citep{Fomalont2015, Leroy2021a}, and soon JWST \citep{Gardner2006}, ELT \citep{Gilmozzi2007}, and SKA \citep{Hall2007} are promising to transform our understanding of how galaxies form and evolve. These observational advances will benefit from matched theoretical studies that quantify how the relevant (astro-)physical processes operating on sub-kpc scales shape the properties of galaxies and their interstellar medium \citep{Somerville2014, Naab2017}. This goal of galaxy theory is best approached with numerical simulations given the complexity, interconnectedness, and multi-scale nature of the involved physics (e.g., \citealt{Vogelsberger2020}).

In the past, two main approaches have been employed to simulate the evolution of galaxies in a proper cosmological context. Cosmological volume simulations provide large samples of galaxies with a broad range in properties residing in a variety of cosmological environments (e.g., \citealt{Dubois2014, Vogelsberger2014, Schaye2015, Khandai2015, Dave2016, Pillepich2018a, Dave2019}). Here, physical processes are usually modeled in a simplified, parametrized manner and at a comparably low numerical resolution, e.g., the scale heights of disk galaxies ($\sim{}100$ pc) are typically unresolved. Furthermore, by modeling the relevant physics, especially the effects of stellar feedback \citep{Mayer2008, Scannapieco2012, Crain2015}, on a sub-grid scale, the predictions of the simulations are highly susceptible to model degeneracies. The alternative option are cosmological zoom-in simulations which can reach higher numerical resolution thus enabling them to model baryonic processes in the interstellar medium (ISM) on a more physical basis (e.g., \citealt{Guedes2011, Hopkins2014, Ceverino2014, Feldmann2015, Agertz2015, Wang2015, Wetzel2016, Bellovary2019}). Primary drawbacks of the zoom-in approach are the resulting much smaller sample sizes and/or higher computational costs.

Combining the strengths of both approaches offers the prospect of providing large samples of highly-resolved, realistic galaxies that can then be compared with high-resolution observations to constrain galaxy theory. Recently, first efforts have been undertaken in this direction. These approaches differ in many aspects, e.g., in the implementation and calibration of the baryonic physics and in the numerical method of solving the underlying system of equations. 

One general option is to run a large collection of high-resolution zoom-in simulations of individual galaxies (or small groups thereof) to increase samples sizes (e.g., \citealt{Wang2015, Sawala2016, Feldmann2016, Grand2017, Hopkins2018, Kruijssen2019}). While a powerful method, this approach also has a number of severe shortcomings, e.g., potential selection biases, limited large scale correlations, and the contamination of the refinement region, that limit its applicability. 
Instead, the approach of the {\sc NewHorizon} zoom-in simulation \citep{Dubois2021} is to resolve an ensemble of galaxies in a large refined patch of $\sim{}(16\,{\rm cMpc})^3$. {\sc NewHorizon} is run down to $z=0.25$ with the adaptive mesh refinement code {\sc RAMSES} \citep{Teyssier2002} and makes use of a heavily modified version of the physical model of the {\sc Horizon-AGN} simulation \citep{Dubois2014, Volonteri2016, Kaviraj2017}. High numerical resolution and the modeling of low temperature cooling enable {\sc NewHorizon} to partly resolve the multiphase nature of the ISM.

An alternative approach is to increase the resolution of cosmological volume simulations and improve the employed physical modeling. One advantage of using cosmological volumes over large zoom-ins is that the former can be analyzed more straightforwardly given that the high-resolution region spans the entire cubic volume thus eliminating contamination artifacts.

The TNG50 simulation \citep{Nelson2019}, run with the moving mesh code {\sc AREPO} \citep{Springel2010a}, applies the IllustrisTNG physics model \citep{Pillepich2018} to a $(51.7\,{\rm cMpc})^3$ cosmological box providing a sizable sample of galaxies at a mass resolution (baryonic particle mass $m_{\rm b}\sim{}9\times{}10^4\,M_\odot$) similar to many zoom-ins. Originally calibrated for large volume simulations \citep{Pillepich2018a, Nelson2019a}, the physics model of TNG50 accounts for many baryonic processes in an idealized manner, e.g., the multiphase structure of the ISM is not directly resolved, star formation often takes place in low density gas ($n\geq{}0.11\,{\rm cm}^{-3}$), and galactic outflows are put in by hand and temporarily decoupled from the hydrodynamics. 

The {\sc Romulus25} simulation \citep{Tremmel2017}, run with the smoothed particle hydrodynamics solver {\sc ChanNGa} \citep{Menon2015}, partly addresses some of these shortcomings by adopting a physics model used previously in a large number of high-resolution zoom-in simulations (e.g., \citealt{Governato2007, Governato2010, Shen2010, Guedes2011}) and by applying it, after re-tuning of some of the model parameters, to a $(25\,{\rm cMpc})^3$ cosmological box with a mass resolution of $m_{\rm b}\sim{}2\times{}10^5\,M_\odot$. Specifically, {\sc Romulus25} includes lower temperature gas cooling and a more physical driving of galactic outflows via localized supernova explosions. However, in this model the cooling time of gas heated by supernova feedback is artificially prolonged \citep{Stinson2006}. Furthermore, {\sc Romulus25} does not attempt to trace the dense, star forming (usually molecular) component of the ISM and thus does not properly model the distribution of star formation and stellar feedback in galaxies.

While these recent simulations undoubtedly demonstrate significant progress, a potential concern is the existence of model degeneracies given that their underlying physical models both differ starkly and, in the case of TNG50 and {\sc Romulus25}, are calibrated to observational data. One particularly promising, but challenging, path towards increasing the predictive power of galaxy simulations is to aim for a full accounting of well understood physical processes with only a a minimal number of (ideally zero) tunable parameters.
Implementing this research direction requires a sufficiently high dynamic range to model the relevant physical processes in a fully cosmological context. For instance, identifying the sites of star formation requires a resolution better than a few tens of pc while cosmological accretion and gravitational tides involve scales of tens of Mpc. Furthermore, the adopted physical model should be sufficiently realistic and comprehensive, e.g., the different ISM phases should be reproduced and stellar feedback modeled with as few assumptions as possible. Finally, a sufficiently large (and preferably unbiased) sample of highly-resolved galaxies is needed to compare with observational data across cosmic history.

Fortunately, following this path has now become feasible given the increased computing capacity of supercomputers and algorithmic improvements in modeling galaxies numerically. Most critical, however, is the recent development of more accurate galaxy models that account for the relevant baryonic processes based on physical principles and that minimize the use of ad hoc parametrization (e.g., \citealt{Hopkins2011c, Agertz2013, Hopkins2014, Semenov2016, Kim2017, Li2017, Hopkins2018, Marinacci2019, Kim2020, Hopkins2022}). In particular, the detailed accounting of stellar feedback sources has shown to be paramount for producing galaxies in zoom-in simulations with more realistic properties, e.g., flatter rotation curves, lower stellar masses, and larger mass loading factors of galactic outflows \citep{Guedes2011, Hopkins2014, Muratov2015, Applebaum2021}. 

As a first step on this challenging path, we have designed and run the \fb{} suite of cosmological volume simulations as part of the \emph{Feedback in Realistic Environments} (FIRE) project\footnote{\url{https://fire.northwestern.edu}} \citep{Hopkins2014,Hopkins2018, Hopkins2022}. The primary simulation (\fb{}) of this suite improves over current state-of-the-art in two important aspects. First, \pf{} evolves a cosmological volume of $(22.1\,{\rm cMpc})^3$ down to $z=0$ using a baryonic physics model without explicitly tuned sub-grid parameters. This model (FIRE-2, \citealt{Hopkins2018}) has been used previously in cosmological zoom-in simulations (e.g. \citealt{Angles-Alcazar2017a, Chan2018, Ma2018b, Ma2019, Stern2021, Pandya2021}) but it has not yet been applied to cosmological volumes.
Secondly, \pf{} achieves a dynamic range of $\sim{}10^6$, which is about an order of magnitude higher than TNG50, {\sc NewHorizon}, and {\sc Romulus25}, see section section \ref{sect:CompResolution}. The corresponding high spatial resolution ($\sim{}20$ pc) coupled with the more accurate physical modeling and representative sample size makes \pf{} a unique data set to explore the internal structure of galaxies across cosmic time.
\pf{} is thus well suited to both studying the properties of typical galaxies, e.g., the link between galaxy size and dark matter (DM) halo properties \citep{Rohr2022} or the atomic gas scale heights of Milky-Way analogs \citep{Gensior2022}, to exploring rare galaxy populations, such as low mass, DM deficient galaxies \citep{Moreno2022} or starburst galaxies (Cenci et al. in prep), and to quantifying the properties of the circum-galactic and inter-galactic medium. Furthermore, it can be used as a training set for machine learning based emulators, e.g., to predict the distribution of atomic hydrogen on large scales \citep{Bernardini2021}.

\begin{figure*}
\begin{tabular}{cc}
\includegraphics[width=150mm]{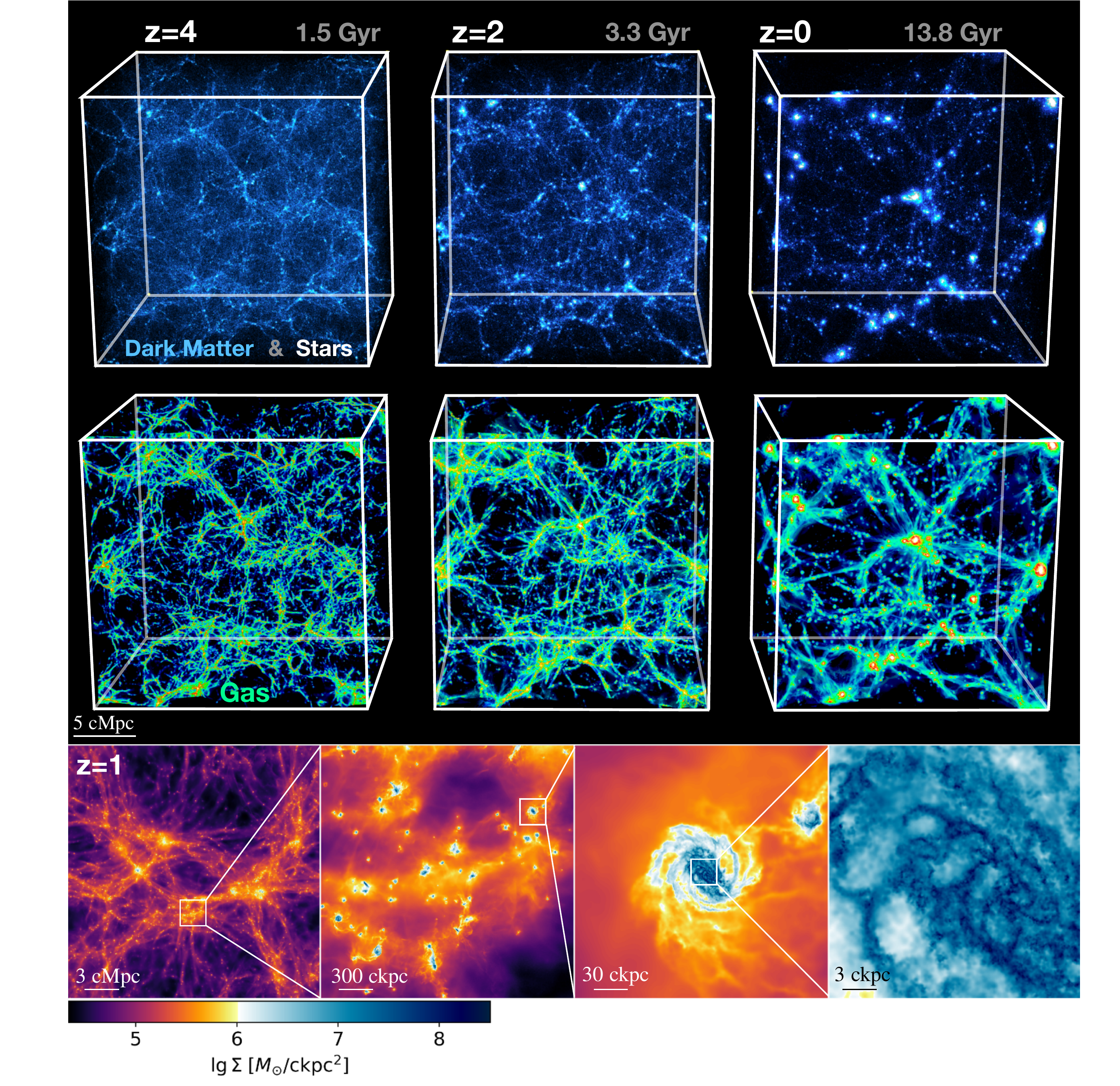}
\end{tabular}
\caption{Visualization of the matter distribution in \pf{}. (Top row) Three-dimensional rendering of the dark matter (blue) and stars (white) in the simulation volume at $z=4$, $z=2$, and $z=0$ (from left to right). Dark matter forms a cosmic web of filamentary structures, overdensities, and voids which evolves with redshift. Stars form at the centers of collapsed dark matter halos. (Middle row) Visualization of the gas distribution which mirrors the cosmic network of dark matter. (Bottom row) Column density maps of the gas projected along the $\sim{}22.1$ cMpc depth of the box at $z=1$. Starting from a view of the gas distribution on cosmological scales, the panels zoom into the interstellar medium of one of many simulated galaxies, illustrating the high dynamic range of \pf{}.}
\label{fig:FBevolution}
\end{figure*}

We highlight the high dynamic range of the simulation in the bottom row of Fig.~\ref{fig:FBevolution}. \pf{} can model both structures on cosmological scales as well as hydrodynamical processes within the dense interstellar medium. Fig.~\ref{fig:FBevolution} also visualizes the distribution of the various matter components in \pf{}. The top row shows the distribution of DM and star particles in the simulation volume at different redshifts, highlighting the formation and subsequent growth of large scale structure. This cosmic web consists of over-dense clusters of DM as well as filaments, sheets, and voids arranged in a complex pattern (e.g., \citealt{Peebles1980, Klypin1983, Davis1985}). Much of the DM in the cosmic web gravitationally collapses into virialized DM halos which then attract gas from their cosmic environments (middle row). Subsequently, stars and galaxies form at the halo centers \citep{White1978}.

A particular feature of the physics-based approach that we follow in this paper is that we intentionally exclude feedback from active galactic nuclei (AGNs) given the large uncertainties involved in its physical modeling. The \fb{} simulation should thus be understood as providing baseline predictions in the absence of AGN feedback.
A comparison between simulation predictions and observations can then be used to make inferences about the role of this feedback channel in galaxy theory. For instance, the low fraction of massive, quiescent galaxies in \fb{}, compared with observations, supports the notion that AGN feedback plays indeed a critical role in galaxy quenching \citep{Springel2005d, Croton2006, Hopkins2006b, Cattaneo2009a}. In contrast, star forming galaxies in \fb{} follow many of their observed global scaling relations indicating that AGN feedback does not strongly affect the latter. We note that understanding the role and impact of this feedback channel is a critical open challenge for galaxy formation and there is increasing evidence that AGN feedback plays an important role not only in massive galaxies (e.g., \citealt{Springel2005d, Dubois2013a, Tremmel2019}) but also in galaxies of lower mass  (e.g., \citealt{Beckmann2017, Dashyan2018, Koudmani2021}). We leave a detailed and more direct analysis of the role of AGN feedback to future work (see \citealt{Wellons2022} for a first exploration of the effects of AGN feedback in a large suite of FIRE-2 zoom-in simulations).

The outline of this paper is as follows. In section \ref{sect:Simulations} we introduce our suite of cosmological volume simulations, including its set-up, the numerical modeling, and various aspects of our post-processing analysis. Subsequently, we focus on the primary simulation (\pf{}). We discuss basic properties of \pf{} galaxies, including various galaxy-scaling relations, in section \ref{sect:PropGals}. Subsequently, in section \ref{sect:CosmicEvolution}, we analyze the evolution of the cosmic star formation rate density, the cosmic gas density, and the column density distribution function of atomic and molecular hydrogen. We summarize our findings in section \ref{sect:Summary}.

\section{Methodology}
\label{sect:Simulations}

\subsection{Initial conditions}
\label{sect:ICs}
 
In contrast to previous FIRE simulations, the \fb{} suite does not use the zoom-in approach to study galaxy evolution but rather it simulates gas, stars, and dark matter in a cubic cosmological volume of $V=(15\,{\rm cMpc}\,h^{-1})^3\sim{}(22.1\,{\rm cMpc})^3$ with periodic boundary conditions.
Initial conditions for all simulations in the \fb{} suite were created with the MUlti-Scale Initial Conditions tool ({\sc MUSIC})\footnote{\url{www-n.oca.eu/ohahn/MUSIC}} \citep{Hahn2011}. Cosmological parameters were taken from Planck-2015 cosmic microwave background measurements combined with baryon acoustic oscillation data as well as supernova and cepheid observations, see \citep{PlanckCollaboration2015a}: $\Omega_{\rm m}=0.3089$, $\Omega_\Lambda=1-\Omega_{\rm m}=0.6911$, $\Omega_{\rm b}=0.0486$, $h=0.6774$, $\sigma_8=0.8159$, $n_{\rm s}=0.9667$. Transfer functions for baryons, cold DM, and total matter were calculated for the same cosmology via the Code for Anisotropies in the Microwave Background ({\sc CAMB})\footnote{\url{camb.info}} \citep{Lewis2000} with $z_{\rm init}=120$ as starting redshift.

The specific initial conditions for the \fb{} suite were chosen by first running a suite of 27 low-resolution ($128^3$) collision-less $N$-body simulations of the chosen volume. Subsequently, one of the boxes was selected and corresponding higher-resolution initial conditions with and without baryonic matter were created. The objective of this manual selection was to obtain a realization of the halo mass function that is close to average for most redshifts. In addition, the selection was weighted towards boxes that do not contain a halo of exceptionally high mass at $z=0$ to avoid the associated higher computational cost and to reduce cosmic variance.

\subsection{Gravity and baryonic physics}
\label{sect:BaryonicPhysics}

The selected cosmological volume was evolved down to $z=0$ both with and without baryonic physics with the combined hydrodynamics and gravity solver {\sc gizmo}\footnote{\url{http://www.tapir.caltech.edu/~phopkins/Site/GIZMO.html}} \citep{Hopkins2015a}. {\sc gizmo} calculates gravitational forces between particles with a heavily modified version of the tree gravity solver of GADGET-3 \citep{Springel2005a, Springel2008} and it models hydrodynamical processes with the meshless-finite-mass (MFM) method \citep{Hopkins2015a}.

Baryonic processes, such as gas cooling and heating, star formation, and stellar feedback, are accounted for via the FIRE-2 physics model \citep{Hopkins2018}. Supermassive black holes and AGN feedback are not included, see below. We now briefly review the most important details of the FIRE-2 model.

The temperature of the gas is calculated over the $T\sim{}10-10^{10}$ K range by modeling free-free, Compton, photo-electric, photo-ionization, metal-line, molecular, fine-structure, dust collisional, and cosmic ray heating (but not cosmic ray transport) and/or cooling processes both from local sources and from a redshift dependent, spatially uniform ultraviolet background \citep{Faucher-Giguere2009}. Ionization states and cooling rates of Hydrogen and Helium are calculated following \cite{Katz1996a} with the fitting functions by \cite{Verner1996}. The simulations follows 15 species (H, He, C, N, O, Ne, Mg, Si, S, Ca, Fe, and 4 tracker species for $r$-process elements) and includes sub-grid metal diffusion from unresolved turbulence \citep{Su2017a, Escala2018}. Each gas particle starts with a metallicity of $2\times{}10^{-6}$, i.e., about $10^{-4}$ solar. Metal cooling uses the rates by \cite{Wiersma2009b} for high temperature gas ($>10^4\,{\rm K}$) and pre-tabulated rates calculated with CLOUDY \citep{Ferland1998} at low temperatures ($\leq{}10^4\,{\rm K}$). Self-shielding from both local sources and the cosmic UV background is accounted for via a Sobolev-length approximation based on the density gradient calibrated on radiative transfer experiments \citep{Gnedin2009a, Faucher-Giguere2010, Rahmati2013a}.

Star formation takes place in self-gravitating, dense ($n\geq{}300$ cm$^{-3}$ for \pf, see Table~\ref{tab:Sims}), Jeans unstable, molecular (self-shielding) gas with a 100\% efficiency per local free-fall time. The molecular-to-neutral gas ratio is calculated via an analytic model \citep{Krumholz2008a, Krumholz2009a, McKee2010} assuming photo-dissociation and two-phase equilibrium. This model requires as inputs the metallicity $Z$ and the dust optical depth for Lyman-Werner photons $\tau$, see \cite{Krumholz2011c}. The metallicity is known for each particle and the dust optical depth is estimated via a local Sobolev-length approximation. Specifically,  $\tau = 434.8\,{\rm cm}^2\,{\rm g}^{-1}\Sigma_{\rm gas}[0.1 + Z/0.02]$ where $Z$ is the metallicity and $\Sigma_{\rm gas}=\rho[d+\rho/\vert{}\vec{\nabla}\rho\vert{}]$ is the gas mass surface density. Furthermore, $d$ is the inter-particle separation which is closely related to the kernel length of the given gas particle \citep{Hopkins2015a, Hopkins2018}.

Stellar feedback includes energy, momentum, mass, and metal injections from supernovae (type II and type Ia) and stellar winds (OB and AGB stars). The ejecta energy per supernova is $10^{51}\,{\rm erg}$. Most feedback quantities are taken from tabulated stellar population models ({\sc Starburst99}; \citealt{Leitherer1999a}) for a \cite{Kroupa2001} initial stellar mass function (IMF). In addition, SN Ia rates are taken from \cite{Mannucci2006} and yields from \cite{Iwamoto1999c}. SN II yields are from \cite{Nomoto2006} and yields for OB/AGB stars follow \cite{Wiersma2009}. Radiative feedback in the form of photo-ionization and photo-electric heating as well as radiation pressure is also included. Radiative transfer effects are accounted for in the Locally Extincted Background Radiation in Optically thin Networks (LEBRON) approximation \citep{Hopkins2012b, Hopkins2014, Hopkins2018, Hopkins2019a}.

\begin{table*}
\begin{tabular}{lccccccccccc}
Name & Comment & N & $z_{\rm final}$ & $L$ & $n_{\rm SF}$ & $m_{\rm b}$ & $m_{\rm DM}$ & $d_{\rm gas, SF}$ & $\epsilon_{\rm gas, min}$ &  $\epsilon_{\rm star}$ & $\epsilon_{\rm DM}$ \\
           &  &   &                       & (cMpc)  &   (cm$^{-3}$) & ($10^4$  $M_\odot$)  &  ($10^5$ $M_\odot$)    &   (pc)               &  (pc)                       & (pc)   & (pc)  \\ \hline
{\bf FB1024} & {\bf \pf} & ${\bf 2\times{}1024^3}$ &  {\bf 0} & {\bf 22.1} & {\bf 300} & {\bf 6.26} & {\bf 3.35} &  {\bf 20.4} & {\bf 1.5} & {\bf 12} & {\bf 80}  \\
FB512 & lower res. re-run & $2\times{}512^3$ &  0 & 22.1 & 100 & 50.1 & 26.8 & 58.8 & 4 & 32 & 160  \\
FB256 & lower res. re-run & $2\times{}256^3$ &  0 & 22.1 & 10 & 401 & 215 & 253 & 16 & 128 & 320 \\ \hline
FB1024-DM & \pfdm & $1024^3$ &  0 & 22.1 & - & - & 3.98 & - & - & - & 80  \\
FB512-DM & lower res. $N$-body & $512^3$ &  0 & 22.1 & - & - & 31.8 & - & - & - & 160  \\
FB256-DM & lower res. $N$-body & $256^3$ &  0 & 22.1 & - & - & 255 & - & - & - & 320  \\
\hline
\end{tabular}
\caption{The FIREbox simulation suite. A systematic name and a short description for each run are provided in the first two columns. Columns three to six list the number of particles at the start of each simulation, the redshift reached by each simulation, the box size, and density threshold for star formation. The final six columns provide the masses of baryonic (gas and star) particles, the masses of dark matter (DM) particles, the inter-particle spacing of gas particles at the star formation threshold, the minimum gravitational softening length of gas particles, and the gravitational softening lengths of star and DM particles. For comparison with the literature, the force resolution is stated in equivalent Plummer softening lengths. The corresponding spline softening lengths are larger by a factor of $\sim{}1.4$. For every hydrodynamical simulation, there is a corresponding collisionless $N$-body simulation with particle masses $m=m_{\rm b} + m_{\rm DM}$ and gravitational softening lengths $\epsilon=\epsilon_{\rm DM}$. This suite is complemented with a higher resolution collisionless simulation FB2048-DM, see \protect\cite{Lazar2021}. The main focus of the present work is the FB1024 hydrodynamical simulation (\pf) listed in the top row.}
\label{tab:Sims}
\end{table*}

None of the current \fb{} runs include a model for AGN feedback. We plan to add cosmic ray physics \citep{Chan2019, Hopkins2019} and AGN feedback \citep{Wellons2022} in future \fb{} simulations to explicitly study the differential impact of these additional physical processes.

\subsection{Numerical resolution}

In all runs, gravity is softened with a cubic spline kernel. 
The force resolution of gas particles is adaptive and set to the gas inter-particle spacing $h = (m_{\rm b}/\rho_{\rm b})^{1/3}$ subject to a lower limit ($\epsilon_{\rm gas, min}$). This lower limit is chosen such that the highest gravitationally-resolved gas density $n^{\rm max}=m_{\rm b}/(\epsilon_{\rm gas, min})^3/m_{\rm H}$ exceeds the star formation threshold density $n_{\rm SF}$ by a factor of $\sim{}1000$, see \cite{Hopkins2018}. 
The force softenings of star and dark matter (DM) particles are non-adaptive. The softening length of star particles was chosen to be similar to the softening length of gas particles at the star formation threshold. Newly formed star particles have thus a similar softening length as the gas particles that spawned them. The Plummer equivalent softening length of DM particles is set to $\sim{}20\,{\rm pc}\,(m_{\rm DM} / 5000 M_\odot)^{1/3}$ to avoid over-softening of the central DM halo profile while also minimizing $N$-body relaxation due to particle scattering \citep{Hopkins2018}. The value of $\epsilon_{\rm gas, min}$ and the softening lengths of star and DM particles are kept fixed in physical (comoving) coordinates at $z\leq{}9$ ($z\geq{}9$).

\pf{} (FB1024), the primary simulation discussed in this paper, contains $1024^3$ gas and $1024^3$ DM particles at the starting redshift with masses $m_{\rm b}=6.3\times{}10^4$ $M_\odot$ and $m_{\rm DM}=3.3\times{}10^5$ $M_\odot$, respectively. A new star particle inherits the mass of the gas particle from which it was created. However, as a result of supernova explosions and stellar winds, star particles lower their mass over time to $\sim{}0.7m_{\rm b}$. The mass resolution in \pf{} is $\approx8\times$ lower than FIRE zooms of Milky-Way analogs \citep[e.g.,][]{Wetzel2016,Hopkins2018}. The minimum gas softening length (Plummer equivalent) is $\epsilon_{\rm gas, min}=1.5$ pc. A more representative measure of the spatial resolution of hydrodynamical processes in the ISM is the inter-particle spacing of gas particles eligible for star formation ($\lesssim{}20$ pc in \pf). Star particles (DM particles) have a Plummer equivalent softening length of $\epsilon_{\rm star}=12$ pc ($\epsilon_{\rm DM}=80$ pc). Mass and force resolution of the FB512 (FB256) runs are correspondingly lower, see Table \ref{tab:Sims}.
The completion of FIREbox required approximately 5 million compute core hours and a wall-clock time of about 3 months.

\begin{figure*}
\begin{tabular}{cc}
\includegraphics[width=85mm]{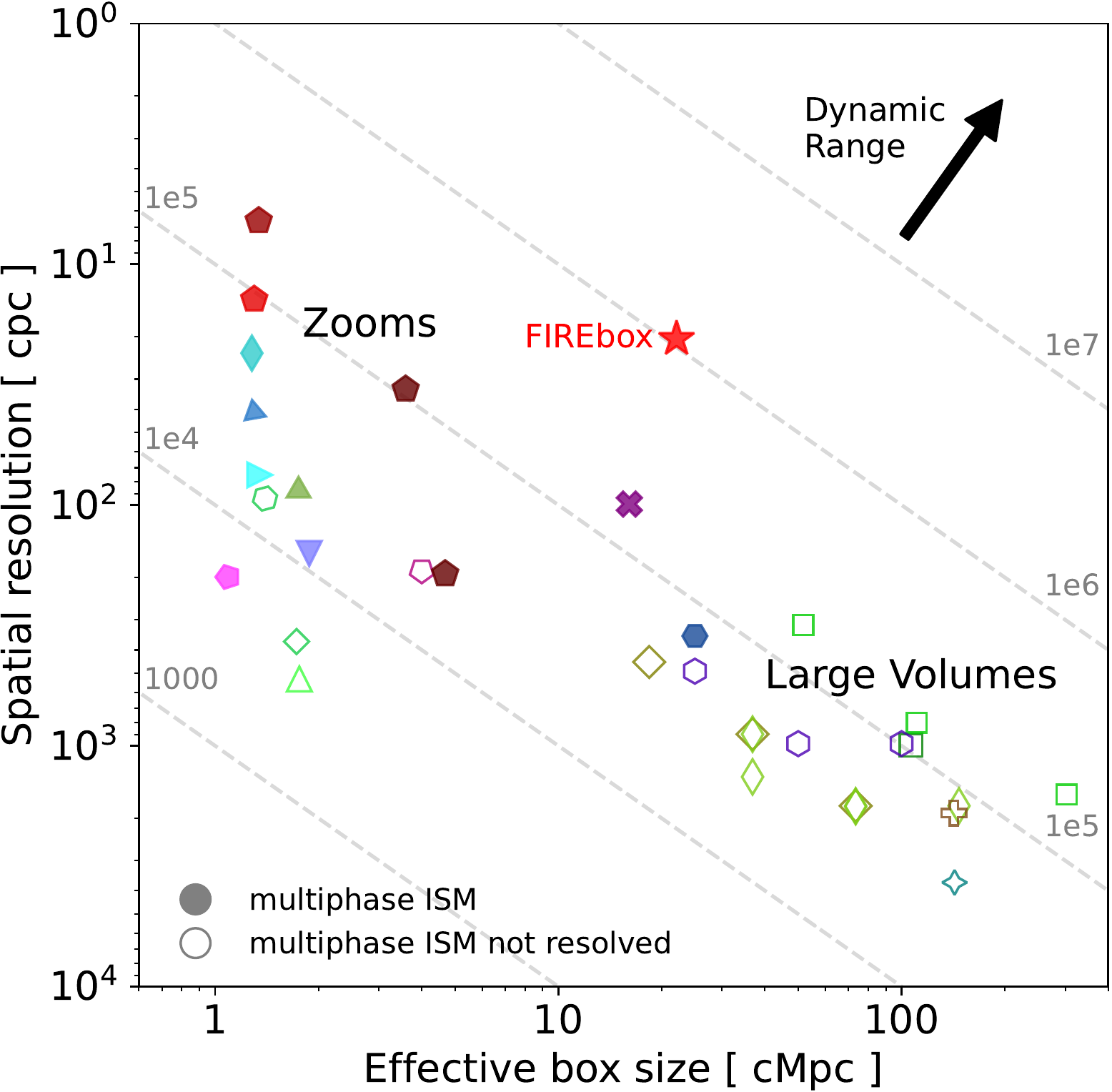} &
\includegraphics[width=85mm]{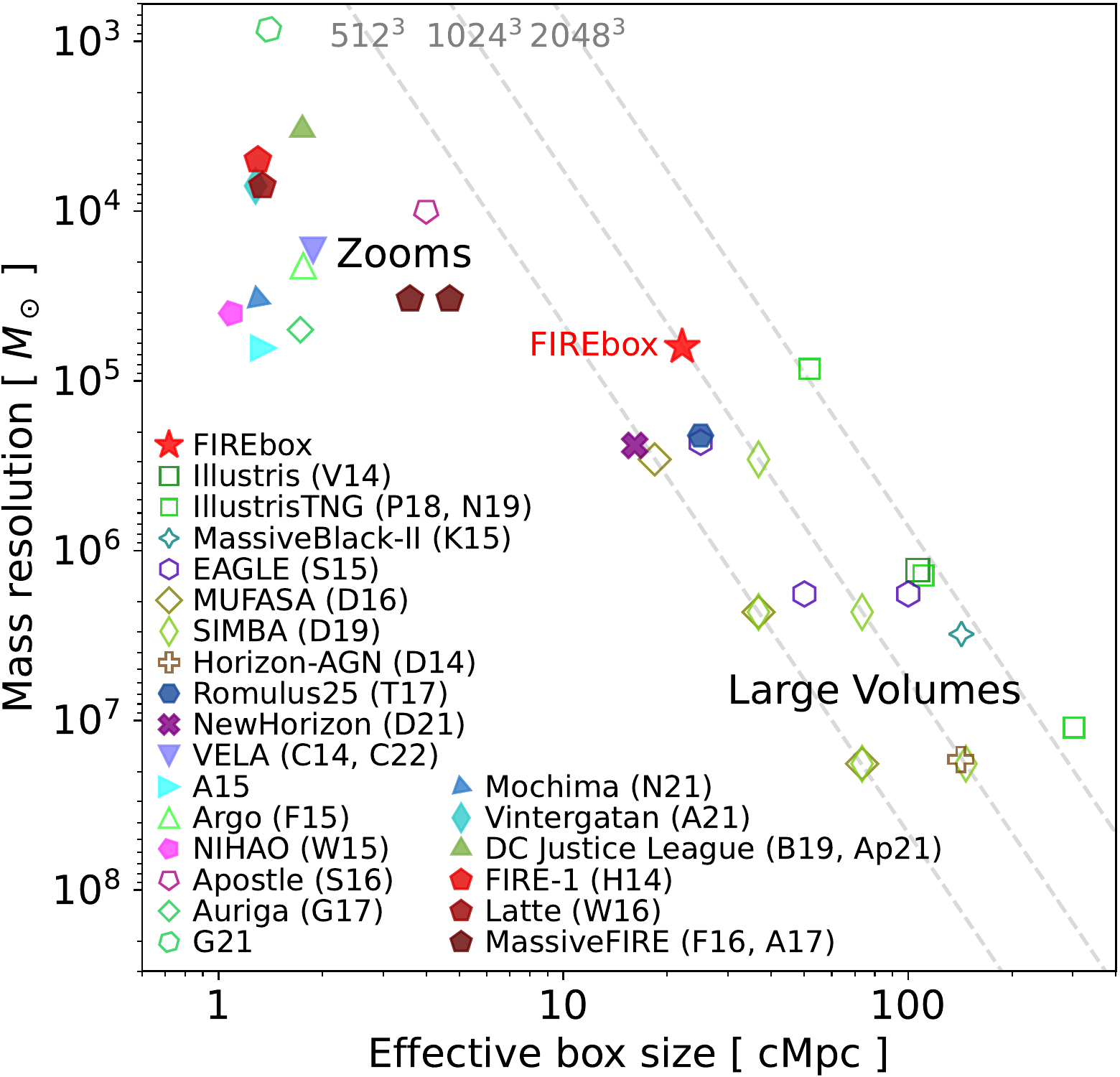} 
\end{tabular}
\caption{Hydrodynamic resolution and box size of FIREbox compared with a representative selection of contemporary cosmological galaxy formation simulations from \protect\citealt{Vogelsberger2014} (V14), \protect\citealt{Pillepich2018a} (P18), \protect\citealt{Nelson2019} (N19), \protect\citealt{Khandai2015} (K15), \protect\citealt{Schaye2015} (S15), \protect\citealt{Dave2016, Dave2019} (D16, D19), \protect\citealt{Dubois2014} (D14), \protect\citealt{Tremmel2017} (T17), \protect\citealt{Dubois2021} (D21), \protect\citealt{Ceverino2014, Ceverino2022} (C14, C22), \protect\citealt{Agertz2015, Agertz2021} (A17, A21), \protect\citealt{Feldmann2015} (F15), \protect\citealt{Wang2015} (W15), \protect\citealt{Sawala2016} (S16), \protect\citealt{Grand2017, Grand2021} (G17, G21), \protect\citealt{Nunez-Castineyra2021} (N21), \protect\citealt{Bellovary2019} (B19), \protect\citealt{Applebaum2021} (Ap21), \protect\citealt{Hopkins2014} (H14), \protect\citealt{Wetzel2016} (W16), \protect\citealt{Feldmann2016} (F16), and \protect\citealt{Angles-Alcazar2017a} (A17).
In each panel, \pf{} is shown by a red star.
(Left) Spatial resolution vs effective box size. The former refers to the \emph{typical} resolution in the star forming gas of a given simulation. Specifically, for particle-based hydrodynamics codes, the spatial resolution is defined as the larger of the minimum gravitational softening length of gas particles and the inter-particle distance at the star formation threshold density. For grid-based codes with a Lagrangian refinement strategy, the spatial resolution is defined similarly based on the minimum cell size and the gas density at the star formation density threshold (see text). The effective box size equals the comoving box length for cosmological volume simulations. For zoom-in simulations, the effective box size is set to 5 times the comoving virial radius of the largest halo in the zoom-in region at the final snapshot. Not shown are zoom-in simulations that do not resolve at least one Milky-Way mass halo or that have a baryonic mass resolution worse than $10^5$ $M_\odot$. Diagonal dashed lines show the resolved dynamic range of a simulation, i.e., the ratio between the effective box size and the spatial resolution (both in comoving units). \pf{} is the first cosmological galaxy formation simulation run to $z=0$ with a dynamic range of more than one million.
(Right) Baryonic mass resolution vs effective box size. Diagonal dashed lines show the approximate number of hydrodynamic resolution elements in the simulation volume. 
In each panel, filled (empty) symbols indicate simulations with (without) a resolved multiphase interstellar medium, see text. \pf{} is able to capture the multiphase structure of gas in and around galaxies in a fully cosmological context and across cosmic history.}
\label{fig:Resolution}
\end{figure*}

\subsection{Comparison with state-of-the-art galaxy formation simulations}
\label{sect:CompResolution}

Fig.~\ref{fig:Resolution} compares the hydrodynamic mass and spatial resolution of \pf{} with a compilation of cosmological galaxy formation simulations of intermediate-to-high mass galaxies reaching $z<2$. Zoom-in simulations that specifically target very low mass galaxies (e.g., \citealt{Fitts2017, Revaz2018, Wheeler2019, Munshi2019}) can reach a higher numerical resolution and are not included in this comparison.
While not an exhaustive list, the compilation includes the state-of-the-art in galaxy formation simulations and it covers a range of hydrodynamics solvers, such adaptive mesh refinement (AMR), smoothed particle hydrodynamics (SPH), moving mesh hydrodynamics, and mesh-less hydrodynamics as well as both zoom-in runs and large-volume simulations. Specifically, the compilation includes cosmological volume simulations from the Illustris \citep{Vogelsberger2014, Genel2014}, IllustrisTNG \citep{Pillepich2018a, Nelson2019}, EAGLE \citep{Schaye2015}, MUFASA \citep{Dave2016}, SIMBA \citep{Dave2019}, MassiveBlack-II \citep{Khandai2015}, Romulus \citep{Tremmel2017}, and Horizon-AGN \citep{Dubois2014} projects, representative zoom-in simulations from the FIRE project \citep{Hopkins2014, Hopkins2018}, such as Latte \citep{Wetzel2016} and MassiveFIRE \citep{Feldmann2016, Feldmann2017b, Angles-Alcazar2017a}, and zoom-in simulations by \cite{Agertz2015} as well as from the Apostle \citep{Sawala2016}, Argo \citep{Feldmann2015}, Auriga \citep{Grand2017, Grand2021}, DC Justice League \citep{Bellovary2019, Applebaum2021}, Eris \citep{Guedes2011}, Mochima \citep{Nunez-Castineyra2021}, NewHorizon \citep{Dubois2021}, NIHAO \citep{Wang2015}, VELA \citep{Ceverino2014}, and Vintergatan \citep{Agertz2021} projects.

Given the intrinsic ambiguity in defining mass and spatial resolution across such a variety of models, we adopt the following operational definitions. For particle-based hydrodynamics codes, the mass resolution is defined as the typical gas particle mass in the simulation. Adopting a more physics-based definition, e.g., using the minimal resolved Jeans mass, would favor even more simulations, such as \pf{}, that directly model the multiphase ISM (shown by filled symbols) compared with simulations that do not (empty symbols), i.e., those that prevent gas from cooling to low temperatures ($T<1000$ K) and/or those that model the ISM with an effective equation of state resulting in highly pressurized and comparably smooth gas disks.
The spatial resolution is set to the larger of the gas inter-particle spacing $d_{\rm SF}$ at the star formation threshold, $d_{\rm SF}=74\,{\rm pc}(m_{\rm b, 4}/n_{\rm SF, 0})^{1/3}$ with $m_{\rm b, 4}=m_{\rm b}/(10^4\,M_\odot)$ and $n_{\rm SF, 0}=n_{\rm SF}/{\rm cm}^{-3}$, and the minimum spline gravitational softening length of gas particles. While dynamical processes may be resolved on scales smaller than $d_{\rm SF}$, those scales are affected by the physics of sink particle formation.
For grid-based simulations with a quasi-Lagrangian refinement scheme, we adopt $m_{\rm b}=\Omega_{\rm b} / (\Omega_{\rm m} - \Omega_{\rm b})m_{\rm DM}$ as hydrodynamic mass resolution, while the spatial resolution is defined as the larger of $d_{\rm SF}$ and the minimum cell size.
In each case, we calculate the spatial resolution in comoving pc at the redshift of the final simulation snapshot. The spatial resolution can substantially exceed the minimum cell size or the minimum gravitational softening length of gas particles, e.g., $\sim{}20$ pc vs 1.5 pc for FIREbox, $\sim{}100$ pc vs 34 pc for NewHorizon, and $\sim{}300$ pc vs 74 pc for TNG-50.

The figure also shows the effective box size of the simulations. The effective box size equals the comoving box length for cosmological volume simulations. For zoom-ins, it is set to 5 times the comoving virial radius of the most massive halo in the highest resolution region at the final simulation redshift to approximately reproduce the typical extent of the zoom-in region uncontaminated by low-resolution dark matter particles. Only the largest simulation is considered when calculating the effective box size for simulations suites consisting of multiple independent runs of similar resolution, such as Apostle, Auriga, or MassiveFIRE. Zoom-in simulations that do not resolve at least one Milky-Way mass halo or that have a baryonic mass resolution worse than $10^5$ $M_\odot$ are not included in the figure. The ratio between the effective box size and the spatial resolution of a simulation defines its dynamic range.

\pf{} opens a new frontier in studying the evolution of galaxies with hydrodynamical simulations given its unique combination of high numerical resolution (comparable to state-of-the-art zoom-ins) and accurate physical modeling in a cosmological volume of $(22.1\,{\rm cMpc})^3$. Specifically, \pf{} is able to both directly resolve the thermodynamic state of the ISM (by enabling self-consistent cooling down to $\sim{}10-20$ K) and to accurately account for multiple stellar feedback channels tied to stellar population synthesis models (see section \ref{sect:BaryonicPhysics}). The underlying FIRE-2 physics model has been employed previously in zoom-in simulations to study the multiphase nature of the ISM,  e.g., the overall properties of massive giant molecular clouds \citep{Benincasa2020, Guszejnov2020} and the vertical pressure profiles and scale heights of galactic disks \citep{Gurvich2020}. With \pf{}, we can study galaxies and their ISM with larger, representative samples from a contiguous cosmological volume enabling a proper statistical analysis and a study of cosmological environments. While the dynamic range of \pf{} ($\gtrsim{}10^6$) already exceeds significantly those of contemporary galaxy formation simulations, higher-resolution follow-up simulations combined with dedicated zoom-ins promise to further extend this frontier towards larger samples of better resolved galaxies.

\subsection{Simulation output}

The properties of gas, star, and dark matter particles are saved as Gadget HDF5 files in (semi-)regular intervals for subsequent analysis. All \fb{} runs, except FB2048-DM, use 1201 save-points that are approximately equally spread in cosmic time between $z_{\rm init}=120$ and $z_{\rm final}=0$, resulting in a close to $11$ Myr average save-point intervals. For FB2048-DM, fewer save-points are used at $z<2$ to mitigate its high storage footprint. Furthermore, three out of every four save-points are stored at reduced resolution (`snipshots') to reduce the overall storage cost of the simulation suite, see below. In addition, all save-points (both snipshots and regular snapshots) are stored in gzip compressed format. 

A save-point is stored either as a level 0 snapshot, a level 1 snapshot, or a level 2 `snipshot'. Level 0 snapshots are compressed in a loss-less manner but are otherwise identical to the original HDF5 output files. The compression can reduce the file size by up to a factor of 2. Level 1 snapshots are identical to level 0 snapshots except that the abundances of individual elements (but not the total metallicity) are stored at reduced precision (1 byte) in hydrodynamical simulations. Level 2 snipshots differ significantly from snapshots. Densities, electron abundances, neutral hydrogen abundances, helium abundances, total metallicities, internal energies, softening lengths, and velocities of gas particles are stored at reduced resolution (typically as a half-precision float and after a log transform for non-negative fields). Additionally, the individual abundances of elements heavier than Helium are dropped. Gas particle masses and coordinates are kept at full resolution, however. In addition, star particles keep the same information as for level 1 snapshots. Dark matter particles are downsampled randomly by a factor 8 with the help of a scrambled Xorshift generator \citep{Marsaglia2003, Vigna2016} such that the same particles are removed (or kept) in all snipshots. The storage footprint of a level 0 snapshot (level 1 snapshot, level 2 snipshot) of FIREbox at $z\sim{}0$ is 207 GB (97 GB, 42 GB).

\subsection{Halo and galaxy catalogs}
\label{sect:halos}

We identify dark matter halos and catalog their various properties, including halo positions, masses, radii, and whether or not a given halo is a sub-halo or a main halo, with the help of the AMIGA Halo finder (AHF; \citealt{Gill2004, Knollmann2009}). We include only halos with at least 100 particles in the subsequent analysis. Halo masses ($M_{\rm halo}$) and radii ($R_{\rm vir}$) are calculated based on the virial overdensity definition \citep{Bryan1998} and include baryonic matter and halo sub-structures. Growth histories for individual halos are constructed with the AHF MergerTree tool by linking halos in subsequent snapshots via the identification numbers of their DM particles.

Intuitively, `sub-halos' are DM halos that reside within other DM halos. More quantitatively, AHF identifies a DM halo of radius $R$ as a `sub-halo' of another, more massive DM halo of radius $R^\prime$ if the distance between the two halos is less than $R^\prime+0.5\,R$. Halos that are not sub-halos are `main halos'. Galaxies in sub-halos are called satellite galaxies, while the primary galaxy of a main halo is called its central galaxy. 

With the help of the AHF particle files, we identify both the direct host halo (which can be a sub-halo or a main halo) for each particle as well as the main halo containing the particle. Subsequently, we use this information to calculate a variety of particle based properties, e.g., stellar masses, star formation rates (SFRs), and gas masses in various three-dimensional spherical apertures, and store them in HDF5 files for subsequent analysis. Halo properties are measured within a sphere of radius $R_{\rm vir}$, while a smaller radius $R_{\rm g}$ (see below) is used to measure galaxy properties. For sub-halos, $R_{\rm vir}$, as reported by AHF and used below, refers to the smaller of the virial and the tidal radius.

The total radius $R_{\rm g}$ and the stellar half mass radius $R_{\rm half}$ of galaxies are defined using two different approaches based on the cumulative spherical stellar mass profile $M_{\rm star}(<R)$ and the virial radius. The first approach follows \cite{Hopkins2018}. Starting from an initial choice for $R_{\rm g}$ of $0.15\,R_{\rm vir}$, the half mass radius is computed as $M_{\rm star}(<R_{\rm half})=0.5M_{\rm star}(<R_{\rm g})$ and the total radius is updated as $R_{\rm g}=3\times{}R_{\rm half}$. The latter steps are repeated until the relative change in $R_{\rm g}$ between one iteration and the next is less than $10^{-5}$. The second approach sets $R_{\rm g}=0.1\,R_{\rm vir}$ and subsequently calculates $R_{\rm half}$ from the stellar mass profile within $R_{\rm g}$. Unless stated otherwise, stellar masses, SFRs, and other properties of galaxies refer to integrated properties within $R_{\rm g}$ computed as in the first method.

\subsection{Gas fractions and temperatures}
\label{sect:gasmethod}

The mass  $m_{\rm gas}$ of each gas particle can be divided into the mass of ionized ($m_{\rm H_{II}}$), atomic ($m_{\rm H_I}$), and molecular hydrogen ($m_{\rm H_2}$) as well as the mass in Helium and in the various metals. The total hydrogen mass of a gas particle is $m_{\rm H}=m_{\rm H_I}+m_{\rm H_2}+m_{\rm H_{II}}=Xm_{\rm gas}$, where the hydrogen mass fraction $X$ can vary from particle to particle. The neutral hydrogen fraction $f_{\rm H_I+H_2}=(m_{\rm H_I}+m_{\rm H_2})/m_{\rm H}$ of each particle is calculated during the run-time of the simulation as described in \cite{Hopkins2018} and is provided in the simulation snapshots. At the level of individual gas particles, $f_{\rm H_I+H_2}$ is also the neutral gas fraction $(m_{\rm atm}+m_{\rm mol})/m_{\rm gas}$ provided we define atomic and molecular gas masses of particles as $m_{\rm atm}=m_{\rm H_I}/X$ and $m_{\rm mol}=m_{\rm H_2}/X$.
The molecular gas fraction $f_{\rm H_2}=m_{\rm H_2}/m_{\rm H}=m_{\rm mol}/m_{\rm gas}$ of each particle, which is also calculated at run-time, is not part of the simulation output, however. We thus recalculate it based on the snapshot data. Specifically, we first calculate the molecular-to-neutral gas ratio $f_{\rm H_2}/f_{\rm H_I+H_2}$ based on its dust optical depth and metallicity following the same approach \citep{Krumholz2011c} as for the run-time calculation described above, see section \ref{sect:BaryonicPhysics}. Given the various assumptions entering this approach, the resulting estimate of the molecular-to-neutral gas ratio should be understood as an approximation that may be highly inaccurate under certain conditions, e.g., at metallicities below $0.01\,Z_\odot$. We then convert the molecular-to-neutral gas ratio to the molecular gas fraction by multiplying the former with $f_{\rm H_I+H_2}$. The atomic gas fraction of a particle $f_{\rm H_I}=m_{\rm H_I}/m_{\rm H}=m_{\rm atm}/m_{\rm gas}$ is calculated as $f_{\rm H_I+H_2}-f_{\rm H_2}$.

Gas temperatures are re-calculated from the internal energy per unit mass $\epsilon$, electron abundance $f_{\rm e}=n_{\rm e}/n_{\rm H}$, Helium abundance $Y$, metallicity $Z$, all of which are provided in the simulation output and from the molecular gas fraction $f_{\rm H_2}$ calculated as described above. The gas temperature is given as $T = \epsilon (\gamma -1) \mu / k_{\rm B}$ with the mean molecular weight $\mu=m_{\rm H}/[X(1-0.5f_{\rm H_2}) + Y/4 + f_{\rm e}X + Z/16]$ and with $X=1-Y-Z$. \pf{} employs a floor in specific internal energy that amounts to a temperature floor of $\sim{}10$ K in atomic gas and $\sim{}18$ K in molecular gas.

\section{Properties of \fb{} galaxies{}}
\label{sect:PropGals}

\begin{figure*}
\begin{tabular}{c}
\includegraphics[width=140mm]{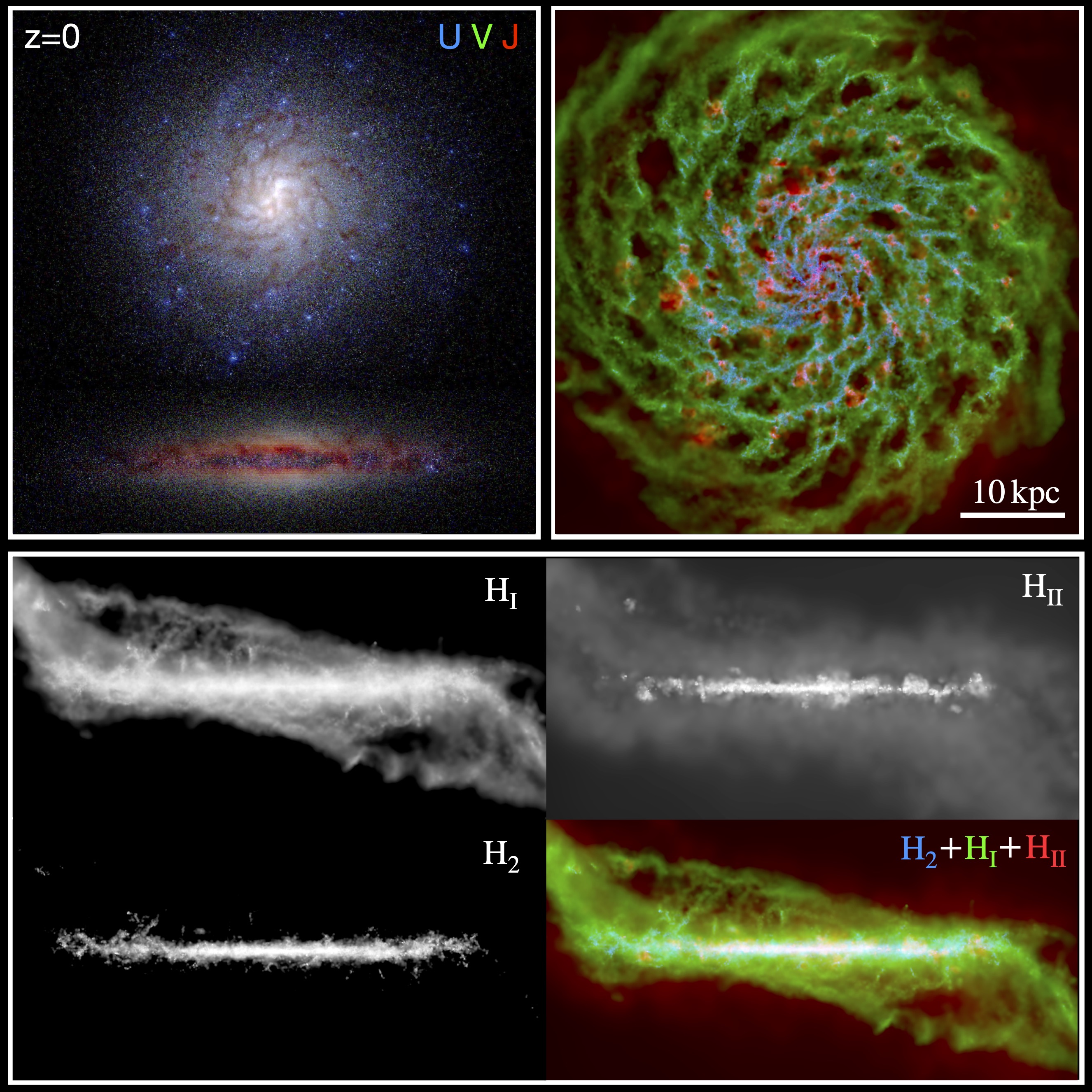}
\end{tabular}
\caption{Visualization of the multiphase structure of baryons in a Milky Way like galaxy at $z=0$ in \pf{}. (Top left) Color composite image in U (blue), V (green), and J (red) broad-bands, created with the radiative transfer code SKIRT \citep{Baes2011, Camps2015}, showing the stellar and dust components of the depicted galaxy in a face-on and edge-on view. This galaxy has an overall disky morphology. (Top right) Face-on, color composite image of the galaxy's molecular (${\rm H_2}$, blue), atomic (${\rm H_I}$, green), and ionized (${\rm H_{II}}$, red) hydrogen content. (Bottom) Edge-on view of the atomic, ionized, molecular, and combined hydrogen content (see legend). The neutral interstellar medium (ISM) resides in an extended, but thin gas disk with a complex internal structure (see text). The neutral ISM is embedded in a thick, but comparably smooth, disk of ionized gas. Ionized gas is also present in regions of various sizes within the plane of the ISM disk. The locations of these ionized regions often coincide with those of young stellar clusters shown in blue in the UVJ image. The ${\rm H_I}$ and ${\rm H_{II}}$ disks are strongly warped at large radii. All images have the same physical scale (50 kpc from left to right), see legend, and show quantities on a logarithmic stretch. The center of the galaxy is shifted vertically in the top left panel. Many of the shown features, e.g., some of the ${\rm H_2}$ spiral arms and ${\rm H_{II}}$ bubbles, are less than 100 parsecs across.}
\label{fig:Multiphase}
\end{figure*}

In this section, we analyze basic properties of \pf{} galaxies. Our main focus lies in comparing our simulation predictions to available observational data. We will demonstrate that many basic galaxy scaling relations predicted by the simulation, e.g., the relations between galaxy stellar mass and their star formation rates, gas content, and metallicity, agree reasonably well with observations. Other properties, such as the stellar mass functions (SMF) and the galaxy stellar mass -- halo mass relation (SHMR) are not reproduced as well. Here, the predictions of \pf{} are more in line with recent stellar mass estimates based on non-parametric panchromatic spectral energy distribution (SED) modeling.

Throughout this section, a \cite{Chabrier2003} IMF is adopted for all observational data. Specifically, we lower stellar masses and SFRs by 0.25 dex when converting from \cite{Salpeter1955} IMF to \cite{Chabrier2003} IMF (see, e.g., \citealt{Lee2006, Gallazzi2008, Herrmann2016}). We ignore the small shift between a \cite{Chabrier2003} IMF and the \cite{Kroupa2001} IMF adopted by \pf{}.

\subsection{The multiphase interstellar medium}

One of the main goals of the \fb{} project is to study the distribution of the various gas phases in and around galaxies at high spatial resolution. As such, it aims to provide a theoretical counterpart to the large number of observational efforts currently being undertaken to map the gas content of galaxies on sub-galactic (few hundreds of parsecs or better) scales, such as THINGS \citep{Walter2008}, LITTLE THINGS \citep{Hunter2012}, HI-MaNGA \citep{Masters2019}, ALMAQuest \citep{Lin2020}, PHANGS-ALMA \citep{Leroy2021a}, and PHANGS-MUSE \citep{Emsellem2021}.

We illustrate the ability of \pf{} to model and spatially resolve the multiphase ISM in Fig.~\ref{fig:Multiphase}. Here, we show gas maps as well as color-composite images of stellar light for a \pf{} galaxy at $z=0$. The halo mass of this chosen galaxy ($1.3\times{}10^{12}$ $M_\odot$) matches the estimated halo mass of the Milky Way (MW, \citealt{Bland-Hawthorn2016}). Overall, this galaxy is a fairly typical example of a MW analog in \pf{}. We will discuss the properties of MW analogs in \fb{} more generally in section \ref{sect:MWanalogs}.

\begin{figure*}
\begin{tabular}{c}
\includegraphics[width=160mm]{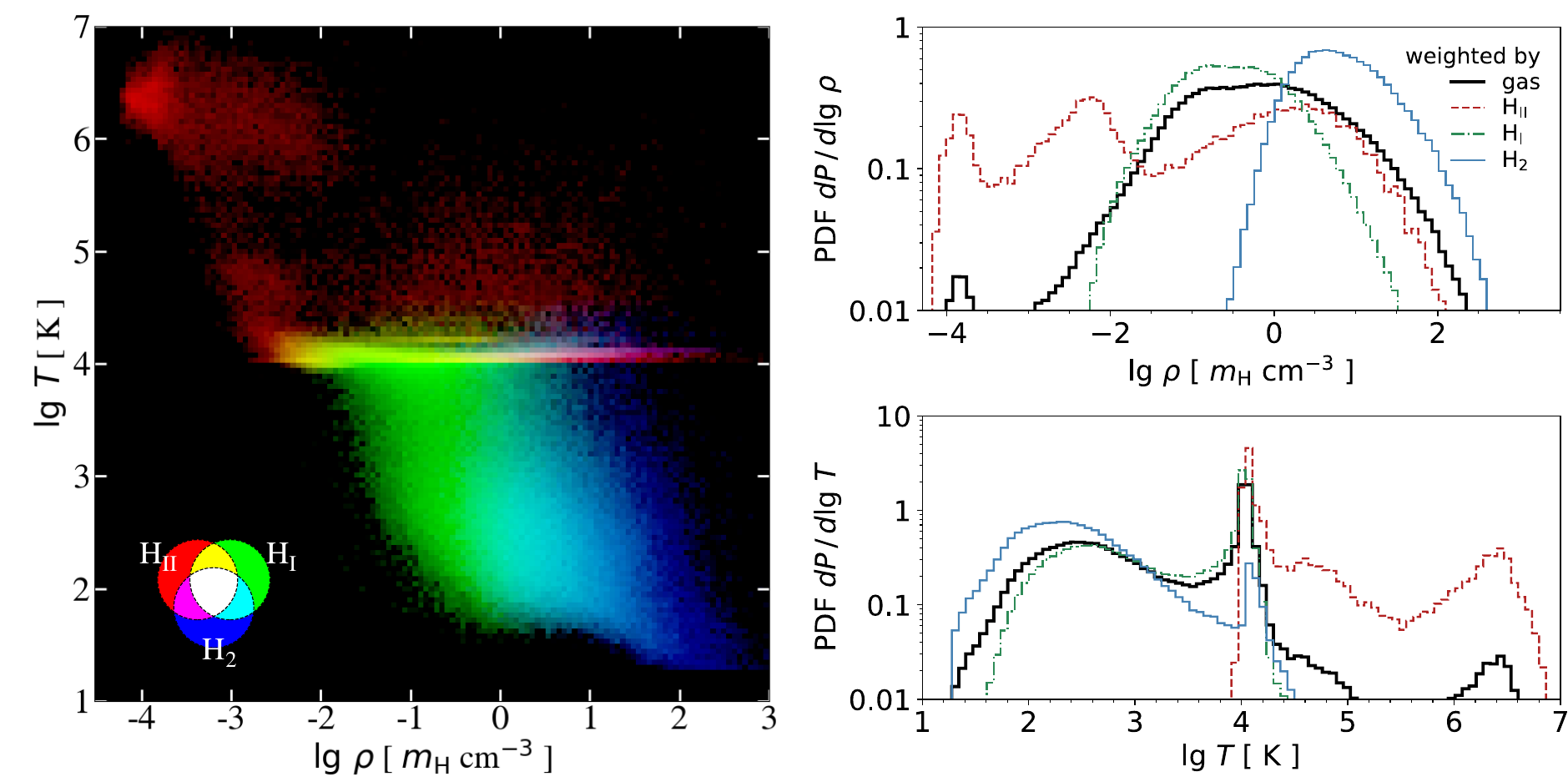}
\end{tabular}
\caption{Thermal properties of the interstellar medium (ISM) of a Milky Way like galaxy at $z=0$ in \pf{}. (Left) Phase diagram of hydrogen gas within a $0.1R_{\rm vir}\sim{}29$ kpc radius of the center of the galaxy. Densities ($\rho$) and temperatures ($T$) vary over many orders of magnitude ($\sim{}10^{-4}-10^3$ m$_{\rm H}$ cm$^{-3}$ and $10-10^{7}$ K) highlighting the computational challenge for galaxy formation simulations. The phase diagram is color-coded according to the hydrogen phase, see legend. The hydrogen phase is not uniquely determined by gas density and temperature alone. 
(Top right) Fraction of gas within $0.1R_{\rm vir}$ having density $\rho$ per unit $\lg{}\rho$. (Bottom right) Fraction of gas within $0.1R_{\rm vir}$ having temperature $T$ per unit $\lg{}T$. In both right hand panels, particle number fractions are weighted by gas mass (thick black solid line), ionized hydrogen mass (red dashed line), atomic hydrogen mass (green dot-dashed line), and molecular hydrogen mass (blue solid line). The small fraction of molecular gas with temperatures near $10^4$ K is an artifact of the approximate treatment of separating neutral gas into atomic and molecular components, see section \ref{sect:BaryonicPhysics}.
Neutral hydrogen consists of a combination of cold ($\sim{}100-1000$ K) and warm ($\sim{}10^4$ K) gas, while ionized gas consists of a hot, dilute ($\sim{}10^{-4}$ m$_{\rm H}$ cm$^{-3}$) phase filling most of the volume (the diffuse hot halo), a warm / hot, low density ($\sim{}10^{-3}-10^{-2}$ m$_{\rm H}$ cm$^{-3}$) phase which forms a disky layer around the neutral ISM disk (see Fig.~\protect\ref{fig:Multiphase}), and warm ionized, relatively dense gas located in the plane of the ISM disk.}
\label{fig:Multiphase2}
\end{figure*}

According to Fig.~\ref{fig:Multiphase}, this galaxy has a well-defined disk morphology. Face-on and edge-on images in U, V, and J broad-band filters, created with the help of the Monte Carlo radiative transfer code \texttt{SKIRT}\footnote{\url{http://www.skirt.ugent.be.}} \citep{Baes2011, Camps2015}, show young clusters of stars, patchy lanes of dust, and an underlying stellar disk that extends out to about 15-18 kpc. Face-on gas maps of the molecular, atomic, and ionized components of the ISM reveal a complex and intricate structure. Neutral hydrogen forms a relatively thin, but flocculent, disk with much of the molecular component residing in the inner, denser regions of the gas disk. Significant amounts of ${\rm H_I}$ gas can be found at large distances ($>25$ kpc) from the galaxy center, i.e., far beyond the extent of the stellar disk, see also \cite{Trapp2022}.

The vertical scale heights of the ${\rm H_I}$ and ${\rm H_2}$ disks at $R\sim{}8$ kpc for this simulated galaxy are approximately $200\pm{}50$ pc and $130\pm{}70$ pc when measured by fitting the vertical density profile in annulus sectors with a Gaussian \citep{Gensior2022}. These scale heights are comparable (within a factor of 2-3) with estimates for the Milky Way \citep{Bacchini2019}, M33 \citep{Combes2012}, and nearby star forming galaxies \citep{Bacchini2019a}. The ${\rm H_I}$ disk remains thin and regular out to about 20 kpc and shows warping at larger radii. 

Most of the ionized gas surrounding this galaxy is part of a diffuse, hot circum-galactic medium filling much of the volume of the DM halo. However, ionized gas can also be found in a puffed-up disky layer surrounding the neutral ISM, possibly a rotating cooling flow that replenishes the disk with gas \citep{Hafen2022}, as well as in a thin disk within the plane of the neutral gas disk, often near the locations of young star clusters.

Fig.~\ref{fig:Multiphase2} explores further the thermal properties and phase structure of the gas in the selected MW analog. The gas within $0.1R_{\rm vir}\sim{}29$ kpc varies broadly in density and temperature ($\sim{}10^{-4}-10^3$ m$_{\rm H}$ cm$^{-3}$ and $10-10^{7}$ K) and consists of ionized, atomic, and molecular phases. In this example galaxy, most of the hydrogen gas within $0.1R_{\rm vir}$ is atomic (70\%). Molecular and ionized hydrogen contribute at the 22\% and 8\% level, respectively. When split by temperature, gas with $T<6000$ K is predominantly neutral, while gas with $T>15000$ K is predominantly ionized. The ionized gas is made up of 3 sub-components: a hot, dilute ($\sim{}10^{-4}$ m$_{\rm H}$ cm$^{-3}$) phase filling most of the volume (the diffuse hot halo), a warm / hot, low density ($\sim{}10^{-3}-10^{-2}$ m$_{\rm H}$ cm$^{-3}$) phase which forms a smooth disky layer around the neutral ISM disk, and a warm ionized phase of relatively dense gas near the center plane of the ISM disk.

\subsection{The star forming sequence}
\label{sect:SFsequence}

\begin{figure*}
\begin{tabular}{cc}
\includegraphics[width=80mm]{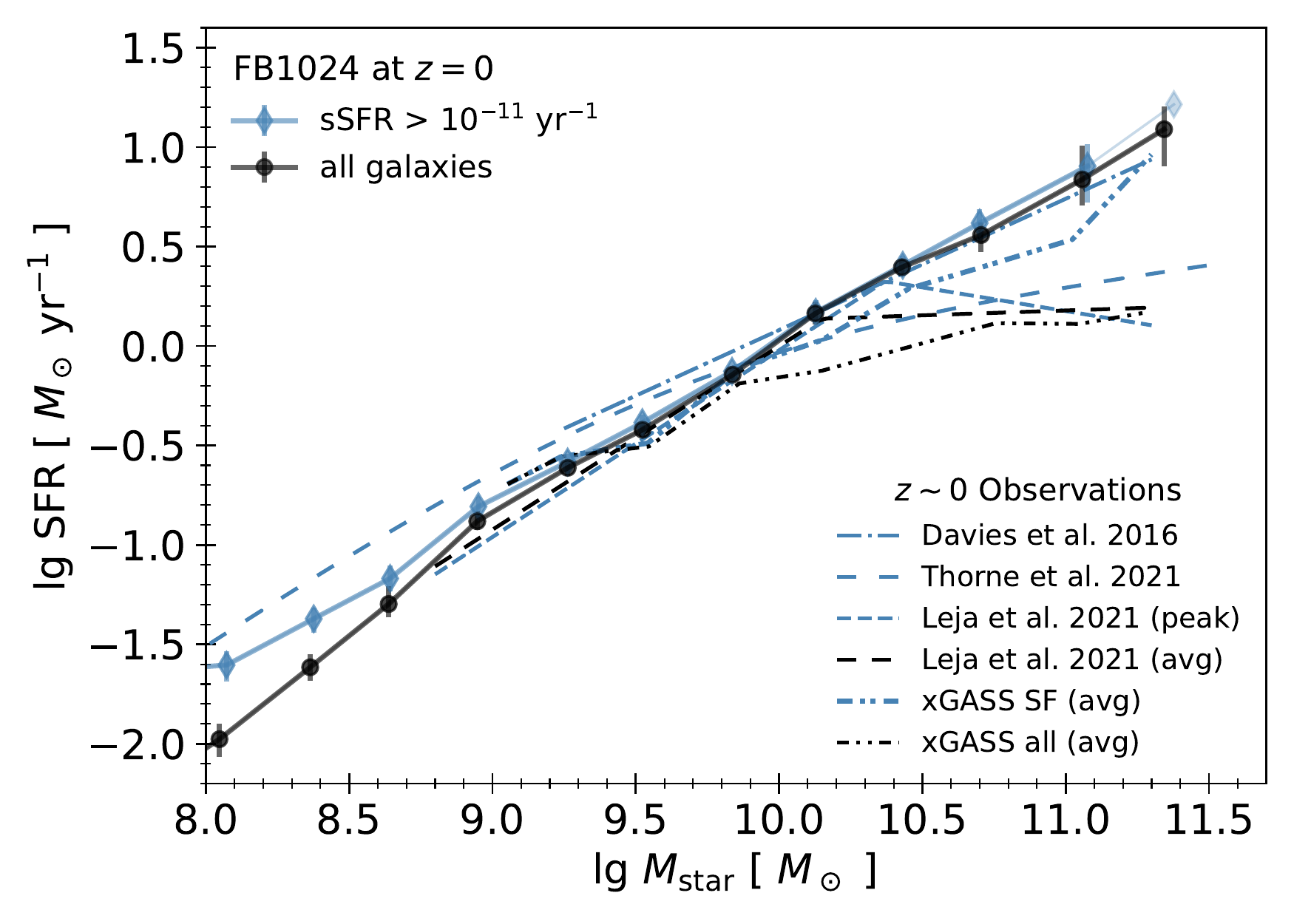} &
\includegraphics[width=80mm]{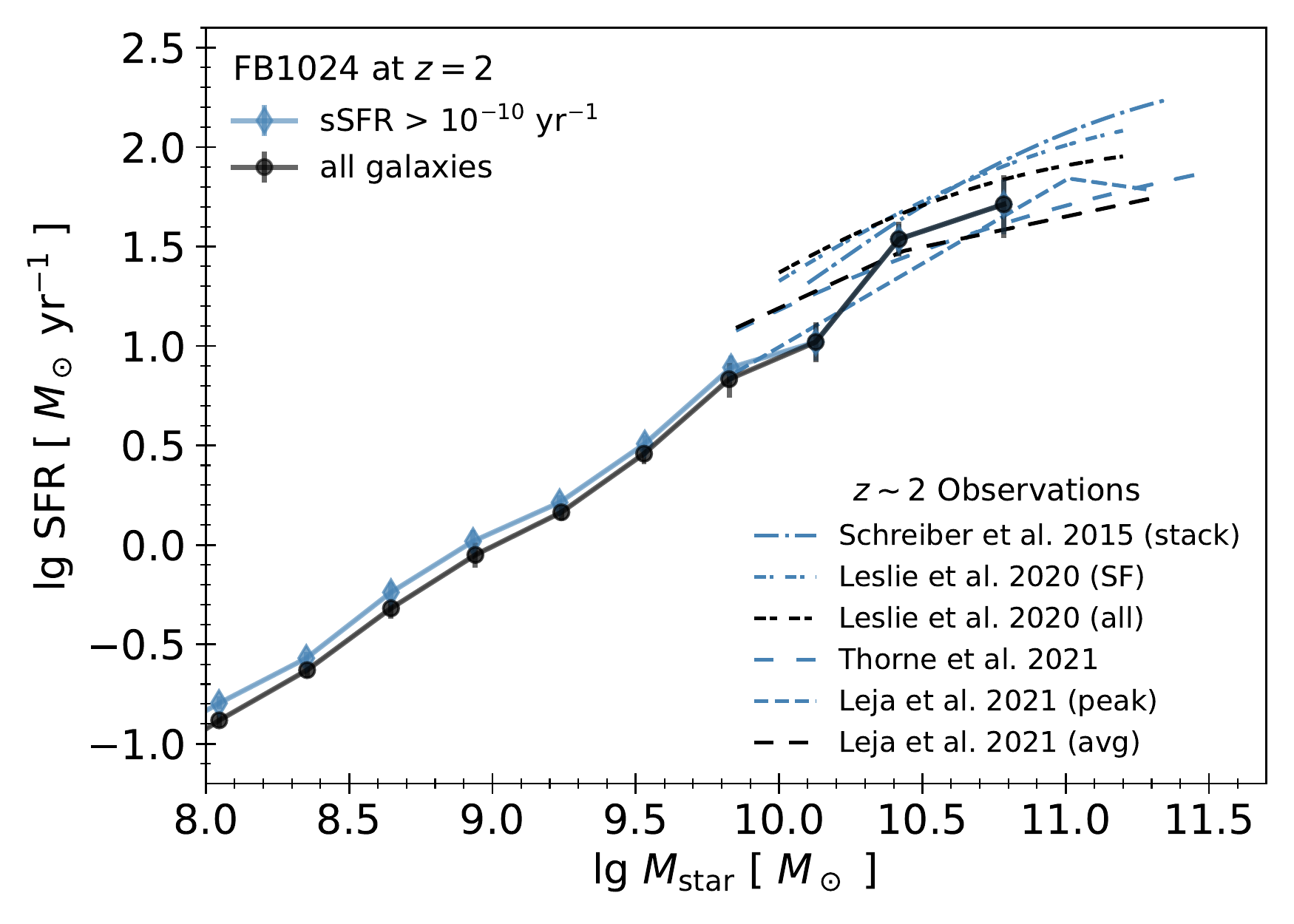}
\end{tabular}
\caption{Star forming sequence in \pf{} and in observations at $z=0$ (left) and $z=2$ (right). Symbols show the logarithm of the average SFR in bins of stellar mass for all galaxies (black circles) and for star forming galaxies (blue diamonds) in \pf{}. The latter population of galaxies is defined as having specific SFRs exceeding $10^{-11}$ yr$^{-1}$ at $z=0$ and $10^{-10}$ yr$^{-1}$ at $z=2$. SFRs are averaged over the past 20 Myr. Error bars refer to 16-84\% percentiles in each bin obtained via bootstrapping. Light shaded symbols without error bars indicate bins containing fewer than four galaxies. Double dot-dashed lines show the star forming sequence for a representative sample of $z\sim{}0$ galaxies from the xGASS survey \citep{Catinella2018} with updated stellar masses as presented in \protect\cite{Feldmann2020}. Dashed and dot-dashed lines show results of recents observational studies \citep{Schreiber2015b, Davies2016, Leslie2020, Thorne2021, Leja2021}, see legend. 
Stellar masses by \protect\cite{Leslie2020} are shifted by 0.2 dex to account for the known systematics of their stellar mass catalog. \pf{} predicts average SFRs of star forming galaxies in good agreement with observations.}
\label{fig:MainSequence}
\end{figure*}

SFRs and stellar masses of star forming galaxies are tightly correlated with a redshift dependent normalization \citep{Brinchmann2004b, Noeske2007d, Elbaz2007}. This empirical relation, the star forming ``main sequence'', links the star formation history of a galaxy (via its stellar mass) to its current star formation activity making it an important empirical constraint for theoretical models. The precise functional form of the star forming sequence is still somewhat uncertain given that it has been measured with a variety of different observational techniques and for galaxy samples subject to different selection effects \citep{Speagle2014, Davies2016}. However, advances in recent years, e.g., access to multi-band (UV to FIR) photometry (e.g., \citealt{Schreiber2015b, Davies2016}) and improved modeling techniques \citep{Chevallard2016, Leja2017, Johnson2021}, have resulted in more robust determinations of stellar masses and SFRs across cosmic history (e.g., \citealt{Thorne2021, Leja2021}). In principle, accurate measurements of different star formation tracers even allow constraints on the short-timescale variability of SFRs \citep{Sparre2017, FloresVelazquez2021}.

Fig.~\ref{fig:MainSequence} compares the star forming sequence in \pf{} with the observed one in today's Universe ($z=0$) and at Cosmic Noon ($z=2$). Specifically, we plot the logarithm of the average SFR in bins of stellar mass both for galaxies in \pf{} (``all'') as well as for those galaxies that are actually star forming (``SF''). The latter are defined to exceed a specific star formation rate (sSFR) of $10^{-11}$ yr$^{-1}$ at $z=0$ and $10^{-10}$ yr$^{-1}$ at $z=2$. These limits approximately remove ``quiescent'' galaxies, i.e., galaxies with SFRs that are an order of magnitude or more below the star forming sequence at the considered redshifts. SFRs in \pf{} are averaged over the past 20 Myr. We find only minimal changes ($<0.05$ dex) for the slope and normalization of the star forming sequence of $M_{\rm star}>10^9$ $M_\odot$ galaxies when we adopt a 5 Myr or 100 Myr averaging time instead.

We compare these theoretical predictions with fits to observational data reported in recent studies \citep{Schreiber2015b, Davies2016, Leslie2020, Thorne2021, Leja2021}. We also analyze a representative sample of low redshift galaxies from the xGASS survey \citep{Catinella2018} with updated stellar masses as presented in \cite{Feldmann2020}. SFRs smaller than their measurement uncertainties are set to their measurement uncertainty. We refer to \cite{Feldmann2020} for a more systematic, parametric approach that simultaneously constrains the slope of the star forming sequence and the corresponding atomic and molecular gas sequences.

When we look at the main sequence of star forming galaxies in \pf{}, we find generally good agreement with observational data at both $z=0$ and $z=2$. Being able to reproduce the slope and normalization of the star forming sequence is a significant achievement of the FIRE-2 model given that the simulation is not tuned to reproduce this (or any other) relation. Our finding also agrees qualitatively with a similar result for galaxies in FIRE-2 zoom-in simulations at $z=0$ \citep{Gandhi2022}.

The star forming sequence in \pf{} at $z=0$ and $z=2$ is well described by a linear function (in log-log space) over a broad range in stellar mass, i.e., 
\begin{equation}
y = A+\alpha_1(x-10),
\label{eq:Linear}
\end{equation}
where $y=\lg{}\langle{}{\rm SFR}/(M_\odot\,{\rm yr}^{-1})\rangle{}$ is the logarithm of the average SFR of galaxies and $x=\lg{}(M_{\rm star}/M_\odot)$ is the logarithm of the stellar mass. Fit results are listed in Table \ref{tab:ScalingRelations}.

\begin{table}
\setlength{\tabcolsep}{5pt}
\begin{tabular}{lcccccc}
Selection & Mass & A & $\alpha_1$ & $x_{\rm b}$ & $\alpha_2$  & $\Delta{}$ \\
\hline
\multicolumn{7}{c}{Star forming sequence at $z=0$}\\
\hline
${\rm sSFR}>10^{-11}{\rm yr}^{-1}$ & $9-11$ & 0.03 & 0.85 & - & - & -  \\
all galaxies & $9-11$ & 0.01 & 0.84 & - & - & -  \\
\hline
\multicolumn{7}{c}{Star forming sequence at $z=2$}\\
\hline
${\rm sSFR}>10^{-10}{\rm yr}^{-1}$ & $8-11$ & 1.00 & 0.94 & - & - & -  \\
all galaxies & $8-11$ & 0.98 & 0.97 & - & - & -  \\
\hline
\multicolumn{7}{c}{Atomic hydrogen sequence at $z=0$}\\
\hline
within 30 kpc & $7-11.5$ & 8.86 & 0.85 & 8.55 & 0.37 & 0.21  \\
within $0.1R_{\rm vir}$ & $7-11.5$ & 8.41 & 1.15 & 8.10 & 0.41 & 0.39  \\
\hline
\multicolumn{7}{c}{Molecular hydrogen sequence at $z=0$}\\
\hline
within 10 kpc & $7-11.5$ & 8.68 & 1.59 & 9.82 & 0.26 & 0.39  \\
within $0.1R_{\rm vir}$ & $7-11.5$ & 8.52 & 1.59 & 9.65 & 0.56 & 0.30  \\
\hline
\multicolumn{7}{c}{Gas-phase oxygen abundance at $z=0$}\\
\hline
within 3 kpc & $6.5-11.5$ & 9.29 & 0.58 & 10.28 & 0.19 & 0.07  \\
within $0.1R_{\rm vir}$ & $6.5-11.5$ & 9.15 & 0.57 & 10.21 & -0.18 & 0.18  \\
\hline
\multicolumn{7}{c}{Stellar iron abundance at $z=0$}\\
\hline
within 3 kpc & $6.5-11.5$ & 7.39 & 0.53 & 10.05 & 0.13 & 0.23  \\
within $0.1R_{\rm vir}$ & $6.5-11.5$ & 7.37 & 0.51 & 10.28 & 0.01 & 0.37  \\
\hline
\end{tabular}
\caption{Parameters of galaxy scaling relations in \pf{}. The first column refers to the selected galaxy population or 3-dimensional aperture, see sections \ref{sect:SFsequence}, \ref{sect:gascontent}, and \ref{sect:MZR}. The second column provides the stellar mass range over which the fit was performed. In each case we fit $\lg{}\langle{}Q\rangle{}$ as a function of $\lg\,M_{\rm star}$, where $\langle{}Q\rangle{}$ is the average SFR, atomic hydrogen mass, molecular hydrogen mass, gas-phase oxygen abundance, or stellar iron abundance for all considered galaxies in the given stellar mass bin. The star forming sequence is well fit by a linear function (equation \ref{eq:Linear}) over the quoted mass regime with normalization $A$ (column 3) and slope $\alpha_1$ (column 4). Similarly, the gas and metallicity sequences are well fit by a broken linear function (equation \ref{eq:BrokenLinear}) over the quoted mass regime with the parameters listed in columns 3-7.}
\label{tab:ScalingRelations}
\end{table}

Focusing on $z=0$ galaxies with ${\rm sSFR}>10^{-11}$ yr$^{-1}$ and $M_{\rm star}=10^{9}-10^{11}\,M_\odot$, we obtain a slope of $\alpha_1=0.85$ and a normalization of $A=0.03$. Excluding satellite galaxies reduces the slope slightly to 0.80. Either slope is somewhat steeper than the analogously calculated slope of $\sim{}0.68$ for star forming galaxies with $M_{\rm star}=10^{9-11}$ $M_\odot$ in the xGASS sample. The normalization of star forming sequence in xGASS is very similar, however, differing by only about 0.1 at the $M_{\rm star}=10^{10}$ $M_\odot$ pivot mass.

Both the normalization and the slope of the star forming sequence depend on redshift. The normalization of the star forming sequence increases by about one order of magnitude when going from $z=0$ to $z=2$ while the slope steepens, becoming near linear ($\sim{}0.94$) at $z=2$ (or $0.95$ if satellites are excluded). A linear slope would imply a mass-independent star formation timescale $M_{\rm star}/{\rm SFR}$ \citep{Schreiber2015b} and could help explain the invariant shape of the stellar mass function of star forming galaxies \citep{Peng2010}. Furthermore, as discussed in \cite{Feldmann2020}, the slope of the star forming sequence is naturally linked to the evolution of gas masses in galaxies and it becomes linear if the gas mass histories of galaxies have all the same shape. A sufficient but not necessary condition for the latter scenario is that galaxies are close to `equilibrium' \citep{Bouche2010, Dave2012a}, i.e., the masses of their ISM evolve only mildly with redshift as frequently seen in models across broad redshift and mass ranges \citep{Finlator2008}. A non-linear slope (as found at low $z$) may instead suggest `downsizing' of the gas mass, i.e., more massive galaxies reach their maximum gas masses at earlier times and subsequently have faster declining gas masses at late times. We plan to analyze the link between gas masses and star formation rates in more detail in future work.

At $z=2$, a single power-law describes the star forming sequence well both for the ``all'' and the ``SF'' sample down to $M_{\rm star}=10^8\,M_\odot$. In contrast, at $z=0$ we observe a steepening of the slope for the ``all'' sample below $M_{\rm star}=10^9$ $M_\odot$. The difference between the $z=0$ ``all'' and ``SF'' samples at low masses is a consequence of a significant number of low mass, central galaxies with very low or vanishing sSFR in \pf{}.

\begin{figure}
\begin{tabular}{c}
\includegraphics[width=80mm]{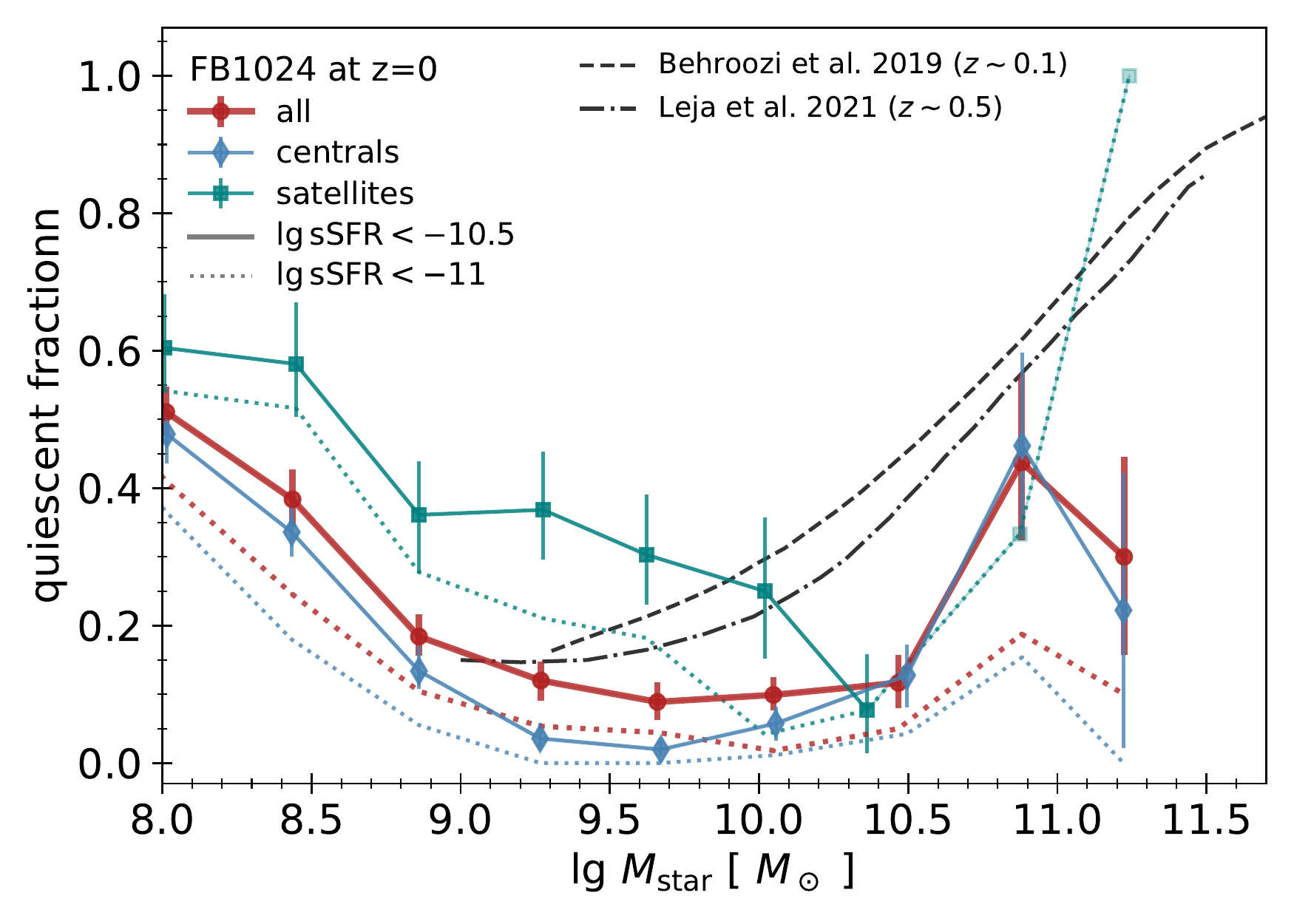}
\end{tabular}
\caption{The fraction of quiescent galaxies in \pf{} at $z=0$. Symbols and solid lines show the fraction of galaxies (red circles), central galaxies (blue diamonds), and satellite galaxies (green squares) with specific SFRs below $10^{-10.5}$ yr$^{-1}$. Dotted lines are the corresponding results for a specific SFR threshold of $10^{-11}$ yr$^{-1}$. 
Error bars refer to 16-84\% percentiles in each bin obtained via bootstrapping. Light shaded symbols without error bars indicate bins containing fewer than four galaxies. Dot-dashed and dashed lines are observational and empirical estimates of the quiescent fraction by \protect\cite{Leja2021} and \protect\cite{Behroozi2019}. SFRs of \pf{} galaxies are averaged over the last 100 Myr as in \protect\cite{Leja2021}.
At $M_{\rm star}\lesssim{}10^{10}$ $M_\odot$, satellite galaxies have a higher quiescent fraction than central (or all) galaxies, presumably as a result of environmental effects (e.g., \citealt{Simha2009, Feldmann2011a, Peng2012, Wetzel2012, Samuel2022}).
\pf{} generally underpredicts the quiescent fraction among massive galaxies when compared with observations. The difference is most severe at the highest masses $M_{\rm star}>10^{11}$ $M_\odot$ but a significant difference is also seen in galaxies of intermediate mass $M_{\rm star}\sim{}10^{9.5-10.5}$ $M_\odot$. Lowering the threshold from $10^{-10.5}\,{\rm yr}^{-1}$ to $10^{-11}\,{\rm yr}^{-1}$ reduces the quiescent fraction significantly which shows that most quiescent, massive galaxies in \pf{} are not fully quenched.}
\label{fig:QuiescentFraction}
\end{figure}

An important difference with observational data is the low fraction of massive, quiescent galaxies in \pf{}, see Fig.~\ref{fig:QuiescentFraction}. Consequently, the average SFR (at fixed stellar mass) of all galaxies in \pf{} is very similar to the average SFR of star forming galaxies alone (except at the lowest masses). While the quiescent fraction is indeed low at early cosmic times, e.g., $\sim{}70\%-80\%$ of galaxies with $M_{\rm star}\sim{}10^{11}$ $M_\odot$ are star forming at $z=2$ \citep{Behroozi2019}, massive galaxies ($M_{\rm star}\sim{}10^{11}$ $M_\odot$) are usually ($\sim{}65\%$) quiescent in today's Universe \citep{Muzzin2013a, Moustakas2013, Behroozi2019, Leja2021}. 

Fig.~\ref{fig:QuiescentFraction} shows the quiescent fraction in \pf{} both for central galaxies, satellites, and the full sample. Here, a galaxy is defined as quiescent at $z=0$ if its sSFR averaged over the last 100 Myr is below a threshold of either $10^{-10.5}$ yr$^{-1}$ or $10^{-11}$ yr$^{-1}$. We also include the data from \cite{Leja2021} for the case of a 100 Myr SFR averaging time and a sSFR-based cut of $10^{-10.5}$ yr$^{-1}$ to separate quiescent and star forming galaxies. Additionally, we plot the predictions of an empirical model by \cite{Behroozi2019} based on low $z$ observational data \citep{Bauer2013, Muzzin2013a}.

In \pf{}, 10-20\% of moderately low mass galaxies ($M_{\rm star}\sim{}10^9$ $M_\odot$) are quiescent in agreement with observational data. However, \pf{} significantly underpredicts the quiescent fraction in more massive galaxies. For instance, 30\% of $M_{\rm star}\sim{}10^{11.2}$ $M_\odot$ galaxies are quiescent in \pf{} (for a $10^{-10.5}$ yr$^{-1}$ sSFR cut), compared with 70-80\% in observations \citep{Muzzin2013a, Leja2021}. Hence, stellar feedback alone (at least if modeled as in FIRE-2) is not sufficient to reproduce the observed fraction of massive, quiescent galaxies at $z\sim{}0$. Evidently though, some massive, quiescent central galaxies can form even without additional feedback sources. However, these quiescent galaxies should be seen as an extension of the star forming sequence towards low SFRs and not as truly passively evolving (`quenched') galaxies, given that the majority of them have sSFR between  $10^{-10.5}$ yr$^{-1}$ and $10^{-11}$ yr$^{-1}$. Perhaps they are related to the observed transition galaxies \citep{Fang2018}. We conclude that alternative forms of feedback, such as cosmic ray feedback (e.g., \citealt{Booth2013a, Salem2014, Chan2019, Hopkins2019}) and AGN feedback (e.g., \citealt{Springel2005c, Croton2006, Vogelsberger2013a, Wellons2022}) are needed to reproduce observational data. Indeed, recent cosmological simulations with AGN feedback reproduce well the observed quiescent fraction at $z=0$ (e.g.,\citealt{Furlong2015a, Donnari2019}).

\begin{figure*}
\begin{tabular}{cc}
\includegraphics[width=80mm]{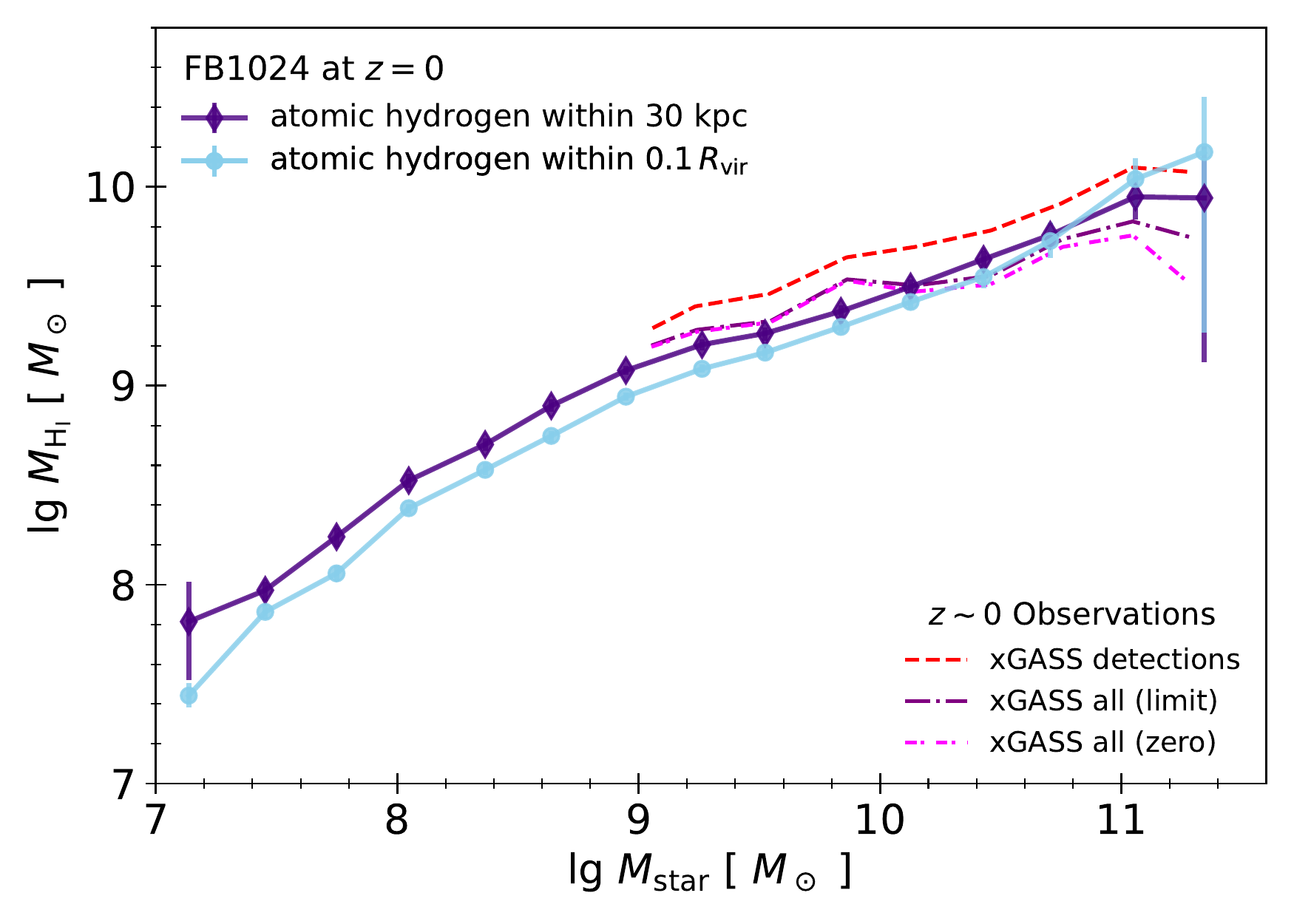} &
\includegraphics[width=80mm]{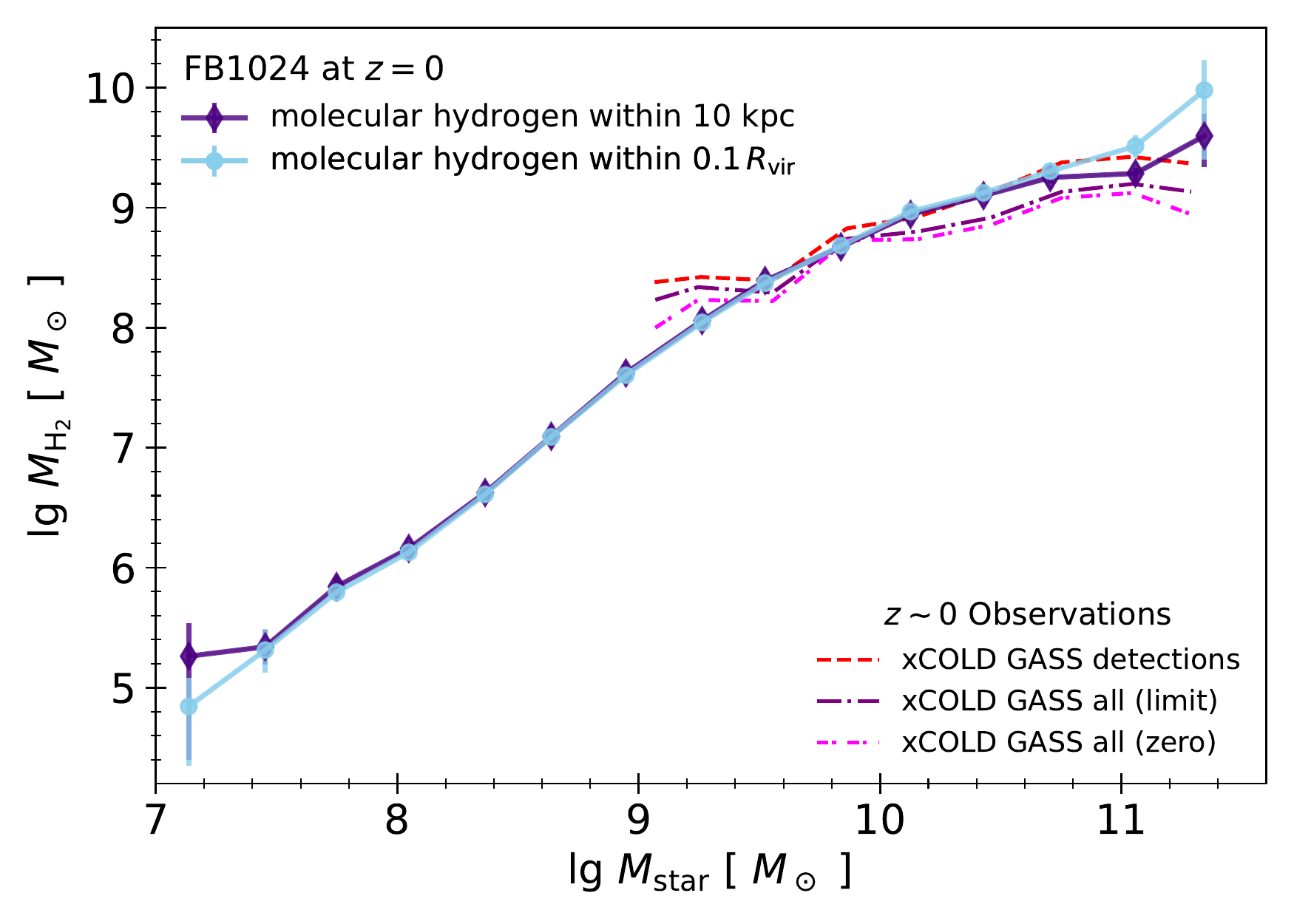}
\end{tabular}
\caption{Relationship between the mass of atomic hydrogen ($M_{\rm H_I}$, left) or molecular hydrogen ($M_{\rm H_2}$, right) and galaxy stellar mass at $z=0$ in \pf{} and in observations. Symbols and solid lines show the logarithm of the average ${\rm H_I}$ or ${\rm H_2}$ mass in bins of stellar mass for \pf{} galaxies within a fixed physical radius (purple diamonds) and within $0.1\,R_{\rm vir}$ (light blue circles). Error bars refer to 16-84\% percentiles of the logarithm of the average gas mass obtained via bootstrapping. Red dashed lines show analogously computed results for galaxies with detected gas masses ($5-\sigma$ in ${\rm H_I}$, $3-\sigma$ in ${\rm H_2}$)  from xGASS \citep{Catinella2018} and xCOLD GASS \citep{Saintonge2017} with updated stellar masses as presented in \protect\cite{Feldmann2020}. Purple (Magenta) dot-dashed lines show corresponding xGASS and xCOLD GASS results when including non-detections by setting the gas mass to the detection limit (to zero). \pf{} predicts average atomic and molecular gas masses in good agreement with these observations.
A broken-linear dependence captures well the scaling of $M_{\rm H_I}$ and $M_{\rm H_2}$ with stellar mass, see Table \ref{tab:ScalingRelations}.}
\label{fig:GasSequence}
\end{figure*}

At low stellar masses ($10^8$ $M_\odot$ $<M_{\rm star}<10^9$ $M_\odot$), the quiescent fraction increases with decreasing mass. For instance, \pf{} predicts that for our chosen 100 Myr averaging time of star formation, 40-50\% of all $M_{\rm star}\sim{}10^8$ $M_\odot$ galaxies are quiescent (the number reduces slightly to 35-45\% if only central galaxies are considered), see also \cite{Feldmann2017}. The quiescent fraction is higher among satellite galaxies than centrals (for $M_{\rm star}<10^{10}$ $M_\odot$), likely as a result of environmental quenching processes (e.g., \citealt{Simha2009, Feldmann2011a, Peng2012, Wetzel2012, Samuel2022}).
The high quiescent fraction at low stellar masses ($M_{\rm star}<10^9\,M_\odot$), especially among centrals, may be in tensions with observations \citep{Geha2012}. We find that the averaging time of the SFR has a significant impact on the quiescent fraction at the low mass end. For a $10^{-11}$ yr$^{-1}$ sSFR threshold, increasing the averaging time to 500 Myr reduces the quiescent fraction of centrals with $M_{\rm star}=10^{8.5}\,M_\odot$ from 16\% to 4.8\% (and from 37\% to 12\% for $M_{\rm star}=10^8\,M_\odot$ centrals). In contrast, reducing the averaging time to 20 Myr, increases the quiescent fraction to 26\% at $M_{\rm star}=10^{8.5}\,M_\odot$ and to 60\% at $M_{\rm star}=10^8\,M_\odot$. The dependence of the quiescent fraction on the averaging time is likely a consequence of the bursty nature of star formation in low mass FIRE galaxies \citep{Sparre2017, FloresVelazquez2021}. Numerical resolution may also play a role here, resulting in excessive burstiness at low stellar masses \citep{Hopkins2018, Samuel2022}.

\subsection{The gas content of galaxies}
\label{sect:gascontent}

Atomic and molecular gas masses of galaxies are correlated with their stellar masses (e.g., \citealt{Catinella2010, Saintonge2011f}). Inferring the shape of these gas sequences is challenging because of a variety of measurement systematics and selection effects. The availability of `representative' (in terms of ${\rm H_I}$ and ${\rm H_2}$ content), purely stellar mass selected galaxy samples \citep{Saintonge2017, Catinella2018} substantially simplifies this challenges but biases may still arise from incorrect modeling assumptions and from the treatment of non-detections (e.g., \citealt{Feldmann2020}).

We compare the gas content of \pf{} galaxies at $z=0$ with observational data from xGASS  \citep{Catinella2018} and xCOLD GASS \citep{Saintonge2017} in Fig.~\ref{fig:GasSequence}. Specifically, we compare the average atomic and molecular hydrogen masses ($M_{\rm H_I}$ and $M_{\rm H_2}$) in bins of stellar mass. We lowered the molecular gas masses reported in \cite{Saintonge2017} by a factor of 1.36 to exclude the contribution from Helium and metals.
The gas masses of simulated galaxies are measured in 3-dimensional spheres of fixed physical radius\footnote{The precise values of these radii are somewhat arbitrary but we chose them for the following reasons. Low mass galaxies ($M_{\rm star}<10^{10}$ $M_\odot$) in xGASS and xCOLD GASS are at redshifts $z=0.01-0.02$, while the redshift range of more massive galaxies is $z=0.025-0.5$. The 3.1-3.5 arcminute half power beamwidth of the Arecibo telescope at the relevant frequencies translates into an aperture radius of $21-41$ kpc at $z=0.01-0.02$ and a radius of $51-100$ kpc at $z=0.025-0.05$. A $30$ kpc fixed radius is thus a sensible choice for galaxies with $M_{\rm star}<10^{10}$ $M_\odot$, while for more massive galaxies we could adopt a larger radius. However, we find that even including all the atomic hydrogen in the virial radius of a $M_{\rm star}=10^{11}$ $M_\odot$ galaxy ($R_{\rm vir}\sim{}280$ kpc) would increase the average ${\rm H_I}$ mass by only 0.3 dex (and by significantly less in galaxies of lower stellar mass) compared to the 30 kpc fixed radius. The IRAM telescope has a beam width of 22 arcseconds at the frequency of the CO (1-0) line, which corresponds to aperture radii of $2.3 - 4.6$ kpc for $z=0.01-0.02$ and $5.7-11.1$ kpc for $z=0.025-0.05$. Adopting a fixed radius of 3 kpc instead of 10 kpc has only a small impact on the inferred ${\rm H_2}$ mass of low mass \pf{} galaxies but misses a large fraction of the molecular gas mass in massive galaxies, e.g., $M_{\rm H_2}$ is lowered by 0.6 dex on average for a $M_{\rm star}=10^{11}$ $M_\odot$ galaxy.
Furthermore, the reported CO line luminosities in xCOLD GASS are aperture corrected to include contributions at larger radii belonging to the ISM.} $r$: $10$ kpc for ${\rm H_2}$ and $30$ kpc for  ${\rm H_I}$.

Gas masses ($M_{\rm H_I}$ or $M_{\rm H_2}$) are not detected in a significant fraction of the galaxies in xGASS and xCOLD GASS. This raises a subtle issue for the comparison with \pf{}. Instead of attempting a full forward modeling, we consider three basic possibilities of dealing with undetected sources. 
First, we include all galaxies in the observational catalog but assume that undetected sources have gas masses that correspond to their detection limit (`all-limit'). Our second analysis is similar to the first but we assign undetected sources a gas mass of zero (`all-zero'). The average gas mass calculated via these two approaches brackets the true value. Finally, we also calculate average gas masses for only the detected sources (`detections').  

According to Fig.~\ref{fig:GasSequence}, the atomic and molecular hydrogen masses of \pf{} galaxies agree well (to usually better than 0.2 dex over the $M_{\rm star}=10^{9}-10^{11}$ $M_\odot$ mass range) with those of galaxies (`all-limit' or `all-zero') in xGASS and xCOLD GASS. The figure also shows the average atomic and molecular hydrogen masses of \pf{} galaxies within 10\% of the virial radius. The latter masses differ usually only by a small amount (0.2 dex) from the average gas masses calculated in the chosen fixed physical radii.

\pf{} offers a prediction of how the atomic and molecular gas sequences scale towards low stellar masses. We find that a broken-linear scaling (in log-log space) captures the general behavior quite well over a broad range in stellar masses ($M_{\rm star}=10^7-10^{11.5}$ $M_\odot$). Specifically, we adopt the following fitting function:
\begin{equation}
y = A + \alpha_1 (x - x_{\rm b}) + (\alpha_2-\alpha_1)\left[\ln\left(1+e^{\frac{x - x_{\rm b}}{\Delta}}\right)-\ln{}2\right]\Delta{},
\label{eq:BrokenLinear}
\end{equation}
where $y = \lg\langle{}M_{\rm H_I}/M_\odot\rangle{}$ ($y = \lg\langle{}M_{\rm H_2}/M_\odot\rangle{}$) is the logarithm of the average atomic (molecular) hydrogen mass in galaxies of a given stellar mass with $x=\lg{}(M_{\rm star}/M_\odot)$. This function has 5 fit parameters: an overall amplitude ($A$), a break stellar mass ($x_{\rm b}$), a slope at low stellar masses ($\alpha_1$), a slope at high stellar masses ($\alpha_2$), and a parameter determining the smoothness of the transition from the low mass to the high mass regime ($\Delta{}$). For $x\ll{}x_{\rm b}$, $y\propto{}\alpha_1{}x$, while for $x\gg{}x_{\rm b}$, $y\propto{}\alpha_2{}x$. The fit parameters are listed in Table~\ref{tab:ScalingRelations}. 

Both gas sequences have a steeper slope at low stellar masses than at high stellar masses. For atomic hydrogen, we find a low mass slope near 1 (0.85 for $r=30$ kpc, 1.15 for $r=0.1\,R_{\rm vir}$), while for molecular hydrogen the slope is super-linear (1.6 for both $r=10$ kpc and $r=0.1\,R_{\rm vir}$). We speculate that the steeper slope in low mass galaxies is a consequence of stellar feedback and the UV background more strongly regulating their gas content (see e.g., \citealt{VandeVoort2016, Fitts2017, Hafen2019, Pandya2020}). Furthermore, since lower mass galaxies tend to have lower ISM metallicities (e.g., \citealt{Tremonti2004a, Finlator2008}) and lower dust-to-metal ratios (e.g., \citealt{Remy-Ruyer2014d, Feldmann2015a}), a smaller fraction of the neutral ISM is in molecular form (e.g., \citealt{Krumholz2008a, Gnedin2011a}) thus explaining the steeper slope of the molecular gas sequence compared with the atomic gas sequence. At the high mass end, slopes are sub-linear (0.4 for atomic hydrogen and 0.3-0.6 for molecular hydrogen) qualitatively consistent with the decline of the atomic and molecular gas to stellar mass ratios with increasing stellar mass found observationally (e.g., \citealt{Saintonge2017, Catinella2018}) and in models (e.g., \citealt{Dave2020}). The break stellar mass between the two regimes is $10^{8.1-8.6}$ $M_\odot$ for ${\rm H_I}$ and $10^{9.6-9.8}$ $M_\odot$ for ${\rm H_2}$. We find that galaxies with a stellar mass near the break stellar mass of the atomic (molecular) hydrogen sequence have an atomic (molecular) hydrogen content of $M_{\rm H_I}\sim{}10^{8.4-8.9}$ $M_\odot$ ($M_{\rm H_2}\sim{}10^{8.5-8.7}$ $M_\odot$).

\subsection{The gas content of Milky Way analogs -- where are the ``missing'' baryons?}
\label{sect:MWanalogs}

A growing number of observational and theoretical studies attest that galaxies like the Milky Way (MW) contain fewer baryons in their halos than expected based on the cosmic average (e.g., \citealt{Maller2004, Anderson2010, McGaugh2010, Crain2010, Feldmann2013b, Schaller2015, VandeVoort2016, Suresh2017, Tumlinson2017, Bregman2018}). 
In this section we provide a census of the baryons in MW-mass halos from \pf{} and compare it, for illustrative purposes, with measurements of the various mass components in the Galaxy and its halo.
To this end, we selected all 23 \fb{} main halos with virial masses between $7.5\times{}10^{11}$ $M_\odot$ and $2.5\times{}10^{12}$ $M_\odot$ at $z=0$. One system, a late stage, galaxy major merger, was excluded from the analysis below.  The average halo mass of this sample is $1.3\times{}10^{12}$ $M_\odot$ matching the current consensus estimate of the virial mass of the Milky Way \citep{Bland-Hawthorn2016}. The average virial radius of the sample is $279$ kpc. 

The MW contains about $8\times{}10^{9}$ $M_\odot$ of atomic hydrogen \citep{Kalberla2008, McMillan2017, Cautun2020} and $\sim{}(1\pm{}0.3)\times{}10^{9}$ of molecular hydrogen \citep{Heyer2015, McMillan2017}. Both mass estimates are subject to large modeling uncertainties and are reported here without contributions from metals and Helium. While they only account for the gas in the MW disk and center, the neutral hydrogen mass in MW satellites is relatively low. The Large Magellanic Cloud (Small Magellanic Cloud) contributes about $5\times{}10^8$ $M_\odot$ in atomic hydrogen \citep{Kim1999} ($4\times{}10^8$ $M_\odot$, \citealt{Stanimirovic1999}) and $\sim{}5\times{}10^7$ $M_\odot$ in ${\rm H_2}$ \citep{Fukui1999} with some additional neutral gas in the Magellanic bridge, stream, and leading arm (e.g, \citealt{Nidever2010, Besla2012}). In comparison, \pf{} predicts an average atomic hydrogen mass of $6.37_{-1.05}^{+1.27}\times{}10^9$ $M_\odot$ and a molecular hydrogen mass of $1.79_{-0.27}^{+0.22}\times{}10^9$ $M_\odot$ within 10\% of $R_{\rm vir}$, both in reasonable agreement with observations. Here, sub- and superscripts refer to 16th and 84th percentiles of the averages obtained via bootstrapping. For the mass of neutral hydrogen \pf{} predicts $8.16_{-1.36}^{+1.54}\times{}10^9$ $M_\odot$ which can be compared with the observed value of $\sim{}9\times{}10^{9}$ $M_\odot$.

The mass of the warm-hot and hot gaseous halo (corona) around the Milky-Way has been constrained to $2.5\pm{}1\times{}10^{10}$ $M_\odot$ via a variety of independent observables \citep{Bland-Hawthorn2016}, including X-ray emission \citep{Miller2015b}, pulsar-based dispersion measures from the Large Magellanic Cloud (LMC) \citep{Anderson2010}, $H_\alpha$ emission from the Magellanic $H_{\rm I}$ stream \citep{Bland-Hawthorn2016}, and ram-pressure effects on the LMC gas disk \citep{Salem2015a}. A more recent estimate by \cite{Bregman2018} based on a combined analysis of a variety of observational data is $2.8\pm{}0.5\times{}10^{10}$ $M_\odot$ of hot gas within 250 kpc of the MW.
\pf{} predicts an average mass of gas with $T>2\times{}10^5$ K of $3.24_{-0.62}^{+0.64}\times{}10^{10}$ $M_\odot$ within the virial radii of MW-like halos ($3.18_{-0.60}^{+0.64}\times{}10^{10}$ $M_\odot$ within $0.1-1\times{}R_{\rm vir}$), in good agreement with the observational estimates. These gas masses include contributions from Helium and metals. This warm-hot and hot gas amounts to $58\%$ of the total gas mass in such halos and it exceeds the $35\%$ fraction of cooler gas ($M_{\rm gas}(T<2\times{}10^4K)=1.9\times{}10^{10}$ $M_\odot$). Gas at intermediate temperatures ($2\times{}10^4\leq{}T/{\rm K}\leq{}2\times{}10^5$) contributes only about $7\%$, as expected from the high cooling rate in this temperature range.

Numerical models predict that hot halos around galaxies are strongly affected by galactic winds driven by feedback (e.g., \citealt{VandeVoort2016, Hafen2019, Stern2021a, Vijayan2021}). Observations with future X-ray telescopes may be able to measure the diffuse halo gas in $L_*$ galaxies out to moderate redshifts \citep{Kaastra2013, Simionescu2021} thus potentially providing a sensitive probe of the physics of feedback processes. We plan to study the formation and evolution of hot halos in \fb{} galaxies in future work.

As mentioned above, the observed baryonic content of the Milky Way halo falls short of the amount expected from the universal baryon fraction \citep{McGaugh2010}. Current observational estimates provide a baryon fraction ($M_{\rm b}/M_{\rm halo}$) of only 7\% \citep{Bland-Hawthorn2016}, i.e., less than half of $\Omega_{\rm b}/\Omega_{\rm m}=15.7\%$. Whether significant amounts of halo baryons have evaded detection so far or whether they are truly `missing' from the halo is still debated. In \pf{}, the baryon fraction of MW-like halos at $z=0$ is $11.6_{-0.4}^{+0.5}\%$, i.e., only about 25\% of the cosmic baryons are missing from MW halos, i.e., reside outside the halo either because they were removed at some point or never accreted in the first place.

The remaining ``extra'' baryons, compared with observations, are distributed among various matter components. 
First, a significant amount ($\sim{}7\times{}10^{9}$ $M_\odot$) of ionized gas in MW-like halos in \pf{} has temperatures below $2\times{}10^5$ K, i.e., it is not in a hot phase. Interestingly, estimates based on modeling of the ${\rm O_{VI}}$ absorption line of $L*$ galaxies predict an even larger average mass of warm ionized gas \citep{Werk2014a}.
Secondly, the hot gas mass (see discussion above) and the galaxy stellar mass in \pf{} are slightly higher than empirical estimates. The average stellar mass of the centrals in our sample of MW analogs is $5.73_{-0.53}^{+0.57}\times{}10^{10}$ $M_\odot$ for $R_{\rm g}=3\,R_{\rm half}$ ($7.64_{-0.63}^{+0.81}\times{}10^{10}$ $M_\odot$ for $R_{\rm g}=0.1\,R_{\rm vir}$) compared with empirically determined stellar mass of $\sim{}5\times{}10^{10}$ $M_\odot$ \citep{Flynn2006, Cautun2020} for the MW. 
Finally, halos of MW analogs in \pf{} harbor a significant amount of stars in a smooth extra-galactic component ($2.4_{-0.3}^{+0.3}\times{}10^{10}$ $M_\odot$ for a galaxy size of $R_{\rm g}=3\,R_{\rm half}$, $5.6_{-1.0}^{+0.8}\times{}10^{9}$ $M_\odot$ for $R_{\rm g}=0.1\,R_{\rm vir}$) and in satellite galaxies ($\sim{}9\times{}10^9$ $M_\odot$). This extra-galactic stellar component exceeds current observational estimates of the `stellar halo' of the MW ($\sim{}1.4\times{}10^{9}$ $M_\odot$ \citealt{Deason2019, MacKereth2020}). However, the latter estimates involve various modeling and selection steps that will need to be properly taken into account, e.g., via forward modeling of our simulated galaxies, to allow for a direct, quantitative comparison.

\subsection{Mass-metallicity relation}
\label{sect:MZR}

The metallicity of the ISM is set by a complex network of processes including metal injections from supernovae \citep{Woosley1995, Nomoto2006}, star formation, galactic outflows that remove metals from galaxies, and inflows of comparably metal-poor gas from the cosmic environment \citep{Almeida2014, Muratov2015, Muratov2016}. The observation of a correlation between ISM metallicity and the stellar mass of galaxies, the mass--metallicity relation (MZR, \citealt{Tremonti2004a}), may thus provide insights into the role these processes play in galaxy evolution. Various physical mechanisms have been proposed to explain the MZR including the ejection of metal-rich gas from low mass halos by supernova feedback \citep{Dekel1986b, Dekel2003}, inefficient star formation (due to feedback in the ISM) in low mass galaxies \citep{Brooks2006}, and the potential under-abundance of massive stars in low mass galaxies as a result of clustered star formation \citep{Koppen2007}. 

In the equilibrium model of galaxy formation \citep{LARSON1972a, Finlator2008, Dave2012a, Feldmann2013, Lilly2013c, Dekel2014}, the ISM metallicity is set by the present balance of metal enrichment, removal, and dilution processes with any memory of the past enrichment level erased over a few gas depletion times. In this model, the star formation activity in a galaxy adjust such that stellar feedback driven outflows roughly balance any gas inflows resulting in gas and metal masses in the ISM that are approximately constant in time. Low mass galaxies tend to have large mass loading factors \citep{Muratov2015, Angles-Alcazar2017b, Pandya2021}, and thus require only small SFRs to achieve this balance, resulting in low equilibrium metallicities \citep{Finlator2008}. The self-regulatory feature of this model also helps to explain why the MZR has such a small scatter. Furthermore, by allowing for evolving ISM masses, this model naturally introduces a dependence of the ISM metallicity on SFR at fixed stellar mass \citep{Lilly2013c} as potentially observed \citep{Mannucci2010, Sanders2021}.

While the metallicity of the ISM is expected to quickly reach equilibrium values under most circumstances, the metallicity in the photosphere of stars is determined to a large degree by the metallicity of the molecular clouds they formed from. The stellar metallicity of galaxies thus reflects both the past ISM metallicity, the star formation history, and the accretion of stars in galaxy mergers. In addition, it may hold clues to the nature and time scale of galaxy quenching \citep{Peng2015}. Similar to ISM metallicities, the stellar metallicities of observed galaxies are found to correlate with their stellar masses \citep{Gallazzi2005}.

\begin{figure*}
\begin{tabular}{cc}
\includegraphics[width=80mm]{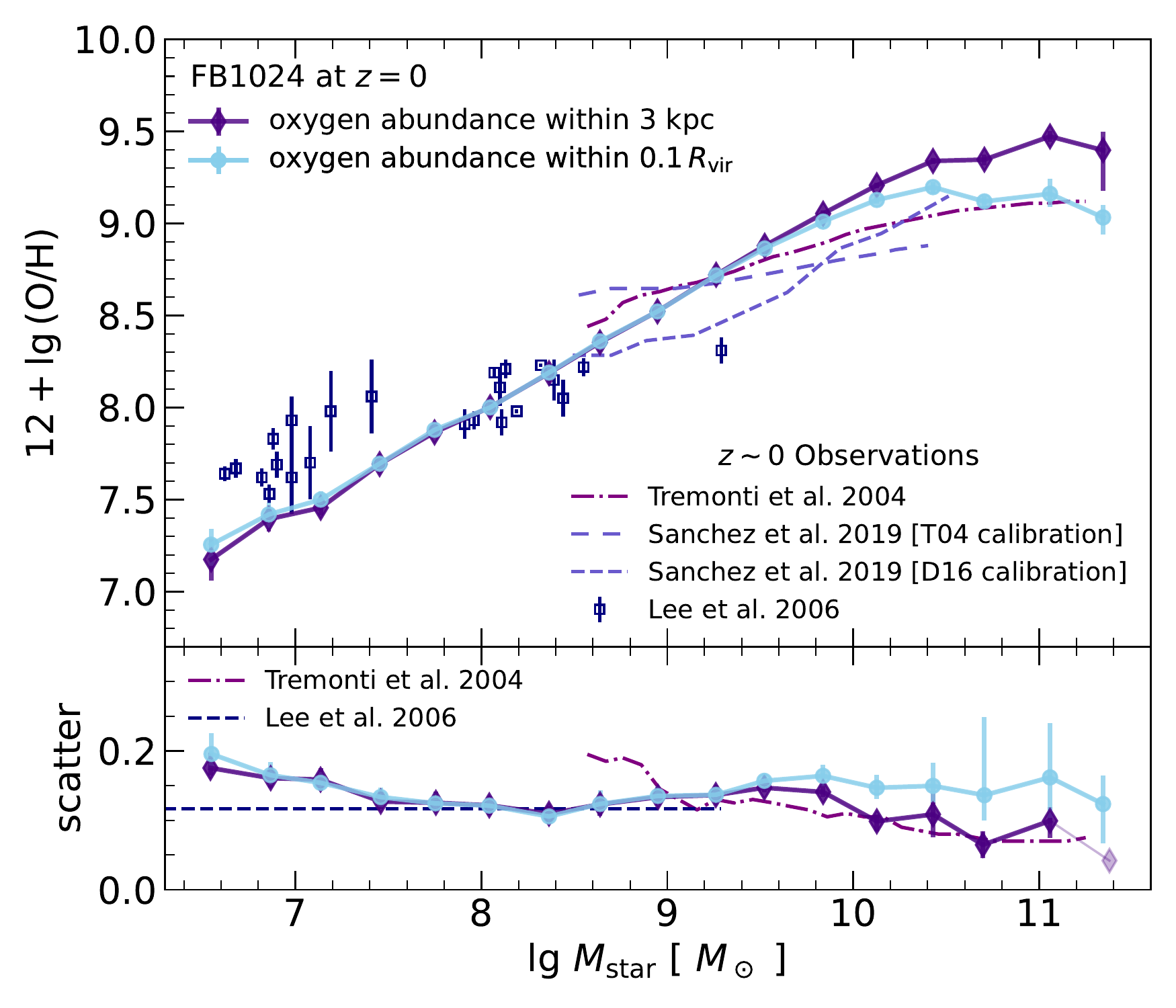} &
\includegraphics[width=80mm]{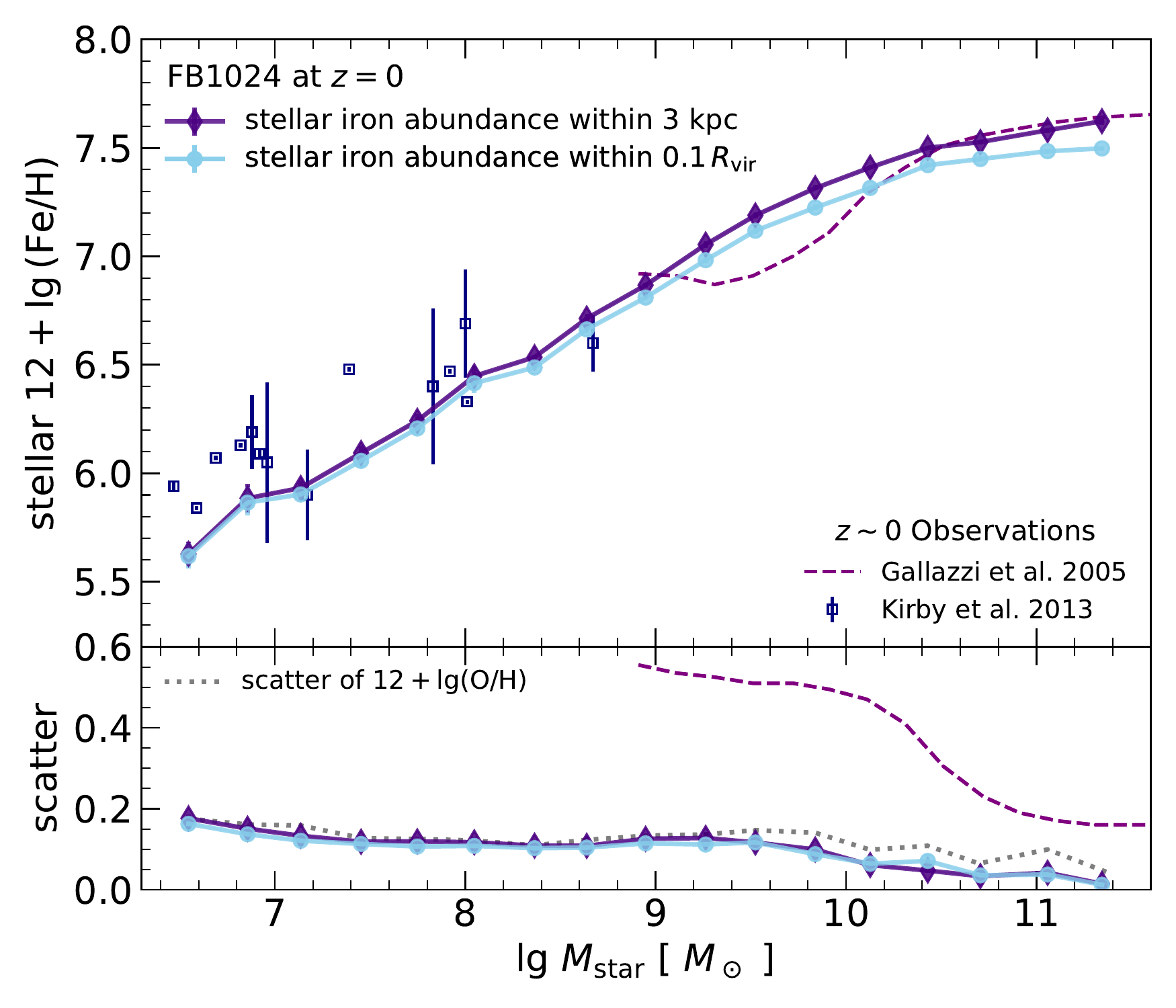}
\end{tabular}
\caption{Relationship between the gas-phase oxygen abundance (left) or stellar iron abundance (right) and galaxy stellar mass at $z=0$ in \pf{} and in observations. 
In the upper left panel, solid lines show the logarithm of the average oxygen abundance plus 12 for \fb{} galaxies in bins of stellar mass for a 3 kpc fixed physical radius (purple diamonds) or within $0.1\,R_{\rm vir}$ (light blue circles). Error bars (16-84 percentiles) are obtained via bootstrapping. Light shaded symbols without error bars indicate bins containing fewer than four galaxies. Observational data \protect\citep{Tremonti2004a, Lee2006, Sanchez2019} are included in the figure as dot-dashed lines, square symbols, and dashed lines. The scatter (one half of the 16-84\% percentile range) of the logarithmic oxygen abundance at a fixed stellar mass is shown in the lower left panel. \pf{} predicts that the stellar mass -- gas phase metallicity relation has low scatter ($\sim{}0.1-0.2$) both in massive and low mass galaxies. The panels on the right hand side are analogous to the panels on the left but for the stellar iron abundance instead of the gas-phase oxygen abundance. Observational data is from \protect\cite{Gallazzi2005} (dashed lines) and \protect\cite{Kirby2013a} (square symbols). Broken-linear dependences capture well the scaling of the gas-phase oxygen abundance and stellar iron abundance with stellar mass in \pf{}, see Table \ref{tab:ScalingRelations}. The scatter of the stellar mass -- stellar metallicity relation in \pf{} increases towards lower stellar masses but it is generally lower than the scatter of the stellar mass -- gas phase metallicity relation (the dotted line in the bottom right panel reproduces the scatter of the latter relation for a 3 kpc aperture radius as shown in the bottom left panel).
}
\label{fig:MZR}
\end{figure*}

We compare the gas phase and stellar metallicities of \pf{} with observational data in Fig.~\ref{fig:MZR}. 
Observed oxygen abundances are reproduced in the figure as originally reported \citep{Tremonti2004a, Lee2006, Sanchez2019}. \cite{Kirby2013a} assumed a Solar iron abundance of 7.52 (consistent with \citealt{Asplund2009}) and we use this value to convert their results from relative to absolute abundances. \cite{Gallazzi2005} measured stellar metallicities (relative to Solar) via stellar absorption indices based primarily on magnesium and iron lines. We equate their reported metallicities with iron abundances relative to Solar and convert to absolute values using again a Solar iron abundance of 7.52.
To aid the comparison with observations, we measure metallicities in \pf{} in a 3-dimensional aperture radius of 3 kpc to match approximately the 1.5 arcsecond radius of SDSS fibers at the median redshift ($z\sim{}0.1$) of the samples of \cite{Tremonti2004a} and \cite{Gallazzi2005}. However, since this radius is too small to include gas in the outskirts of larger galaxies, we also provide a more scale-invariant aperture choice of $0.1R_{\rm vir}$. To account for oxygen depletion inside HII regions, we reduce the oxygen abundance predicted by the simulation by 0.12 dex \citep{Peimbert2010}. 
This quantitative comparison likely suffers from additional systematic uncertainties related to, e.g., the observational metallicity calibration as well as the metal yields and supernova rates adopted by the simulation \citep{Hopkins2018}, all of which are beyond the scope of this paper.

To highlight the general trend between metallicity and stellar mass, we aggregate the metallicities of \pf{} galaxies in 0.3 dex wide bins of stellar mass. Specifically, we show in Fig.~\ref{fig:MZR} the quantities $12 + \lg{}(\langle{}{\rm O/H}\rangle{})$ (top left panel) and $12 + \lg{}(\langle{}{\rm Fe/H}\rangle{})$ (top right panel), where $\langle{}{\rm O/H}\rangle{}$ ($\langle{}{\rm Fe/H}\rangle{}$) represents the average ratio between the number of oxygen and hydrogen nuclei in the gas phase (the average ratio between the number of stellar iron nuclei and stellar hydrogen nuclei) of \pf{} galaxies in 0.3 dex wide bins of stellar mass. In the bottom panels we show the corresponding scatter defined as half the difference between the 84 and 16 percentile of $12 + \lg{}({\rm O/H})$ or $12 + \lg{}({\rm Fe/H})$ for the galaxies in the given stellar mass bin.

\pf{} broadly reproduces the observed MZR at $z=0$ over $\sim{}5$ orders of magnitude in stellar mass \citep{Tremonti2004a, Lee2006, Sanchez2019}, similar to previous results of FIRE-1 zoom-in simulations \citep{Ma2016}. The match is not perfect, however, as \pf{} possibly slightly overpredicts (underpredicts) the oxygen abundance in galaxies with $M_{\rm star}>10^{10}$ $M_\odot$ (with $M_{\rm star}<10^{7.5}$ $M_\odot$). We caution that this comparison is plagued by calibration systematics which can exceed 0.2 dex \citep{Kewley2002, Kewley2008a, Sanchez2019}. As a specific example, we show the MZR reported by \cite{Sanchez2019} for two different metallicity calibrations; one based on [NII], [SII] and H$\alpha$ emission lines \citep{Dopita2016}, the other one using [OII], [OIII], and H$\beta$ \citep{Pagel1979, Tremonti2004a}.

Matching simultaneously both the observed MZR and the star forming sequence of low mass galaxies ($M_{\rm star}\lesssim{}10^9\,M_\odot$) has been pointed out as a major challenge for galaxy formation models \citep{Somerville2014}. The reasonable match between \pf{} and the observational data shown in Fig.~\ref{fig:MainSequence} and Fig.~\ref{fig:MZR} suggests that cosmological simulations with the FIRE-2 physics model are a significant step towards overcoming this challenge.

The MZR in \pf{} shows clear evidence of a flattening at the massive end. This flattening has been observed for many metallicity calibrators (e.g., \citealt{Tremonti2004a, Sanchez2019}). Given its presence in \pf{}, we infer that the flattening of the MZR as reported by observations is likely not merely a consequence of aperture bias \citep{Kirby2013a}. Instead, provided the equilibrium view of galaxy formation is correct, the flattening can be explained by the mass loading factors approaching, and falling below, unity in massive galaxies \citep{Finlator2008, Muratov2015}. The relation between mass loading factor $\eta$ and equilibrium metallicity is $Z_{\rm eq}\propto{}y/(1-R+\eta)$, see \cite{Finlator2008} and \cite{Lilly2013c}, where $y$ is the metal yield and $R\sim{}0.5$ the mass return fraction (\citealt{Krumholz2012}). The equilibrium metallicity is approximately independent of the mass loading factor for $\eta\ll{}1$.

The MZR can be well fit with a broken linear relation given by equation (\ref{eq:BrokenLinear}) with $y = 12 + \lg(\langle{}{\rm O/H}\rangle{})$ and $x=\lg{}M_{\rm star}$. The fit parameters are provided in Table \ref{tab:ScalingRelations}. For the 3 kpc aperture radius, the MZR is sub-linear with a slope of $\sim{}0.6$ at low stellar masses and almost flat with a slope of $\sim{}0.2$ at high stellar masses. The transition between the two regimes occurs at a break stellar mass of $\sim{}10^{10.3}$ $M_\odot$.

Focusing on the scatter of the MZR, \pf{} predicts a value of $\sim{}0.1-0.15$ for much of the probed stellar mass range, in agreement with observations \citep{Tremonti2004a, Lee2006}. Furthermore, the scatter is predicted to decrease slightly at the massive end if the 3 kpc aperture is used in line with results by \cite{Tremonti2004a}. However, as shown in the lower left panel of Fig.~\ref{fig:MZR}, the scatter is almost independent of stellar mass if an $0.1R_{\rm vir}$ aperture radius is adopted.

\pf{} predicts a relation between stellar iron abundance and stellar mass that is in approximate agreement with observational data \citep{Gallazzi2005, Kirby2013a}, except perhaps at the lowest masses ($M_{\rm star}<10^8\,M_\odot$). This overall behavior is consistent with the results of FIRE-2 zoom-in simulations \citep{Gandhi2022}.
The relation between iron abundance and stellar mass follows the same general trend as the MZR and can also be approximated well by a broken linear function (equation \ref{eq:BrokenLinear} with $y = 12 + \lg(\langle{}{\rm Fe/H}\rangle{})$), see Table \ref{tab:ScalingRelations} for the best fit parameters. At low stellar masses, the slope is sub-linear and slightly smaller (0.53 for the 3 kpc aperture radius) than the slope of the MZR. The latter is expected given that the stellar metallicity relation is effectively a convolution of the MZR and the stellar growth history. The stellar metallicity relation significantly flattens (slope 0.13) in massive galaxies, similar to the MZR.

Interestingly, \pf{} predicts a much smaller scatter in stellar metallicities at given stellar mass than reported in the observational study by \cite{Gallazzi2005}. The authors of the latter study point out that their high scatter may reflect, at least partly, the high observational uncertainties in measuring stellar metallicities. In fact, the stellar mass -- stellar metallicity relation predicted by our simulation is even tighter than the MZR with a scatter of less than 0.05 in $M_{\rm star}>10^{10.5}$ $M_\odot$ galaxies. We speculate that this lower scatter is a consequence of the stellar metallicity being a (SFR weighted) time-average of the gas phase metallicity.

\subsection{Galaxy stellar mass functions}

\label{sect:SMF}

\begin{figure*}
\begin{tabular}{cc}
\includegraphics[width=80mm]{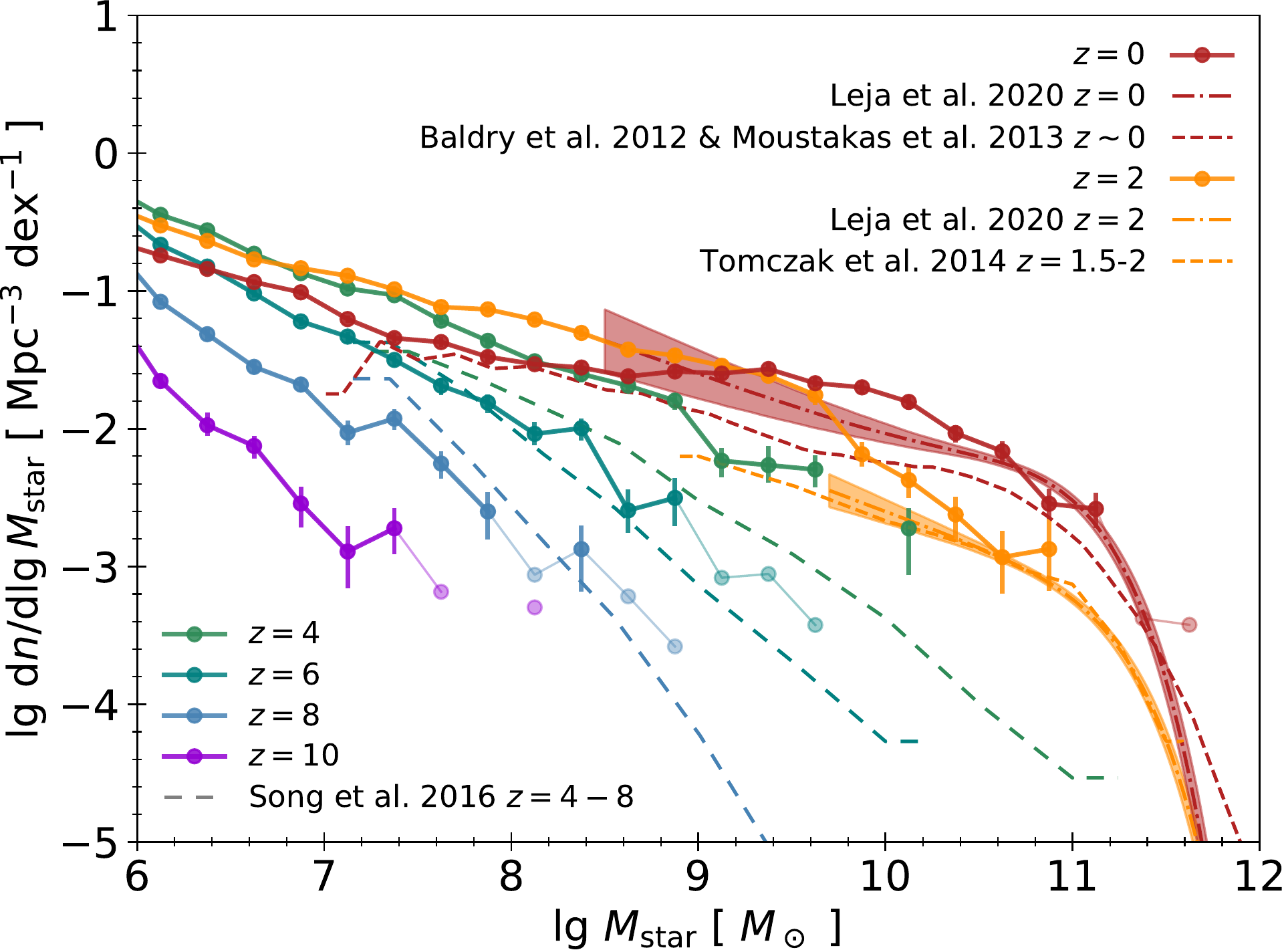} &
\includegraphics[width=80mm]{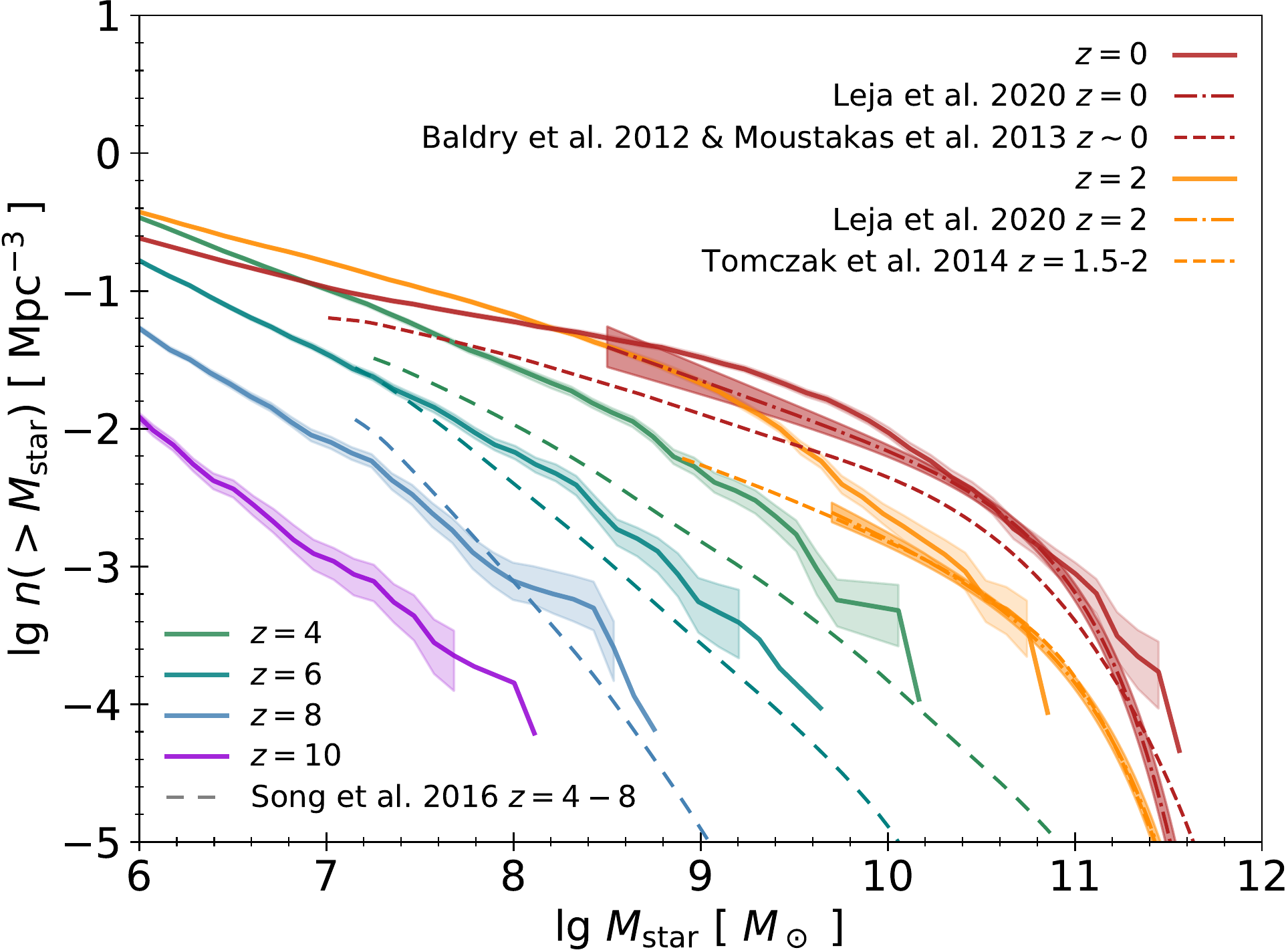}
\end{tabular}
\caption{Stellar mass function (SMF) predicted by \pf{} for $z=0-10$ and a comparison with observational estimates. In the left panel (right panel), circles (solid lines) show the differential (cumulative) SMF of all galaxies with $M_{\rm star}>10^6$ $M_\odot$ in the simulation volume. The abundance of galaxies is re-weighted to account for cosmic variance in the halo mass function, see Appendix \ref{app:reweighting}. Uncertainties (16-84\%) of the differential and cumulative SMFs are calculated via bootstrapping and shown by error bars (left panel) or shaded regions (right panel). Light shaded symbols without error bars indicate bins containing fewer than four galaxies. The panels also show observational estimates of the SMF at various redshifts \protect\citep{Baldry2012, Moustakas2013, Tomczak2014, Song2015, Leja2020}.  \pf{} predicts a SMF at $z=0$ similar to recent estimates by \protect\cite{Leja2020} based on non-parametric modeling except for a moderate excess at both low and high stellar masses. The $z=0$ SMF is generally higher than those based on more traditional stellar mass estimates (e.g., \protect\citealt{Baldry2012, Moustakas2013}). \pf{} struggles in reproducing the differential SMF in galaxies with $M_{\rm star}\sim{}10^{8.5}-10^{10}\,M_\odot$ at $z\leq{}4$ as a result of limited numerical convergence (see text). Interestingly, \pf{} predicts a drop in the SMF at high masses at low $z$, even though AGN feedback is not included.}
\label{fig:SMFz}
\end{figure*}

The stellar mass function (SMF) of galaxies provides an important point of comparison to observational data for galaxy formation simulations. Here, we use observational data spanning $z=0-8$ \citep{Baldry2012, Moustakas2013, Tomczak2014, Song2015} as provided by \cite{Behroozi2019}. In addition, we compare with SMF estimates from the recent work by \cite{Leja2020}. All data is converted (if necessary) to a \cite{Chabrier2003} IMF.

The realized halo mass function (HMF) in \fb{} differs from the true HMF because of the finite box size and limited numerical resolution. Similarly, the realized SMF in \fb{} differs from the SMF that would be obtained if the exact same physical model were applied to an infinitely large cosmological volume. This 'cosmic variance' can become large for small simulated volumes, e.g., the stellar mass density may vary by $\sim{}0.2$ dex for a random selection of initial conditions of a $L=35.5$ Mpc box \citep{Genel2014}. The initial conditions for \pf{} were selected with the objective to reduce the difference between the realized and true HMF as a first mitigation measure, see section \ref{sect:ICs}. In addition, we calculate SMFs and other number-density-based quantities via a re-weighting approach (Appendix \ref{app:reweighting}). The latter increases (decreases) the contribution from galaxies in halos that are under-abundant (over-abundant) relative to a reference halo mass function, here \cite{Behroozi2013c}. To reduce biases of the HMF arising from baryonic effects, we match halo masses in \pf{} with those of the corresponding collisionless simulation \pfdm{} based on cumulative abundances. The main caveat of our re-weighting approach is its reliance on halo mass alone. In its present form, the re-weighting does not correct for secondary trends, e.g., with large-scale environment, halo concentration, or formation time which have been shown to correlate non-trivially with galaxy properties (e.g., \citealt{Matthee2017, Feldmann2019}). Our approach differs from methods to constrain SMFs from observational data (e.g., \citealt{Efstathiou1988, Weigel2016}) in that it aims to correct for variations in halo abundance instead of limits in stellar mass or luminosity.

Fig.~\ref{fig:SMFz} shows both the differential and the cumulative SMF in \pf{} for $z=0-10$. At $z\geq{}6$, the shape and normalization of the SMF in \pf{} agrees reasonably well with observations. The low mass slope of the simulated SMF decreases with decreasing redshift in qualitative agreement with \cite{Song2015}. At $z\leq{}4$, the SMF in \pf{} is higher than observed, especially for galaxies of $M_{\rm star}\sim{}10^9-10^{10}$ $M_\odot$. A similar, but weaker, behavior has been reported in previous SAMs and hydrodynamical simulations (e.g., \citealt{Vogelsberger2014, Somerville2015b}). At low redshift, \pf{} overpredicts the abundance (or stellar masses) compared with traditional SMF estimates \citep{Baldry2012, Moustakas2013} but is in much better agreement with recent studies in which stellar masses are inferred from a non-parametric modeling of the star formation histories of galaxies \cite{Leja2020}, especially for MW analogs with $M_{\rm star}\sim{}10^{10.5}\,M_\odot$. We note that \fb{} simulations are not tuned to reproduce any of the SMFs, in contrast with most other cosmological simulation suites (e.g., \citealt{Vogelsberger2014, Schaye2015, Pillepich2018a}), i.e., our results are predictions directly based on the FIRE-2 physics model.

We re-iterate two main areas of disagreement in Fig.~\ref{fig:SMFz}. First, at low to intermediate stellar masses ($M_{\rm star}\sim{}10^{8.5}-10^{10}$ $M_\odot$), \pf{} overestimates the observed SMF at $z\leq{}4$. A comparison with high resolution ($m_{\rm b}<10^4$ $M_\odot$) FIRE zoom-in simulations shows that galaxies in moderately low mass ($M_{\rm halo}\sim{}10^{11}$ $M_\odot$) halos have lower stellar masses at increased numerical resolution, see Appendix~\ref{app:CompFIRE2}. Unfortunately, this implies that the stellar masses of such galaxies are not converged at the resolution of \pf{}. 

Secondly, at high stellar masses, \pf{} appears to over-predict galaxy abundances. Here the stellar masses are converged, see Appendix~\ref{app:CompFIRE2}. Given that \pf{} does not include AGN feedback, a mismatch at the massive end is not unexpected. However, the shape of the simulated SMF at the massive end ($M_{\rm star}>10^{10.5}$) may still be marginally consistent with the observations by \cite{Leja2020} if we account for the low numbers of massive galaxies in \pf{} and the associated large statistical errors. At low $z$, the SMF in \pf{} shows a turn-over above which the SMF drops quickly with increasing mass. This behavior is qualitatively similar to observations but the turn-over occurs at a lower stellar mass ($M_{\rm star}\sim{}10^{10}\,M_\odot$) in \pf{}. Hence, while galaxy quenching by AGN feedback may be needed to reproduce the exact position and shape of the SMF at the high-mass end, it may not be the primary reason that the SMF shows a break. We discuss the physical origin of this turn-over in more detail in the next section.

Finally, \pf{} suggests that the SMF decreases by up to $\sim{}0.3$ dex with decreasing redshift at the lowest stellar masses ($M_{\rm star}<10^9$ $M_\odot$) between $z=2$ and $z=0$, which is qualitatively similar to the behavior of the low mass end of the HMF over this redshift range. Whether this trend is consistent with observations is currently not known given that $M_{\rm star}\sim{}10^9$ $M_\odot$ is close to the mass completeness limit of galaxy surveys exploring the SMF at $z=1-2$ \citep{Tomczak2014, Leja2020}. Also, semi-empirical models do not necessarily predict this trend \citep{Behroozi2019}. Future, deeper observations may be required to test this prediction of our model.

\subsection{Stellar mass -- halo mass relation}
\label{sect:SHMR}

\begin{figure}
\begin{tabular}{cc}
\includegraphics[width=80mm]{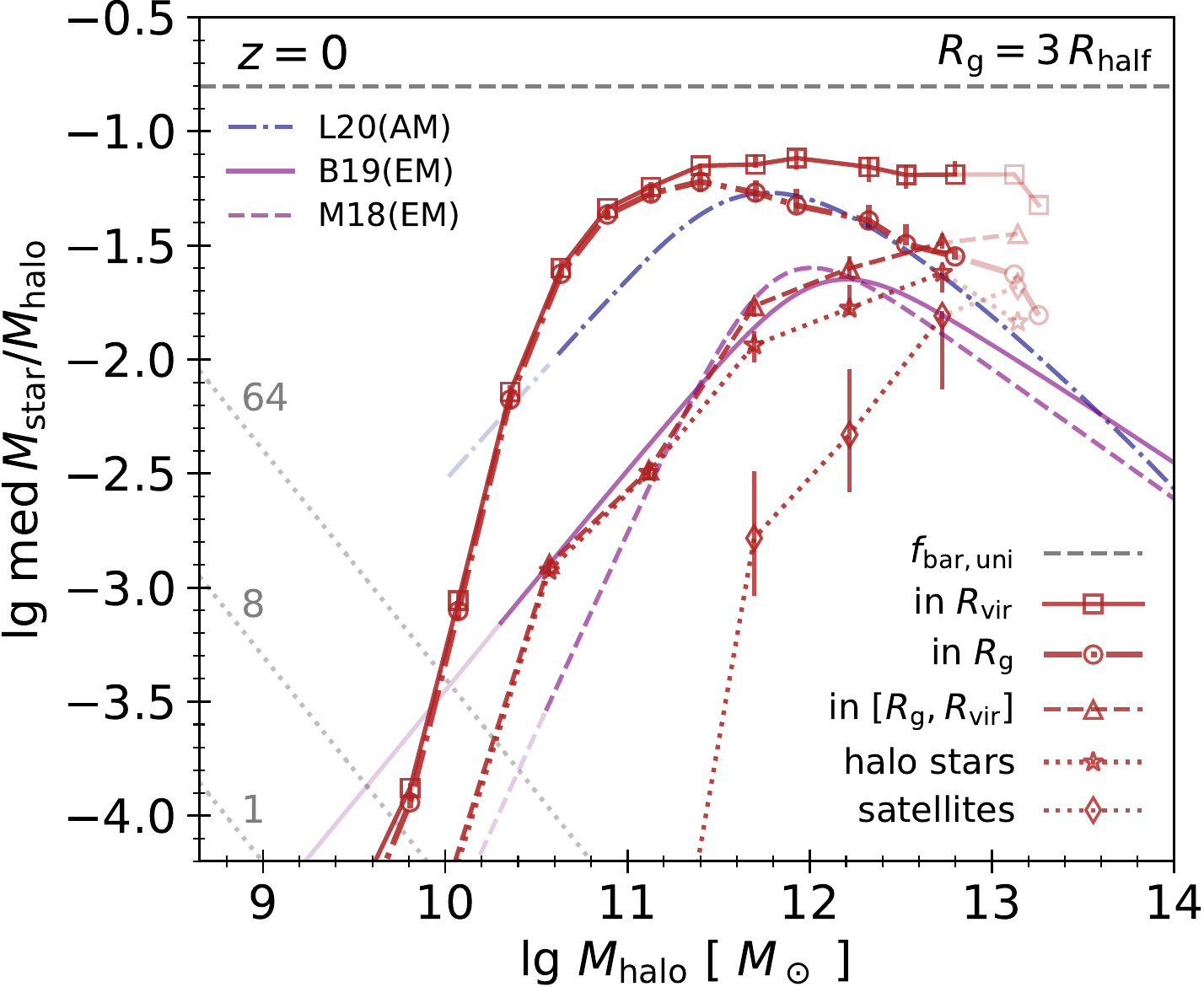}
\end{tabular}
\caption{Median stellar mass fractions of central galaxies and their parent halos in \pf{} at $z=0$.
Different symbols refer to different mass components in the simulation. The galaxy stellar mass -- halo mass relation (SHMR) is shown by circles. The galaxy stellar mass is defined as the stellar mass within 3 times the stellar half-mass radius, see section \ref{sect:halos}. Squares show the total stellar mass -- halo mass relation. The galaxy and total stellar masses differ because of the stellar mass component outside the central galaxy but within the halo (triangles). The latter is further split into the stellar mass within identified sub-halos (diamonds) and stellar mass outside identified sub-halos (stars). Uncertainties (16-84\%) are calculated via bootstrapping and shown by error bars. Light shaded symbols without error bars indicate bins containing fewer than four galaxies.
Gray dotted lines correspond to 1, 8, and 64 star particles of mass $m_{\rm b}=6.3\times{}10^4$ $M_\odot$ in halos of a given mass (from bottom to top).
Estimates of the galaxy stellar mass -- halo mass relation (SHMR) via abundance matching (AM) of the SMF by \protect\cite{Leja2020} (dot-dashed line) and via empirical modeling (EM, \citealt{Moster2018, Behroozi2019}, dashed and solid lines) are also shown. Extrapolations beyond the stellar mass range of the observational data are shown by a light colored line.  
The galaxy stellar fraction decreases with increasing halo mass for $M_{\rm halo}>10^{11.5}$ $M_\odot$, while the total stellar fraction scales only weakly with $M_{\rm halo}$ over the $10^{11.5-13}$ $M_\odot$ mass range. The increase in the stellar mass outside massive galaxies is driven by an increasing halo star contribution and, for $M_{\rm halo}>10^{12}$ $M_\odot$, by a higher lock-up of stars in satellite galaxies.}
\label{fig:SHMR}
\end{figure}

The galaxy stellar mass -- halo mass relation (SHMR) is closely related to the SMF. The latter can be obtained from the former (and vice versa) with the help of the HMF. We use this abundance matching (AM) approach \citep{Kravtsov2004, Vale2004, Behroozi2010a} to calculate the SHMR from the SMFs provided by \cite{Leja2020}. For simplicity of the calculation we ignore the scatter ($\sim{}0.2$ dex, \citealt{Reddick2013, Zu2015a}) of the SHMR. Given that \pf{} matches approximately the $z=0$ SMF of \cite{Leja2020}, we also expect a reasonable agreement with the derived SHMR. An alternative method of estimating the SHMR from observational data is empirical modeling \citep{Moster2018, Behroozi2019}. Here, we expect some level of disagreement, however, as these models are based on the SMFs derived from traditional SED-fitting, see discussion of Fig.~\ref{fig:SMFz}.

The SHMR of central galaxies in \pf{} at $z=0$ is given in Fig.~\ref{fig:SHMR}. The ratio between galaxy stellar mass and halo mass rises quickly with increasing mass for $M_{\rm halo}<10^{11}\,M_\odot$, it reaches a peak near $M_{\rm halo}\sim{}10^{11.4}\,M_\odot$ with a maximum value $\sim{}0.42$ dex below the universal baryon fraction, and then decreases slowly toward larger masses. We obtain a qualitatively similar result if we define the galaxy radius as $R_{\rm g}=0.1\,R_{\rm vir}$ instead of $3\,R_{\rm half}$ (not shown). However, in this case, the decline of the galaxy stellar fraction with increasing halo mass is shallower and the peak is shifted to $M_{\rm halo}\sim{}10^{11.7}\,M_\odot$. 

Overall, the SHMR of \pf{} is in qualitative agreement with the AM prediction based on the \cite{Leja2020} SMF for $\sim{}10^{11.5}-10^{13}\,M_\odot$ halos and for our fiducial choice $R_{\rm g}=3\,R_{\rm half}$. The SHMR in \pf{} peaks at lower halo masses (by about 0.4 dex), however, and galaxies in simulated halos with $M_{\rm halo}\sim{}10^{10.5}-10^{11.4}\,M_\odot$ have higher stellar masses (by about 0.3 dex).
The latter result may partly explain the overestimate of the extra-galactic stellar component around simulated MW analogs discussed in section \ref{sect:MWanalogs} as their stellar halos are largely built from tidally disrupted galaxies in $M_{\rm star}\sim{}10^{8.5}\,M_\odot$ (i.e., $M_{\rm halo}\sim{}10^{11}\,M_\odot$) halos \citep{Purcell2007}.
As expected, empirical estimates \citep{Moster2018, Behroozi2019} differ significantly from the simulation estimates with the former showing overall lower stellar masses in $M_{\rm halo}>10^{10}\,M_\odot$ halos.

We now further investigate the decrease of the galaxy stellar fraction in massive galaxies seen in \pf{}. A similar behavior has been found empirically and it is often attributed to the quenching of star formation by AGN feedback (e.g., \citealt{Croton2006, Martizzi2012, Dubois2013a, Wellons2022}). Given the lack of the latter in \fb{}, there are several remaining possibilities which could explain this result. 
First, more massive halos could lose a larger fraction of their gas, e.g., by stellar feedback driven outflows, before they are converted to stars. We can discount this possibility since $10^{12}$ $M_\odot$ halos contain about 75\% of the universal baryon fraction, see section \ref{sect:MWanalogs}. Furthermore, a more detailed study of the baryon content of \pf{} halos (Feldmann et al. in prep) shows that the baryon fraction of massive halos does not strongly decrease with increasing halo mass.
Secondly, more massive halos could convert a smaller amount of the available baryons into stars potentially due to, e.g., the formation of a stable virial shock which keeps much of baryons in a hot, dilute state \citep{Birnboim2003, Keres2005, Faucher-Giguere2011, Stern2020}. Finally, a similar amount of baryons may be converted into stars but the distribution of the stars could be more extended in more massive halos, e.g., a larger fraction of the stellar mass could reside in satellite galaxies or in a stellar halo potentially build from minor and major mergers (e.g., \citealt{Naab2009, Feldmann2010, Oser2010, Hilz2013, Rodriguez-Gomez2016, Dubois2016}).

To investigate these latter possibilities, we also show in Fig.~\ref{fig:SHMR}, the total stellar mass in halos, the stellar mass in satellite galaxies, and the stellar mass in halo stars (defined as stars within a halo but outside any galaxy).
While the ratio between galaxy stellar mass and halo mass decreases with increasing halo mass for $M_{\rm halo}>10^{11.5}$ $M_\odot$, the stellar mass within the halo (total stellar mass) is an approximately constant fraction of the halo mass over the $M_{\rm halo}=10^{11.5}-10^{12.5}$ $M_\odot$ regime, with potentially a weak decline at the highest halo masses. Hence, we can largely exclude the second possibility mentioned above and conclude that a change in the spatial distribution of the stellar component, rather than a change in the baryonic conversion efficiency, drives the high-mass turn-over in the SMFs seen in Fig.~\ref{fig:SMFz} and the reduction of the galaxy stellar fraction in massive halos seen in Fig.~\ref{fig:SHMR}.

Taking a closer look, we see that stars that do not belong to identified sub-halos (``halo stars'') make up the majority of the stellar mass outside of centrals in $M_{\rm halo}\lesssim{}10^{12}$ $M_\odot$ halos. In more massive halos, stars locked up in satellite galaxies also contribute at a significant level. For Milky-Way like systems ($M_{\rm halo}\sim{}10^{12}$ $M_\odot$), \pf{} predicts that the ratio between the stellar mass outside the central galaxy and the galaxy stellar mass is $\sim{}0.45\pm{}0.05$ (if $R_{\rm g}=3\,R_{\rm half}$) and $\sim{}0.13\pm{}0.03$ (if $R_{\rm g}=0.1\,R_{\rm vir})$, i.e., a sizable, but definition-dependent fraction of the total stellar mass resides outside central galaxies. A similar conclusion was reached by \cite{Pillepich2014} who analyzed the stellar mass outside galaxies for a set of cosmological volume \citep{Vogelsberger2014} and zoom-in \citep{Guedes2011, Marinacci2014} simulations using $R_{\rm g}=2\,R_{\rm half}$. They reported ratios ranging from $\sim{}0.1$ to $0.6$, depending on the simulation suite, for Milky-Way like halos, similar to our findings. We conclude that the decrease of the galaxy stellar fraction with increasing halo mass in \pf{} is driven primarily by an increasing contribution of a smooth halo star component and, at the highest masses, by a higher amount of stars in satellite galaxies.

\subsection{Galaxy sizes}
The sizes of \fb{} galaxies are presented in \cite{Rohr2022}. At $z=0$, the stellar half-mass radii of \fb{} galaxies with $M_{\rm star}\sim{}10^{9.5-10.5}\,M_\odot$ are $\sim{}3-5$ kpc, in broad agreement with effective radii of observed galaxies (e.g., \citealt{Mowla2019, Nedkova2021}). In contrast, massive galaxies ($M_{\rm star}>10^{11}\,M_\odot$) in \fb{} are more compact (by $\sim{}0.2-0.3$ dex) possibly because of the lack of AGN feedback, while low mass galaxies tend to have larger sizes (by $\sim{}0.3$ dex) than observed. The agreement with observations is better at $z=2$, when low mass galaxies ($<10^{9.5}\,M_\odot$) have sizes similar to those of observed star forming galaxies \citep{Mowla2019, Nedkova2021} while more massive \fb{} galaxies ($>10^{10}\,M_\odot$) have typical sizes falling between those of observed star forming and quiescent galaxies. We caution that various systematics affect this preliminary comparison with observations (see, e.g., \citealt{Genel2018a}). A more robust analysis that calculates the sizes of \fb{} galaxies via mock observations as well as a more systematic exploration of galaxy morphology is left for future work.

\section{Cosmic evolution of gas, stars, and star formation}
\label{sect:CosmicEvolution}

A major achievement of observational efforts with the Hubble and Spitzer Space Telescopes, as well as ground-based instruments, has been to map out the cosmic star formation history and stellar mass build-up from the Cosmic Dawn to the present time \citep{Lilly1996, Madau1996, Hopkins2006, Madau2014, Bouwens2015}. In addition, observations of the neutral and molecular hydrogen content have made it possible to study how star formation in galaxies is fueled, see, e.g., \cite{Walter2020}. Clearly, these observations provide an important point of comparison for galaxy models. In this section, we compare the evolution of the cosmic density of SFR, stellar mass, and atomic and molecular gas in \pf{} to observational data to further validate, and explore the limitations of, the FIRE-2 physics model.

\subsection{Cosmic star formation history and stellar mass}
\label{sect:CSFH}

\begin{figure*}
\begin{tabular}{cc}
\includegraphics[width=80mm]{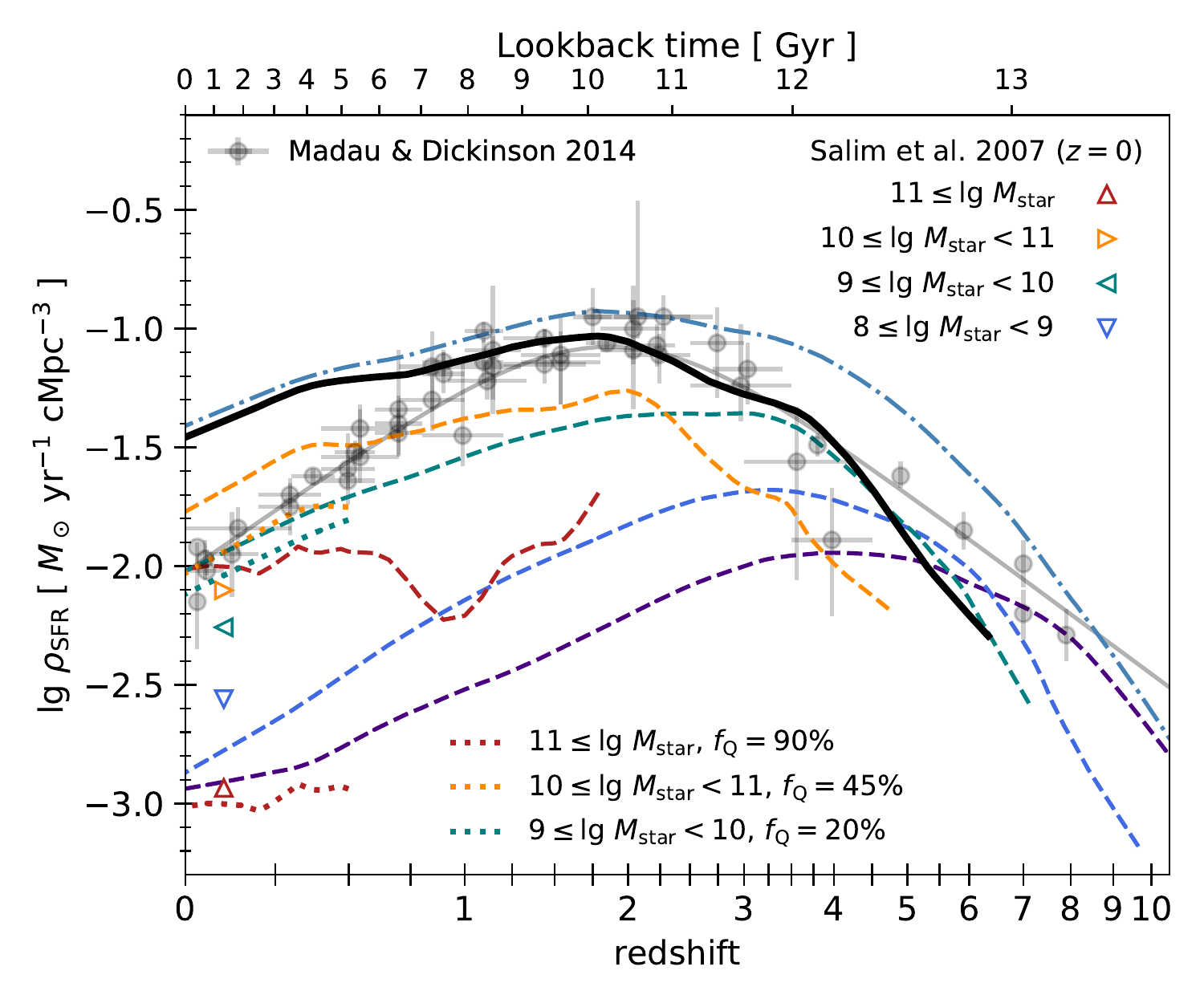} & 
\includegraphics[width=80mm]{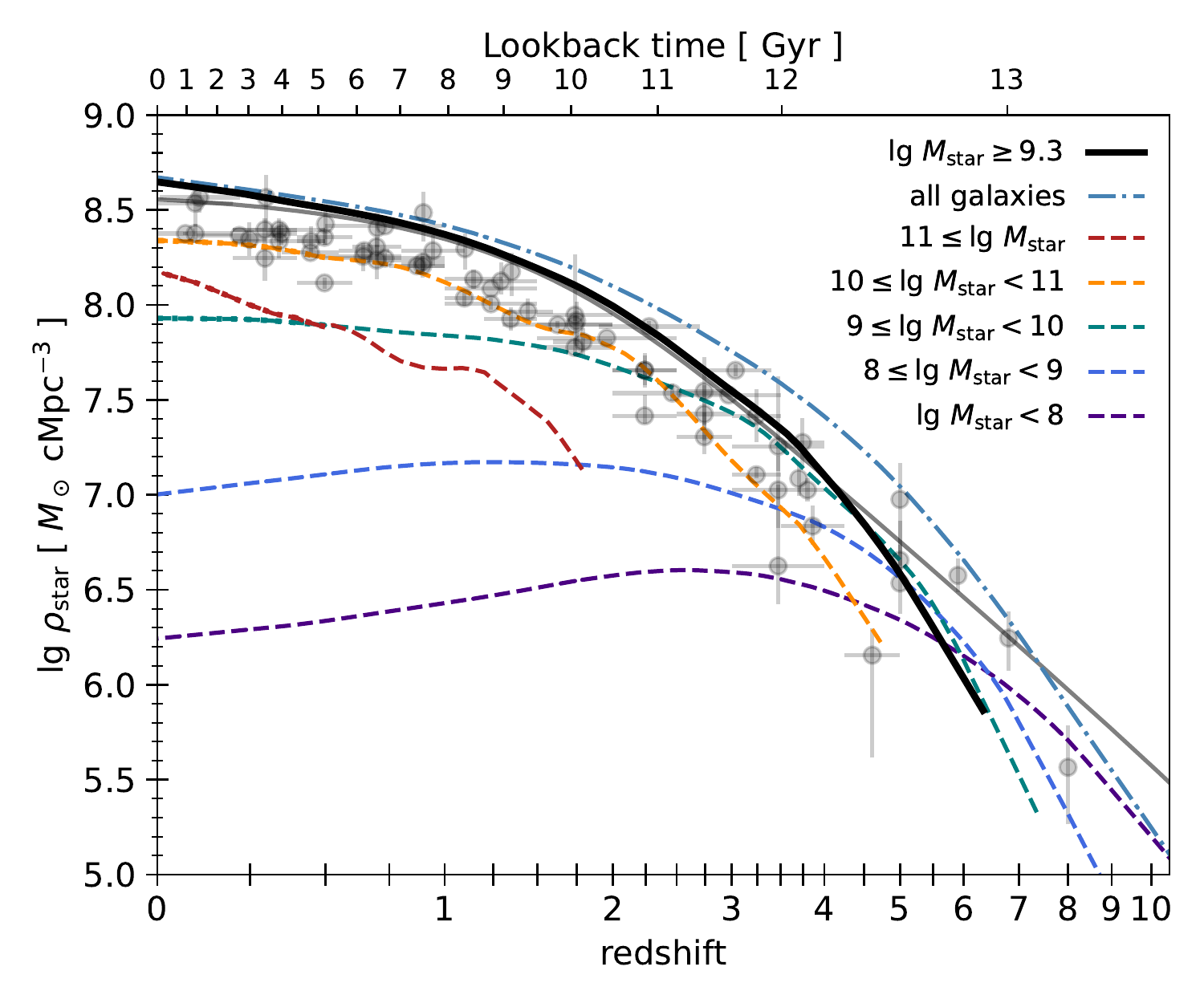}
\end{tabular}
\caption{Cosmic star formation history (CSFH, left panel) and cosmic stellar growth history (CSGH, right panel) in \pf. Blue dot-dashed lines show the CSFH and CSGH of all identified galaxies in the simulation volume, while black solid lines correspond to the case when low mass galaxies with $\lg{}M_{\rm star}/M_\odot<9.3$ are excluded. Star formation rates (SFRs) of simulated galaxies are averaged over the past 20 Myr. The abundance of galaxies is re-weighted to account for cosmic variance in the halo mass function, see Appendix \ref{app:reweighting}. Dashed lines show the contributions of galaxies in various stellar mass ranges (see legend).
Triangles are observational estimates of the cosmic SFR density at $z\sim{}0.1$ for the same stellar mass ranges \protect\citep{Salim2007}.
Gray symbols and lines refer to an observational compilation by \protect\cite{Madau2014}. The gray line in the right panel is the integral of the CSFR reduced by an effective stellar mass loss of 27\%. The observational data is converted to a Chabrier IMF using the conversion factors of \protect\cite{Madau2014}. The observational estimates of the CSFH and CSGH include galaxies above a luminosity threshold of $0.03 L_*$ which corresponds approximately to a mass threshold of $\lg M_{\rm star}/M_\odot\sim{}9.3$ at $z=0-3$.
The CSFH in \pf{} is in good agreement with observations at $z\sim{}1.5-4.5$. At $z<1$, massive, low $z$ galaxies often do not efficiently quench their star formation, resulting in an over-estimation of the CSFH possibly as a result of the lack of AGN feedback in \pf{}. Excluding a fraction $f_{\rm Q}$ of quenched galaxies \protect\citep{Behroozi2019} by hand (dotted lines) results in much better agreement with the observational estimate (triangles).
Intermediate mass galaxies ($9\leq{}\lg M_{\rm star}/M_\odot<10$, teal-colored dashed line) dominate the CSFH and CSGH at $z=3-5$, and galaxies with $10\leq{}\lg M_{\rm star}/M_\odot<11$ (orange dashed line) at $z\lesssim{}2$. In contrast, most of the star formation and stellar mass build-up during the Epoch of Re-ionization takes place in galaxies with low stellar masses ($\lg M_{\rm star}/M_\odot<8$, purple line) and low SFRs ($<0.03$ $M_\odot\,{\rm yr}^{-1}$). 
}
\label{fig:SFH}
\end{figure*}

Fig.~\ref{fig:SFH} analyzes the cosmic star formation history (CSFH) and the cosmic stellar growth history (CSGH) in \pf{}. Specifically, it plots the volume-averaged SFR density and the stellar mass density as a function of redshift both for all identified galaxies in the simulation volume and for sub-sets of galaxies based on their stellar mass. Stellar masses and SFRs of galaxies are measured within $R_{\rm g}=3\,R_{\rm half}$. SFRs are averaged over the past 20 Myr. The abundances of simulated galaxies are re-weighted based on their halo masses as described in Appendix \ref{app:reweighting}. We use a Locally Weighted Scatterplot Smoothing approach \citep{Cleveland1979} to reduce the noise in our predictions for the CSFH and CSGH.

When analyzing the CSFH and CSGH for galaxies of different stellar masses, we find that cosmic star formation and stellar mass are dominated by low mass galaxies at high $z$ ($M_{\rm star}<10^8\,M_\odot$ at $z>7$). With decreasing $z$, more massive galaxies take over as main contributors. Since Cosmic Noon, galaxies with $10^{10}$ $M_\odot$ $<M_{\rm star}<10^{11}$ $M_\odot$ dominate both the CSFH and the CSGH.

We can compare the prediction of our simulation with the compilation of observational data by \cite{Madau2014}. We take the data as is except that we adjust stellar masses and SFRs for the assumption of a \cite{Chabrier2003} IMF using the conversion factors provided by authors. The observational data only includes sufficiently luminous galaxies ($L>0.03L_*$) which corresponds to a stellar mass threshold of approximately $M_{\rm star, lim}=10^{9.3}$ $M_\odot$ over $z=0-3$ \citep{Madau2014}.

We integrate the fit to the CSFH reported by \cite{Madau2014} to obtain the corresponding average CSGH as follows:
\begin{equation}
\rho_{\rm star}(t) = \left[1 - R_{\rm eff}(t)\right] \int_0^t \rho_{\rm SFR}(t')dt',
\end{equation}
where $R_{\rm eff}=0.27$ is the effective mass return fraction\footnote{This value was adopted by \cite{Madau2014} based on the asymptotic mass return fraction $R$ of a \cite{Salpeter1955} IMF. However, $R_{\rm eff}$ generally differs from $R$. In \pf{}, $R_{\rm eff}(z)\sim{}0.35 - 0.11z$ holds for $z=0-6$ if $\lg{}M_{\rm star, lim}=10^{9.3}$ $M_\odot$ is adopted. Given that  $R_{\rm eff}$ evolves with $z$ in a mass threshold dependent manner, we adopt a constant value $R_{\rm eff}=0.27$ for simplicity. For a non-zero mass threshold, $R_{\rm eff}$ can become smaller than zero (i.e., $1-R_{\rm eff}>1$) at high $z$. The reason being that galaxies with masses below the threshold never contribute to $\rho_{\rm SFR}$, but the stellar mass they form is included in $\rho_{\rm star}$ once the masses of their descendants exceed the threshold.}, which depends not only on the IMF but also on $\lg{}M_{\rm star, lim}$.

Comparing the simulation predictions (thick black line) and observations (thin gray line) in Fig.~\ref{fig:SFH}, we find excellent agreement over $z\sim{}1.5-4.5$. Measuring stellar masses and SFRs not within $R_{\rm g}$ but within virial radii increases the CSFH and CSGH by about 0.13 dex and 0.18 dex. At higher redshifts, the CSFH and CSGH of $M_{\rm star, lim}>10^{9.3}$ $M_\odot$ galaxies falls short of the observational data. Here, however, the assumed equivalence between $L=0.03L_*$ and $M_{\rm star, lim}>10^{9.3}$ likely does not hold given the younger stellar ages and lower metallicities of high $z$ galaxies. Generally, the inferred CSFH (or CSGH) decreases much faster with increasing $z$ when galaxies with low stellar masses (here $\lg{}M_{\rm star}<10^{9.3}\,M_\odot$) are excluded given their increased contribution to the total CSFH and CSGH at higher $z$. We defer a detailed analysis of the high redshift properties of \fb{} galaxies to future work. At $z\lesssim{}0.7$, the CSFH in \pf{} differs noticeably from observational data. At $z=0$ the predicted SFR density exceeds observations by $\sim{}0.5$ dex, while the stellar density is too high by $\sim{}0.1-0.2$ dex.

\begin{figure*}
\begin{tabular}{cc}
\includegraphics[width=80mm]{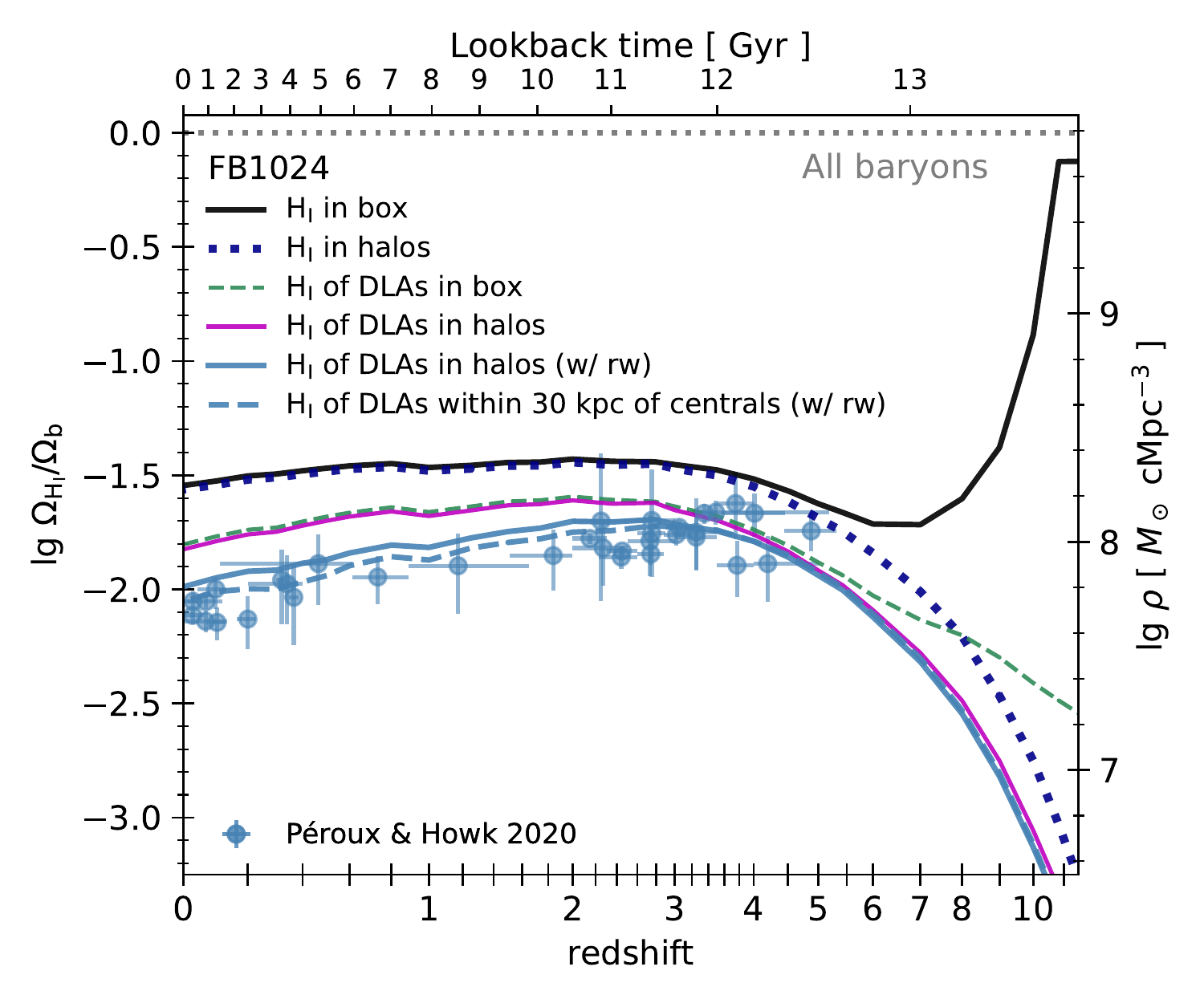} & 
\includegraphics[width=80mm]{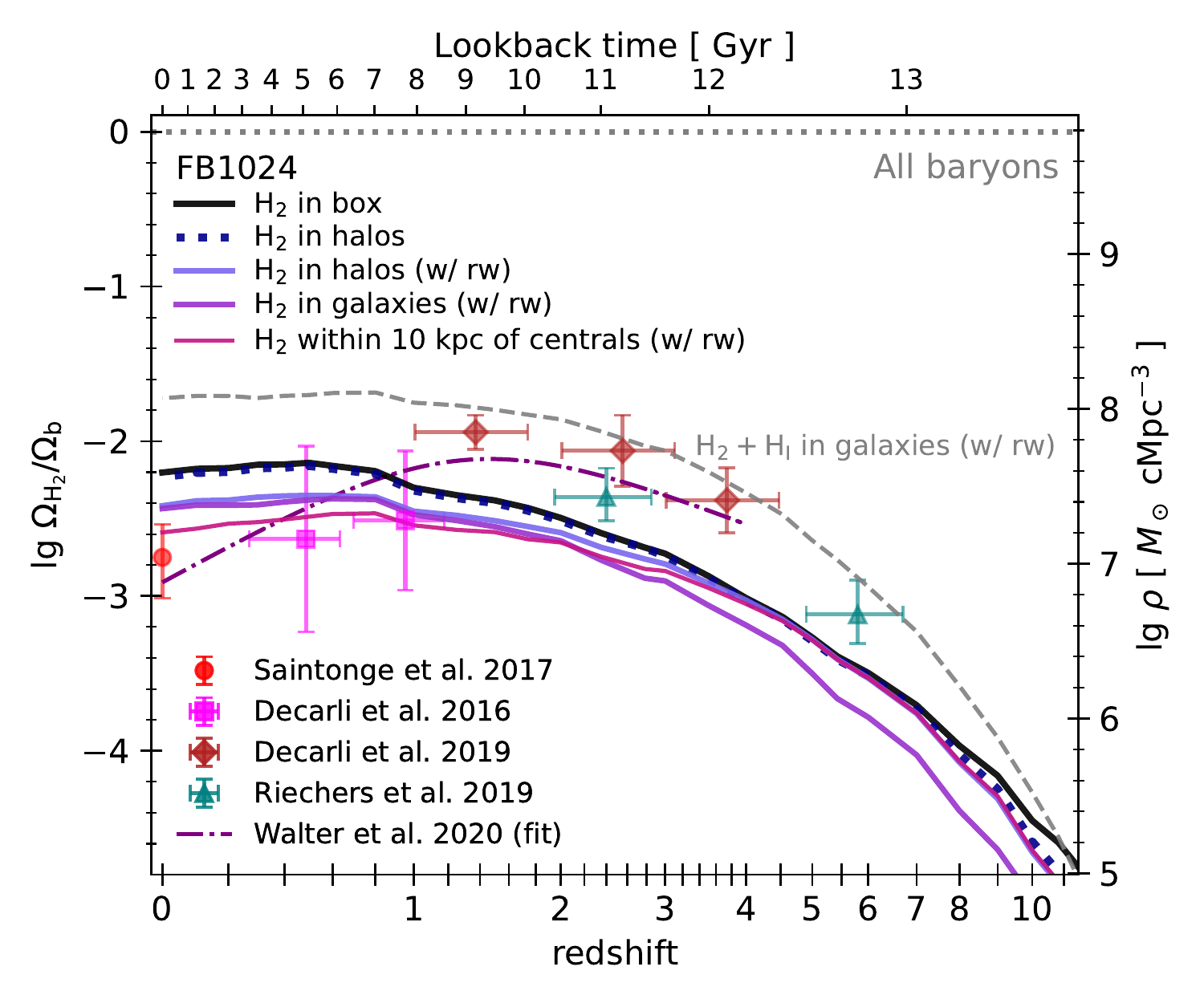}
\end{tabular}
\caption{
Cosmic evolution of the mass fractions of atomic and molecular hydrogen in \pf. (Left panel) The mass fraction of atomic hydrogen (${\rm H_I}$) in the simulation volume is shown by a black solid line. This mass fraction decreases from the beginning of re-ionization ($z\sim{}11$) until $z\sim{}7$. Between $z\sim{}4$ and $z=0$ the fraction in ${\rm H_I}$ in \pf{} remains approximately constant. At those times, the vast majority of atomic gas resides in DM halos (blue dotted line).
A green dashed line shows the fraction of ${\rm H_I}$ with column densities above $2\times{}10^{20}$ cm$^{-2}$ (damped Lyman-$\alpha$ absorbers, DLAs). At high redshift ($z>6$), DLAs can also be found outside the virial radii of dark matter halos identified in \pf. The cosmic ${\rm H_I}$ density in DLAs peaks around $z\sim{}2-3$.
Blue and pink solid lines show the mass fraction of atomic hydrogen in DM halos attributable to DLAs with and without re-weighting (`rw') by halo abundance, see Appendix \ref{app:reweighting}. The blue dashed line shows the ${\rm H_I}$ mass in DLAs within a 30 proper kpc radius around central galaxies. Also included in the panel are a compilation of observational data \protect\citep{Noterdaeme2009, Noterdaeme2012, Delhaize2013, Rhee2013, Zafar2013, Crighton2015, Hoppmann2015, Rhee2016, Sanchez-Ramirez2016, Rao2017, Jones2018, Rhee2018, Bera2019, Hu2019} by \protect\cite{Peroux2020}. Observational estimates for $z>0.4$ were obtained via absorption spectroscopy and are thus typically limited to atomic hydrogen in DLAs. Simulations and observations agree well at those redshifts.
(Right panel) The mass fraction of molecular hydrogen (${\rm H_2}$) in the simulation volume is shown by a black solid line. Other lines show the ${\rm H_2}$ mass fraction in halos, in galaxies (here understood as the molecular hydrogen within a sphere of radius $0.1\,R_{\rm vir}$ located at the (sub-)halo center), and within 10 kpc of central galaxies, with and without re-weighting, see legend. The figure further includes observational estimates \protect\citep{Decarli2016, Saintonge2017, Decarli2019, Riechers2019} based on carbon-monoxide (CO) line emission with a CO-to-${\rm H_2}$ conversion factor $\alpha_{\rm CO}=3.6$ $M_\odot$ (K km s$^{-1}$ pc$^2$)$^{-1}$. The cosmic ${\rm H_2}$ mass fraction in \pf{} increases from early times until $z\sim{}1$ in agreement with observations but does not decline steeply at later times as observations may imply.
Mass fractions are reported in units of the cosmic baryonic density ($\rho_{\rm bar, uni}\sim{}6.2\times{}10^9$ $M_\odot$ cMpc$^{-3}$ for our adopted cosmology; left $y$ axis) or in solar masses per comoving Mpc$^3$ (right $y$ axis).}
\label{fig:OmegaHIH2}
\end{figure*}

To understand the origin of this discrepancy at low $z$ we separate the \pf{} sample into various stellar mass bins and calculate their contribution to the cosmic SFR density. We compare the simulation data with observational estimates of the cosmic SFR density \citep{Brinchmann2004b, Juneau2005, Salim2007}. In the following, we specifically compare with the work by \cite{Salim2007} but we found similar results when using the data by \cite{Brinchmann2004b}. The SFR density in low mass galaxies is in approximate agreement with observations. Specifically, \pf{} predicts a contribution of $10^8$ $M_\odot$ $\leq{}M_{\rm star}$ $<10^{9}\,M_\odot$ galaxies that is about 0.25 dex lower than observed and a contribution from galaxies of intermediate mass ($10^9$ $M_\odot$ $\leq{}M_{\rm star}$ $<10^{10}$ $M_\odot$) is about 0.3 dex higher than observed. In contrast, massive galaxies ($10^{10}$ $M_\odot$ $\leq{}M_{\rm star}$ $<10^{11}$ $M_\odot$ and $10^{11}$ $M_\odot$ $\leq{}M_{\rm star}$)  in \pf{} contribute at much higher levels to the cosmic SFR than found observationally ($\sim{}0.5$ dex and $\sim{}1$ dex). Massive galaxies thus appear primarily responsible for the high cosmic SFR at low $z$, presumably because of the lack of AGN feedback in \pf{}.

To test whether the low quenched fraction in \pf{} can explain the high cosmic SFR, we reduce the cosmic SFR in each stellar mass bin by the expected fraction $f_{\rm Q}$ of quenched galaxies given by \cite{Behroozi2019}. Specifically, we use $f_{\rm Q}=20\%$, $45\%$ and $90\%$ for the stellar mass bins of $10^9$ $M_\odot$ $\leq{}M_{\rm star}$ $<10^{10}$ $M_\odot$, $10^{10}$ $M_\odot$ $\leq{}M_{\rm star}$ $<10^{11}$ $M_\odot$, and $10^{11}$ $M_\odot$ $\leq{}M_{\rm star}$. As shown in Fig.~\ref{fig:SFH}, a reduction by $f_{\rm Q}$ brings the simulation predictions in much better agreement with observations. 

Even though \pf{} underestimates the fraction of massive, quenched galaxies at low $z$, the properties of star forming galaxies themselves appear well reproduced (see section \ref{sect:PropGals}). Hence, we can infer that the physical mechanism(s) responsible for quenching of star formation should not affect the galaxy scaling relations (e.g., the molecular gas sequence or the mass--metallicity relations) of the population of star forming galaxies too severely. 
Furthermore, going back in time, we see that the stellar feedback model in \pf{} explains well the evolution of the average SFR and stellar mass density at $z\gtrsim{}1-5$. In other words, \pf{} does not leave much room for AGN feedback to affect the CSFH and CSGH at early cosmic times. Instead, the role of AGN feedback at $z>1$ may be to turn quiescent galaxies with low, but non-zero sSFRs into the truly passively evolving galaxies observed at those redshifts \citep{Kriek2006, Straatman2016}.

\subsection{Cosmic gas density}

The evolution of the cosmic gas density is connected to the evolution of the cosmic star formation rate and stellar mass density. On the one hand, a larger fraction of the overall baryonic mass gets locked up in stars with increasing cosmic time, thus reducing the total gas density in the Universe. On the other hand, stellar feedback, a natural by-product of star formation, strongly affects the properties of the cosmic gas, in particular the abundance of atomic and molecular hydrogen.

Several previous FIRE studies analyzed the ${\rm H_I}$ content within the zoom-in regions around individual galaxies, but did not fully sample the intergalactic medium (e.g., \citealt{Faucher-Giguere2015, Faucher-Giguere2016, Hafen2017, Stern2021}). With \fb{}, we can more rigorously quantify the integrated neutral hydrogen mass and column density distribution across cosmic history. Caveats include the simplified modeling of local shielding of UV/ionizing photons in FIRE \citep{Hopkins2018} and the dependence of our results on the chosen UV background, here \cite{Faucher-Giguere2009}.

The left panel of Fig.~\ref{fig:OmegaHIH2} compares the evolution of the cosmic ${\rm H_I}$ mass density in \pf{} with observational data compiled by \cite{Peroux2020}. Given the challenge in detecting the 21 cm hyperfine transition emission line of atomic hydrogen beyond $z\sim{}0.4$, the evolution of the ${\rm H_I}$ mass fraction at higher $z$ is primarily constrained by absorption spectroscopy of high column density systems, specifically Damped Lyman-$\alpha$ systems (DLAs). To ease the comparison, Fig.~\ref{fig:OmegaHIH2} thus reports the mass density of atomic hydrogen in \pf{} both restricted to DLAs (i.e., only counting ${\rm H_I}$ with column densities above $2\times{}10^{20}$ cm$^{-2}$) as well as the overall amount. To this end, we estimate the column density of atomic hydrogen for each gas particle as $N_{\rm H_I}=\Sigma_{\rm gas} f_{\rm H_I} X / m_{\rm H}$, with $\Sigma_{\rm gas}$ calculated as described in section \ref{sect:BaryonicPhysics}.

As the figure shows, the cosmic  ${\rm H_I}$ density in DLAs predicted by \pf{} is in good agreement with observational data once we re-weight the halo abundance to account for the finite box size. In particular, we find that the atomic hydrogen density of DLAs associated with halos changes by a factor $\sim{}2$ between $z\sim{}3-4$ and today's Universe. About 2-4\% of the cosmic baryon density $\Omega_{\rm b}$ is in atomic hydrogen at $z<4$. A comparison with Fig.~\ref{fig:SFH} reveals that the mass in stars exceeds the mass in atomic hydrogen at $z\lesssim{}1.5-2$ in agreement with observational estimates \citep{Driver2018}.

Comparing the total amount of atomic hydrogen in \pf{} (solid black line) with the ${\rm H_I}$ contribution by DLAs (dashed green line) we see that out to at least $z\sim{}5$ the majority of the cosmic ${\rm H_I}$ mass (50\%--65\%) is associated with DLAs \citep{Wolfe1986, Lanzetta1991}. The remaining fraction of atomic hydrogen ($\sim{}35-50\%$) is significantly higher than the 10-20\% contribution expected from sub-DLAs ($10^{19}$ cm$^{-2}<N_{\rm H_I}<2\times{}10^{20}$ cm$^{-2}$) at those redshifts \citep{Peroux2005, Zafar2013, Berg2019}. This suggests that atomic hydrogen with low column densities ($N_{\rm H_I}<10^{19}$ cm$^{-2}$) contributes rather significantly to the cosmic  ${\rm H_I}$ density.

The fraction of atomic hydrogen in DLAs decreases noticeably towards higher redshifts in qualitative agreement with observations \citep{StorrieLombardi2000}. For instance, \pf{} predicts that only $\sim{}25\%$ of the cosmic ${\rm H_I}$ mass is hosted by DLAs at $z=8$. While the ${\rm H_I}$ density in DLAs declines with increasing redshift at $z>4$, an even stronger decline is seen for those DLAs that are associated with dark matter halos (blue solid line) during the Epoch of Re-ionization (EoR, here $z\sim{}6-11$). For instance, we predict that the ${\rm H_I}$ density in DLAs associated with halos is lower by over an order of magnitude at $z=9$ compared with $z\sim{}3$. More generally, while almost all of the cosmic atomic hydrogen at $z<5$ resides within halos (blue dotted line), most of the atomic hydrogen at $z>7$ can be found outside halos, see also \cite{Villaescusa-Navarro2018}.
Fully accounting for atomic hydrogen, especially during the EoR, thus requires modeling the contribution outside halos as well as from systems with  column densities below those of DLAs.

The right panel of Fig.~\ref{fig:OmegaHIH2} shows the evolution of the cosmic ${\rm H_2}$ mass density in \pf{}. We compare our model predictions with compilations of observational data \citep{Peroux2020, Walter2020}. Given the tight empirical correlation between molecular hydrogen and star formation rate \citep{Bigiel2008, Genzel2010, Saintonge2017, Feldmann2020}, at least in the local Universe, one might expect that the evolution of the ${\rm H_2}$ mass density mirrors the evolution of the CSFH \citep{Decarli2019, Tacconi2020}, i.e., with a peak near Cosmic Noon and a noticeably decline towards low redshift. However, this is not what we see in Fig.~\ref{fig:OmegaHIH2}. Instead, we find that the cosmic ${\rm H_2}$ density in \pf{} increases with cosmic time until $z\sim{}1$, after which it remains approximately constant down to $z=0$. The latter can be understood as follows. First, the typical molecular depletion time of \pf{} galaxies increases with increasing cosmic time, qualitatively similar to observations \citep{ Tacconi2020}. The cosmic ${\rm H_2}$ mass density thus increases relative to the CSFH with increasing cosmic time, i.e., more molecular gas is required at later times to sustain a given cosmic star formation activity. Secondly, the CSFH in \pf{} declines at late times somewhat less steeply than observations suggest (Fig.~\ref{fig:SFH}). As a consequence, the cosmic ${\rm H_2}$ mass density in \pf{} evolves only weakly at low $z$.

The evolution predicted by \pf{} differs from the findings of recent observational studies \citep{Decarli2019, Walter2020}, even though it may be broadly in line with other observational measurements \citep{Decarli2016, Saintonge2017, Riechers2019}. While this difference may indicate a potential short-coming of the FIRE physics model, we note that molecular gas plays a somewhat limited role in \pf{} given the high density threshold of star formation \citep{Hopkins2018}. Furthermore, the neutral hydrogen density predicted by \pf{} exceeds the observed molecular density at all $z$ suggesting there is sufficient neutral gas in \pf{} galaxies. Therefore, another possibility is that our approximate approach of estimating molecular fractions, see section \ref{sect:BaryonicPhysics}, breaks down at higher $z$. However, \cite{Krumholz2011c} demonstrated that this approach predicts molecular fractions with an absolute error of better than 0.1 for more than 80\% of the ISM mass of galaxies with a range of stellar masses and ISM conditions when compared to a non-equilibrium radiative transfer solution.

Observational biases are yet another concern. The abundance of molecular hydrogen is typically inferred indirectly from the line luminosity of carbon-monoxide (CO) molecules or from the continuum emission of dust grains. The latter method suffers from uncertainties in the dust-to-gas ratios and dust temperatures \citep{Scoville2014c, Liang2018, Liang2019} and includes a contribution from atomic gas (e.g., \citealt{Scoville2014c}), while the former approach requires knowledge of the conversion factor between CO luminosity and ${\rm H_2}$ mass. While this conversion factor is well constrained for molecular gas in the Milky-Way \citep{Solomon1987, Bolatto2013}, it has been shown to vary significantly with galaxy properties such as metallicity and interstellar radiation field \citep{Leroy2011, Feldmann2012e, Bolatto2013}. The conversion factor is thus a significant systematic for molecular gas estimates based on CO data, especially at higher $z$ \citep{Walter2020}. Bringing the ${\rm H_2}$ predicted by \pf{} at $z>1$ in agreement with \cite{Decarli2019} and \cite{Walter2020} would require a conversion factor that is about 0.4 dex lower than the standard value for the Milky Way. Such a reduction in the conversion factor would also help mitigating the tension between the theoretically predicted and observed molecular gas fractions in galaxies at the Cosmic Noon \citep{Narayanan2012e}, see also \cite{Lagos2015, Dave2017, Popping2019, Dubois2021}.

\begin{figure}
\begin{tabular}{c}
\includegraphics[width=80mm]{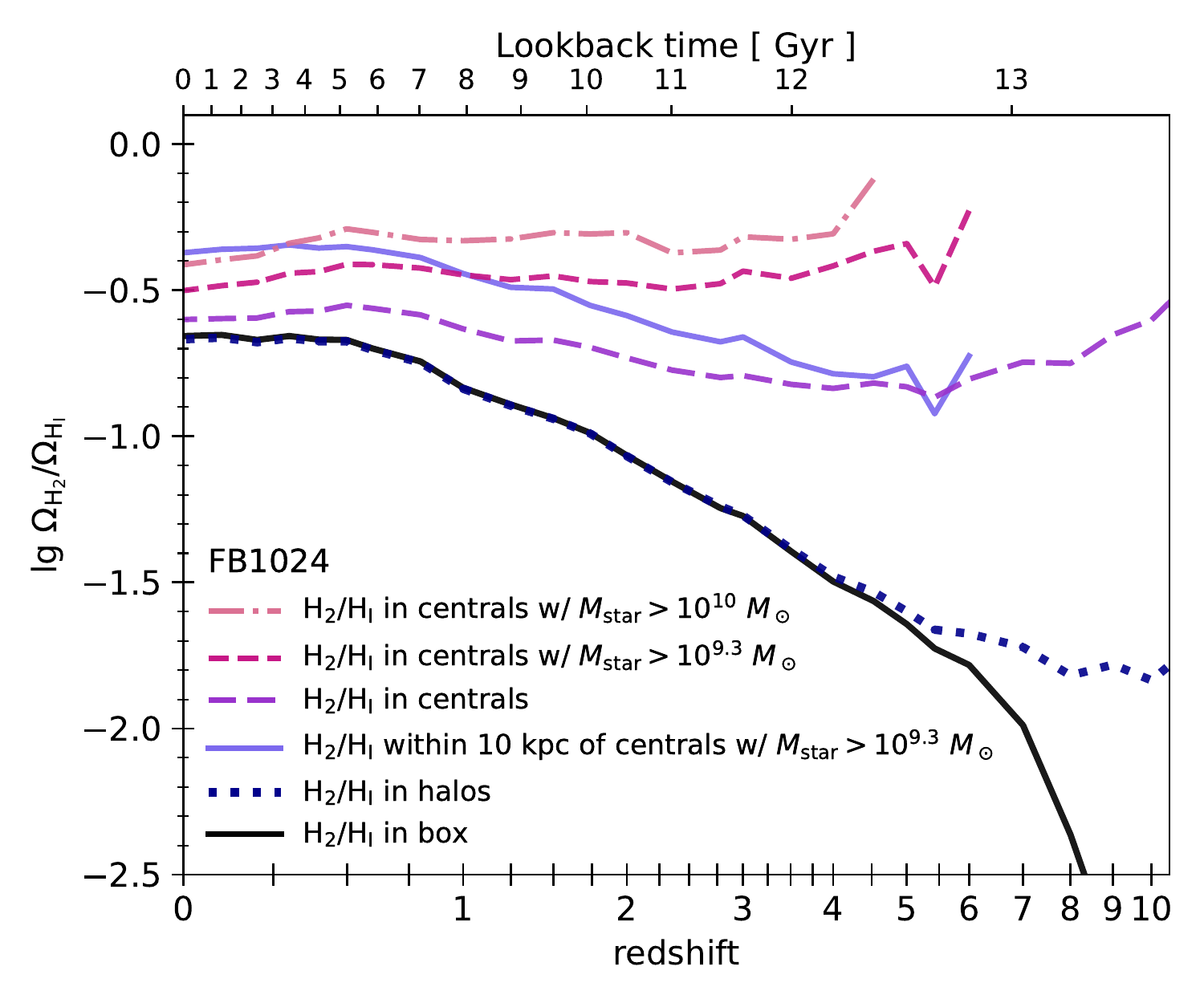}
\end{tabular}
\caption{Evolution of the mass ratio between molecular and atomic hydrogen in \pf. The ${\rm H_2}$-to-${\rm H_I}$ ratio in the simulation volume (in DM halos) is shown by a black solid line (a blue dotted line), while a purple solid line refers to the ratio between $M_{\rm H_2}$ and $M_{\rm H_I}$ within 10 physical kpc around central galaxies with stellar masses exceeding $10^{9.3}\,M_\odot$. The ${\rm H_2}$-to-${\rm H_I}$ ratio within $0.1\,R_{\rm vir}$ of central galaxies (central galaxies with $M_{\rm star}>10^{9.3}\,M_\odot$, central galaxies with $M_{\rm star}>10^{10}\,M_\odot$) is shown by a long-dashed purple line (short-dashed red line, dot-dashed pink line).
Although the cosmic abundance of molecular hydrogen relative to atomic hydrogen decreases towards higher redshift,  the ${\rm H_2}$-to-${\rm H_I}$  ratio in central galaxies of a given stellar mass is relatively constant across most of cosmic history when measured within $0.1\,R_{\rm vir}$. In contrast, the ${\rm H_2}$-to-${\rm H_I}$ ratio measured in a fixed physical radius evolves more strongly with redshift. Atomic hydrogen dominates over molecular hydrogen in the interstellar medium of all but the most massive galaxies both in the present-day Universe and at early cosmic times.}
\label{fig:H2toHI}
\end{figure}

The conversion factor for high $z$ galaxies has been empirically constrained by comparing CO emission and dynamical masses \citep{Daddi2010a}. This approach tacitly assumes, however, that gas in high redshift galaxies is predominantly molecular (e.g., \citealt{Saintonge2013a}). To test this assumption, we plot in Fig.~\ref{fig:H2toHI} the ratio between molecular and atomic hydrogen in \pf{}. The figure offers several insights. 

First, it shows that with the possible exception of the most massive galaxies at high $z$, the cold gas in galaxies is never ${\rm H_2}$ dominated. In fact, less than a third of neutral gas is in molecular form with the mass ratio between molecular and atomic hydrogen near or below 0.5. If taken at face value, this result suggests that the CO to ${\rm H_2}$ conversion factor as inferred from dynamical masses could be significantly overestimated.
Secondly, Fig.~\ref{fig:H2toHI} shows that, when averaged over cosmic scales, the ${\rm H_2}$ to ${\rm H_I}$ mass ratio decreases strongly with increasing redshift. This result holds both for the gas phases in the box as well as the gas residing in DM halos. The latter plateaus at a mass ratio of $\sim{}1.5\%$ at $z\gtrsim{}8$, while the former continues to drop with increasing $z$ during the EoR.

Finally, the ${\rm H_2}$ to ${\rm H_I}$ mass ratio within galaxies is almost independent of $z$. The normalization and redshift evolution of the latter depends on the mass of the selected galaxies (more massive galaxies tend to have a larger ${\rm H_2}$ to ${\rm H_I}$ mass ratio) as well as the radius enclosing the gas components. Calculating the mass ratio within $0.1\times{}R_{\rm vir}$ results in a flatter evolution than using a radius of fixed physical size. Given that the molecular-to-neutral gas ratio depends sensitively on gas column density and metallicity, see section \ref{sect:gasmethod}, these trends in the ${\rm H_2}$ to ${\rm H_I}$ mass ratio are likely driven by both the spatial and the stellar mass dependence of gas densities and metallicities around galaxies.

\subsection{Large scale distribution of atomic and molecular hydrogen}

\begin{figure*}
\begin{tabular}{cc}
\includegraphics[width=160mm]{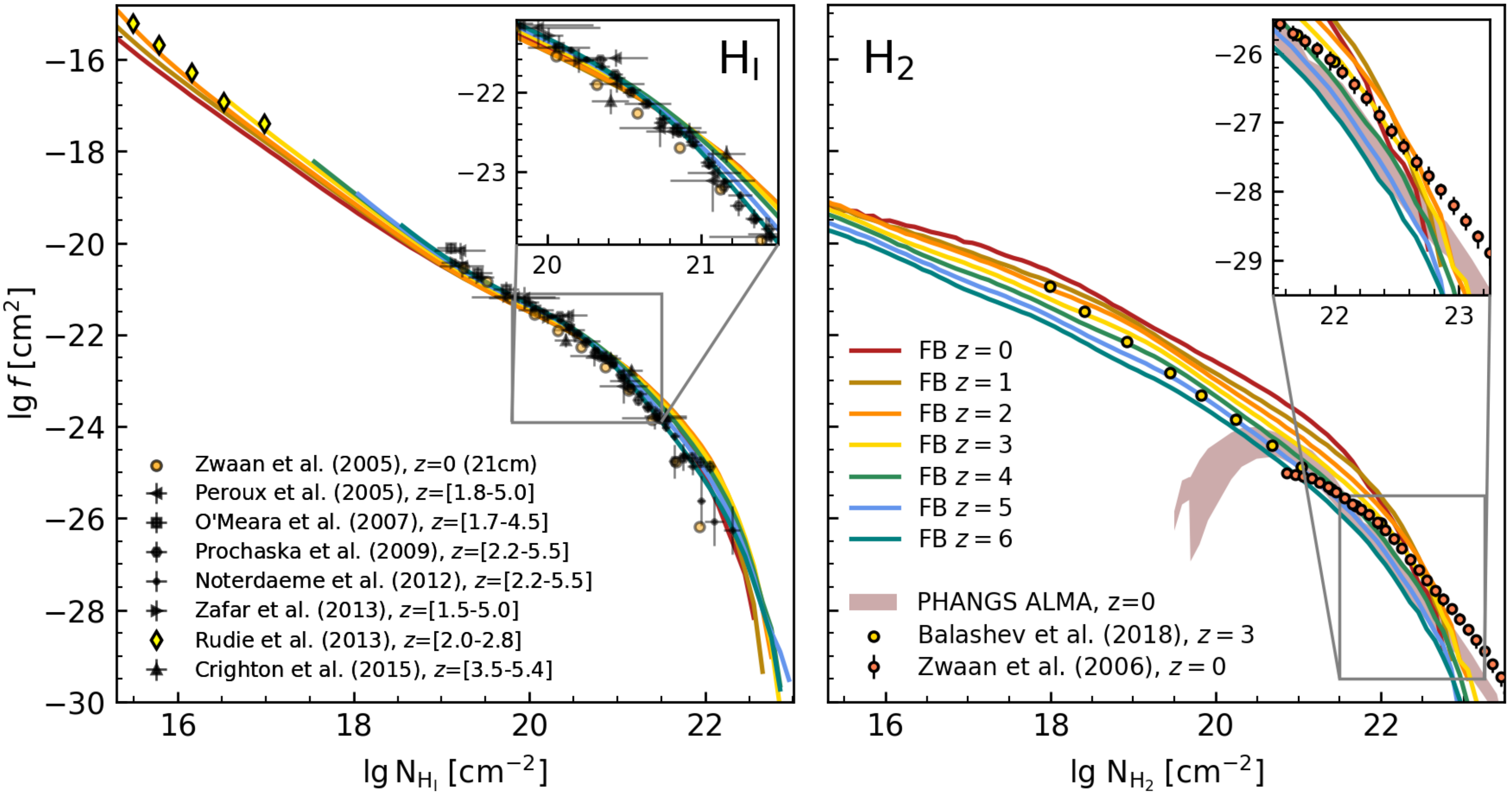}
\end{tabular}
\caption{Column density distribution functions (CDDFs, $f$) of atomic and molecular hydrogen in \pf{}. The CDDF $f(N, z)$ quantifies the number of intervening systems per unit path length $X(z)$ and unit column density $N$. (Left) The CDDF of atomic hydrogen in \pf{} reproduces well the observed ${\rm H_I}$ CDDF from quasar absorption line spectroscopy and 21 cm emission line surveys \protect\citep{Zwaan2005, Peroux2005, OMeara2007, Prochaska2009, Noterdaeme2012a, Zafar2013, Rudie2013, Crighton2015} but with some apparent deviations at $N_{\rm H_I}>10^{21}$ cm$^{-2}$. We note that the observational estimate for $z=0$ assumes optical thin emission which may not hold at high column densities \protect\citep{Zwaan2005}.
(Right) The CDDF of molecular hydrogen in \pf{} and observational estimates by PHANGS \protect\citep{Szakacs2022} and by \protect\cite{Zwaan2006} and \protect\cite{Balashev2018}.
In agreement with the literature, the ${\rm H_I}$ CDDF does not strongly evolve with cosmic time. In contrast, the ${\rm H_2}$ CDDF at $N_{\rm H_2}\leq{}10^{22}\,{\rm cm}^{-2}$ increases by more than one order of magnitude between $z=6$ and $z=0$.
}
\label{fig:CDDF}
\end{figure*}

The column density distribution function (CDDF) measures the number of intervening systems per unit column density $N$ and absorption length $X$ \citep{Bahcall1969}. The CDDF provides an excellent point of comparison for theoretical models given that it probes cosmic gas under a range of physical conditions and in a variety of cosmic environments \citep{Altay2011, McQuinn2011, Rahmati2015, Crain2017, Balashev2018, Szakacs2022}.

At $z<0.4$, the CDDF of atomic hydrogen is observationally accessible via its 21-cm line emission \citep{Zwaan2005, Peroux2020}, while Ly$\alpha$ absorption spectroscopy of background quasars can probe the CDDF at higher redshift (e.g., \citealt{Prochaska2009, Noterdaeme2009, Noterdaeme2012a}). Intervening systems include both Ly$\alpha$ forest absorbers with $N_{\rm H_I}<10^{17.2}$ cm$^{-2}$ (e.g., \citealt{Rauch1998}), Lyman limit systems ($10^{17.2}\leq{}N_{\rm H_I}/{\rm cm^{-2}}<10^{20.3}$, e.g., \citealt{Peroux2003}), and Damped Ly$\alpha$ systems ($N_{\rm H_I}\geq{}10^{20.3}$ cm$^{-2}$, e.g., \citealt{Wolfe2005}). Observationally, the ${\rm H_I}$ CDDF is approximately described by a single power-law over $N_{\rm H_I}\sim{}10^{13}-10^{21}$ cm$^{-2}$ \citep{Tytler1987} with a break at higher column densities ($\sim{}10^{20.5}-10^{21}$ cm$^{-2}$; \citealt{Peroux2003}). The shape of the ${\rm H_I}$ CDDF is almost invariant with redshift and its normalization shows only a moderate change (factor 2 between $z=4$ and $z=2.2$) with redshift \citep{Zwaan2005, Prochaska2005, Prochaska2009}.

To calculate the CDDF, we project the atomic or molecular hydrogen density in the simulation box along a specified axis onto a 2-dimensional grid with resolution of 150 comoving pc which is comparable to the $\sim{}100$ pc resolution of the PHANGS-ALMA survey at $z=0$ \citep{Leroy2021}. 
In more detail, we use a combination of \texttt{smooth} and \texttt{tipgrid} for the deposition of the ${\rm H_I}$ and ${\rm H_2}$ masses onto the grid\footnote{\url{https://github.com/N-BodyShop/smooth}}. First, \texttt{smooth} computes a smoothing length for every particle as half of the distance to the $n$th neighbor particle. We found that $n=80$ provides a good balance between over-smoothing and too high particle noise for this application.
Next, the simulation region is divided into $n_{\rm s}$ equally spaced slabs of depth $\Delta{}L=15/n_{\rm s}\,{\rm cMpc}\,h^{-1}$ for the chosen spatial direction. The advantage of using slices is that it reduces the chance of line-of-sight overlap between separate absorbing systems. However, we find practically little difference in the estimated CDDF over much of the column density and redshift range of interest when varying $n_{\rm s}$ between 1 and 10. In the following, we use $n_{\rm s}=10$ but report the CDDF only if it differs by less than 5\% from the CDDF calculated with $n_{\rm s}=1$.
Next, \texttt{tipgrid} projects particles in the same slab onto a two-dimensional grid by depositing the atomic or molecular hydrogen mass of each gas particle via the SPH scatter approach with a cubic spline kernel and the smoothing lengths calculated beforehand. The CDDF is then obtained from the column density distributions of the pixels of all slabs normalized to $\Delta{}X$, where the absorption distance $\Delta{}X$ is related to the comoving slab depth $\Delta{}L$ via $\Delta{}X = \Delta{}L\frac{H_0}{c}(1+z)^2$.

The left panel of Fig.~\ref{fig:CDDF} compares the CDDF of atomic hydrogen in \pf{} with a compilation of observational data over $z=0-5.5$. Overall the agreement is good, especially at $N_{\rm H_I}<10^{21}$ cm$^{-2}$. \pf{} predicts that the ${\rm H_I}$ CDDF does not strongly evolve with cosmic time in agreement with observations. \pf{} overestimates the incidence of low redshift systems with the highest column densities ($N_{\rm H_I}>10^{21}$ cm$^{-2}$). However, the observational estimate for $z=0$ assumes optical thin emission which may result in an underestimate at the highest column densities \citep{Zwaan2005}.

The right panel of Fig.~\ref{fig:CDDF} shows our prediction for the CDDF of molecular hydrogen ($H_2$) and compares it with observational data \citep{Zwaan2006, Balashev2018, Leroy2021, Szakacs2022}. The ${\rm H_2}$ CDDF is in broad agreement with the observations at $z=3$ but shows some differences at low $z$, in particular a steeper decrease with increasing column density for large $N_{\rm H_2}$ and a higher normalization at low column densities. In contrast to the ${\rm H_I}$ CDDF, the normalization of the ${\rm H_2}$ CDDF shows a noticeable dependence on redshift, increasing by over one order of magnitude from $z=6$ to $z=2$ at all column densities reflecting the overall increase in the cosmic molecular gas density \citep{Peroux2020}, see also Fig.~\ref{fig:OmegaHIH2}. Between $z=2$ and $z=0$, the $H_2$ CDDF slightly decreases at the highest column densities ($N_{\rm H_2}\gtrsim{}10^{22}$ cm$^{-2}$) and increases at lower column densities, leading to a change in its shape. While the ${\rm H_2}$ CDDF is ``bottom-light'' compared with the ${\rm H_I}$ CDDF, it increases monotonically with decreasing column densities down to at least $N_{\rm H_2}\sim{}10^{16}$ cm$^{-2}$, i.e., there is no indication of a turn-over in the ${\rm H_2}$ CDDF as seen in the observational study of \cite{Szakacs2022} presumably due to sensitivity and incompleteness limits.

\section{Summary and Conclusions}
\label{sect:Summary}

We have introduced the \fb{} suite, a set of galaxy formation simulations in a cosmological volume ($L=22.1$ cMpc) run down to $z=0$ with the GIZMO gravity-hydrodynamics solver in mesh-less hydrodynamics mode \citep{Hopkins2015a} and with the FIRE-2 physics model \citep{Hopkins2018}. The \fb{} volume contains about 20-30 Milky-Way analogs as well as over a thousand lower mass galaxies enabling the study of representative samples of highly resolved galaxies.
The main simulation analyzed in this paper (\pf{}), has a baryonic mass resolution of $m_{\rm b}\sim{}6.3\times{}10^4$ $M_\odot$ and a spatial resolution of $\sim{}20$ pc in dense interstellar gas, comparable to state-of-the-art zoom-in simulations. The high numerical resolution combined with the fully cosmological setting results in an unprecedented dynamic range ($\gtrsim{}10^6$) for a galaxy formation simulation. \pf{} is able to capture simultaneously the multiphase structure of the interstellar medium in galaxies and the impact of baryonic physics on cosmological scales.
Importantly, \pf{} is not tuned to specific observational data, such as the stellar mass function, but rather it implements comparably well-understood physical processes in a self-consistent fashion without adjusting model parameters. As such it provides a true prediction of galaxy formation theory in a $\Lambda{\rm CDM}$ Universe. Modeled baryonic processes include gas cooling, star formation, stellar winds, supernova feedback, and radiative feedback (photo-ionization, photo-electric heating, and radiation pressure). Feedback from active galactic nuclei is currently not included.

In this work, we have focused on validating our methodology by comparing basic predictions of \fb{} with observational data across cosmic time. Specifically, we have analyzed various fundamental galaxy scaling relations as well as the cosmic evolution of gas masses, stellar masses, and SFRs, highlighting successes and failures of the FIRE-2 model. Future studies based on \pf{} will discuss, e.g., the morphologies of the simulated galaxies, their star formation rates and depletion times, and the link between galaxy and halo formation. Our main findings are as follows:

\begin{itemize}

\item \pf{} predicts average SFRs of star forming galaxies in good agreement with observations both at $z=0$ and $z=2$ (Fig.~\ref{fig:MainSequence}). The slope of the star forming sequence is slightly sub-linear at $z=0$ ($\sim{}0.85$) and near linear at $z=2$ ($\sim{}0.95$). 

\item  \pf{} underestimates the presence of massive, quiescent galaxies at low z (Fig.~\ref{fig:QuiescentFraction}). While \pf{} naturally accounts for a variety of environmental and stellar feedback driven quenching channels, additional sources, such as AGN feedback, are thus necessary to fully suppress star formation in massive galaxies at low $z$.

\item Simulated galaxies have atomic and molecular gas masses (for a given stellar mass) in good agreement with observational data at $z=0$, see Fig.~\ref{fig:GasSequence}. According to \pf{}, these gas sequences extend down to (at least) $M_{\rm star}\sim{}10^7\,M_\odot$ and they are well described by broken power-laws over 4 orders of magnitude in stellar mass.

\item \pf{} broadly reproduces the observed mass--metallicity relation at $z=0$ over many orders of magnitude both for gas phase metallicities as well as stellar metallicities, see Fig.~\ref{fig:MZR}. In addition, the simulation predicts a low scatter ($\lesssim{}0.1$ dex) for both relations. Both mass--metallicity relations are well fit by broken power-laws.

\item \pf{} predicts a stellar mass function (SMF) at $z=0$ similar to recent estimates by \cite{Leja2020} based on non-parametric modeling except for a moderate excess at both low and high stellar masses (Fig.~\ref{fig:SMFz}). Our predicted $z=0$ SMF is generally higher than those based on more traditional stellar mass estimates (e.g., \citealt{Baldry2012, Moustakas2013}). At intermediate redshifts ($z\sim{}2-4$), \pf{} over-predicts the SMF at low-to-intermediate galaxy masses ($M_{\rm star}\sim{}10^{8.5}-10^{10}\,M_\odot$). A comparison with FIRE-2 zoom-in simulations reveals that reaching a mass resolution of $m_{\rm b}<10^4\,M_\odot$ may be needed to sufficiently lower stellar masses in halos of $M_{\rm halo}\sim{}10^{11}\,M_\odot$. At high $z\geq{}6$, the SMF in \pf{} agrees well with estimates by \cite{Song2015}.

\item The galaxy stellar-to-halo mass ratio in \pf{} increases with increasing halo mass at $M_{\rm halo}<10^{11}\,M_\odot$, peaks near $M_{\rm halo}<10^{11.5}\,M_\odot$, and then declines towards the massive end in qualitative agreement with empirical estimates. The FIRE-2 physics models thus predicts a peak in the galaxy baryonic conversion efficiency even without the inclusion of AGN feedback. However, as our study of the SMF highlights, the stellar masses at the massive end tend to be too high if no additional feedback sources are included, i.e., the decline in the stellar mass -- halo mass ratio is too shallow. The galaxy baryonic conversion efficiency reaches a peak at intermediate halo masses because the fraction of stellar mass residing outside galaxies, i.e., in a stellar halo and in satellite galaxies, increases strongly with increasing halo mass at the massive end (Fig.~\ref{fig:SHMR}). In contrast, the ratio between the stellar mass in the halo and the halo mass declines only weakly at the massive end after  peaking near $M_{\rm halo}=10^{12}\,M_\odot$.

\item The halos of Milky-Way analogs have a baryon fraction of $11.6^{+0.5}_{-0.4}\%$, which is only about 25\% lower than the universal baryon fraction. This percentage is higher than the empirical estimate of 7\% of detected baryons. The observationally `missing' baryons are located in various components including ionized gas with temperatures below $2\times{}10^5$ K and an extra-galactic stellar component.

\item The cosmic star formation history (CSFH) and the stellar mass build-up in \pf{} broadly match observational estimates at $z>1$. At low $z$, \pf{} over-estimates the cosmic SFR density by a factor of $\sim{}3$. This mismatch is driven to a large degree by the under-prediction of the quenched fraction in \pf{} which results in too high a star formation activity in halos hosting $M_{\rm star}>10^{10}\,M_\odot$ galaxies.

\item The cosmic ${\rm H_I}$ density is in broad agreement with observations and shows little evolution with redshift. The cosmic ${\rm H_2}$ density increases monotonically with increasing cosmic time until $z\sim{}1$ after which it remains approximately constant, see Fig.~\ref{fig:OmegaHIH2}. The near constancy of the cosmic ${\rm H_2}$ density at $z<1$ is in tension with some observational data \citep{Walter2020}. This tension could be reduced if higher $z$ galaxies have a lower ${\rm H_2}$ mass per CO luminosity compared with Milky-Way like galaxies in the nearby Universe.

\item Finally, we compare the column density distribution functions (CDDF) of atomic and molecular hydrogen in \pf{} with observations finding good agreement for ${\rm H_I}$, see Fig.~\ref{fig:CDDF}.  In contrast to the ${\rm H_I}$ CDDF, the normalization of the ${\rm H_2}$ CDDF shows a noticeable dependence on redshift, increasing by over one order of magnitude from $z=6$ to $z=2$ at all column densities reflecting the overall increase in the cosmic molecular gas density \citep{Peroux2020}.

\end{itemize}

\pf{} makes it possible to explore the predictions of the FIRE-2 physics model statistically, by providing a representative sample of highly resolved galaxies across cosmic history. However, the current iteration of \fb{} should be understood as a first step in this direction with much work yet to be done. 
While the model is broadly successful in reproducing a number of observational constraints, we also noted various areas of tension or disagreement. In particular, \pf{} is unable to produce massive, quenched galaxies in the appropriate numbers and also predicts a cosmic star formation rate density that is too high at late times. It is possible that the inclusion of feedback from super-massive black holes will remedy these shortcomings \citep{Su2021, Wellons2022}. However, adding AGN feedback also increases the uncertainty of the model predictions as it introduces significant modeling degeneracies. 

Additional work is also needed in both completing the accounting of the relevant processes and in modeling them at the required resolution level. For instance, magnetic fields and cosmic ray pressure may affect the cloud structure on small scales \citep{Hennebelle2019}, accelerate galactic winds \citep{Booth2013a, Salem2014, Girichidis2016, Dashyan2020}, or quench star formation (e.g., \citealt{Su2020}). Recent progress on modeling these physical processes is encouraging (e.g., \citealt{Chan2019, Hopkins2019, Farcy2022}) and we hope to include them in the future. \pf{}, with its focus on comparably well understood physics, provides a robust base-line prediction for such future model extensions.

\section*{Acknowledgements}

The authors thank the referee for insightful comments that helped to improve the paper.
RF thanks Oliver Hahn, Marcel van Daalen, and Jose O\~{n}orbe for help with MUSIC and CAMB.
RF acknowledges financial support from the Swiss National Science Foundation (grant no PP00P2\_157591, PP00P2\_194814, 200021\_188552). 
EQ was supported in part by a Simons Investigator grant from the Simons Foundation and NSF AST grant 2107872.
CAFG was supported by NSF through grants AST-1715216, AST-2108230,  and CAREER award AST-1652522; by NASA through grants 17-ATP17-006 7 and 21-ATP21-0036; by STScI through grants HST-AR-16124.001-A and HST-GO-16730.016-A; by CXO through grant TM2-23005X; and by the Research Corporation for Science Advancement through a Cottrell Scholar Award.
Support for PFH was provided by NSF Research Grants 1911233, 20009234, 2108318, NSF CAREER grant 1455342, NASA grants 80NSSC18K0562, HST-AR-15800. Numerical calculations were run on the allocations AST21010 and AST20016 supported by the NSF and TACC, and NASA HEC SMD-16-7592.
DK was supported by the NSF Grant AST-2108314.
LB, MB, and EC acknowledge financial support from the Swiss National Science Foundation (PP00P2\_194814, 200021\_188552).
JSB was supported by NSF grant AST-1910346.
JG gratefully acknowledges financial support from the Swiss National Science Foundation (grant no CRSII5\_193826).
JM is funded by the Hirsch foundation. Sabbatical leave support for JM was provided by Pomona College and the Harry and Grace Steele Foundation.
AW received support from: NSF via CAREER award AST-2045928 and grant AST-2107772; NASA ATP grant 80NSSC20K0513; HST grants AR-15809, GO-15902, GO-16273 from STScI.
We acknowledge PRACE for awarding us access to MareNostrum at the Barcelona Supercomputing Center (BSC), Spain. This research was partly carried out via the Frontera computing project at the Texas Advanced Computing Center. Frontera is made possible by National Science Foundation award OAC-1818253. This work was supported in part by a grant from the Swiss National Supercomputing Centre (CSCS) under project IDs s697 and s698. We acknowledge access to Piz Daint at the Swiss National Supercomputing Centre, Switzerland under the University of Zurich's share with the project ID uzh18. This work made use of infrastructure services provided by S3IT (www.s3it.uzh.ch), the Service and Support for Science IT team at the University of Zurich. All plots were created with the Matplotlib library for visualization with Python \citep{Hunter2007}. This research has made use of NASA's Astrophysics Data System.

\section*{Data availability}
The data supporting the plots within this article are available on reasonable request to the corresponding author. A public version of the GIZMO code is available at \url{http://www.tapir.caltech.edu/~phopkins/Site/GIZMO.html}. FIRE data releases are publicly available at \url{http://flathub.flatironinstitute.org/fire}.

%%%%%%%%%%%%%%%%%%%%%%%%%%%%%%%%%%%%%%%%%%%%%%%%%%


%\bibliography{main}

%%%%%%%%%%%%%%%%%%%%%%%%%%%%%%%%%%%%%%%%%%%%%%%%%%

%%%%%%%%%%%%%%%%% APPENDICES %%%%%%%%%%%%%%%%%%%%%

\appendix

\section{Re-weighting}
\label{app:reweighting}

We estimate the true (ensemble average) halo mass function (HMF) for our adopted cosmology with the help of HMFcalc\footnote{\url{https://hmf.icrar.org}} \citep{Murray2013}. Specifically, we select the `Behroozi+2013 (Tinker Extension to High-z)'  fitting function \citep{Behroozi2013c} and a CAMB transfer function. We then calculate the HMF for the virial halo criterion \citep{Bryan1998} over a $\lg{}M_{\rm vir}/M_\odot=7-15$ range in steps of $\Delta{}\lg{}M_{\rm vir}=0.05$.

The realized cumulative HMF in the simulation volume $V$ can be estimated as the number of halos above a certain mass $x=\lg{}M_{\rm h}$ divided by the simulation volume, i.e.,
\begin{equation}
\hat{\Phi}(x) = \sum_{\{i:x_i\geq{}x\}} \frac{1}{V}.
\end{equation}

The idea of the re-weighting approach is to replace the equal weights of $1/V$ in the sum above with halo dependent weights. Specifically, approximating the true cumulative HMF $\Phi(x)=\int_{x}^{\infty} \phi(x')\,dx'$ with the following sum over the halos in the simulation volume
\begin{equation}
\Phi(x)\approx{}\sum_{\{i:x_i\geq{}x\}} \phi(x_i)\,\Delta{}x_i,
\label{eq:CHMF}
\end{equation}
suggests that we can replace $1/V$ with weights $w_i= \phi(x_i)\,\Delta{}x_i$. Here, the line elements $\Delta{}x_i$ represent the typical spacing in logarithmic halo mass between halos with $\lg{}M_{\rm h}$ near $x_i$ and $\phi(x)$ is the true differential HMF. We calculate the line elements $\Delta{}x_i$ as $d_i/N_i$ by counting the number of halos ($N_i$) in a top hat kernel of diameter $d_i$ and centered on $x_i$. The diameter is chosen such that the kernel includes a fixed number of halos $N_i=100$ subject to strict lower and upper bounds of  $d_i\geq{}0.05$ and $d_i\leq{}0.5$. Before the weights are calculated, the masses of halos in hydrodynamical simulations are converted to the masses expected for a corresponding collisionless $N$-body simulation by matching the cumulative abundances of halos in \pf{} runs with and without baryonic physics.

The weights $w_i$ exceed $1/V$ for underrepresented halos in the simulation volume thus boosting their contribution and vice versa for overrepresented halos. Once we assign weights $w_i$ to all halos, we can thus calculate re-weighted properties and mass functions in a straightforward manner. For instance, differential stellar mass functions can be obtained via a weighted histogram, while cumulative stellar mass functions sum all the weights of the host halos of galaxies above a certain stellar mass.

In case re-weighting is used, only halos containing more than 300 DM particles obtain updated weights. Halos excluded from re-weighting obtain the standard weight $w_i=1/V$. Sub-halos are assigned the weights of their parent main halos.

\begin{figure}
\begin{tabular}{c}
\includegraphics[width=90mm]{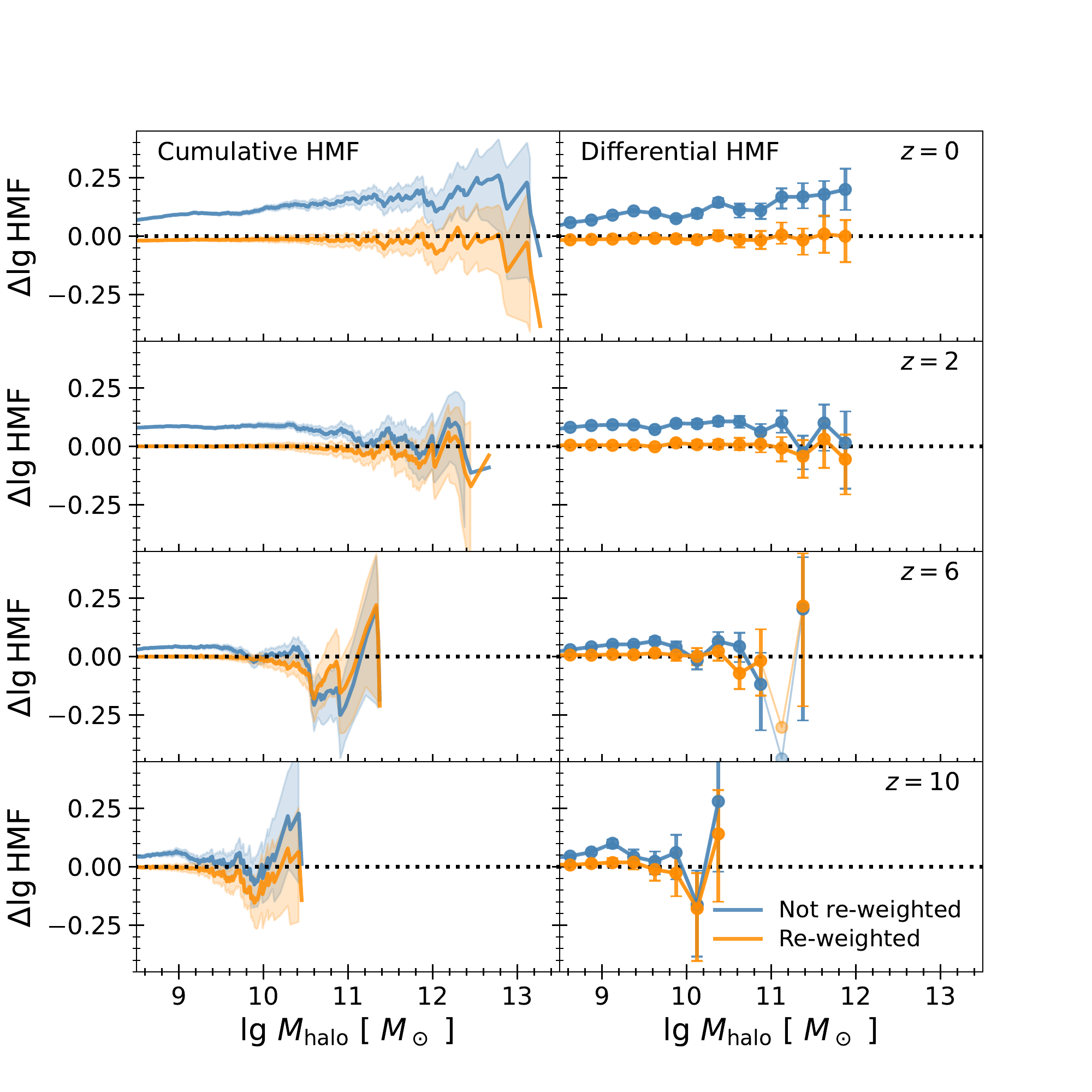}
\end{tabular}
\caption{Halo mass function (HMF) in \pf{} relative to the reference HMF at $z=0-10$ with and without re-weighting. Halo masses in the simulation are first converted to halo masses in a corresponding collisionless $N$-body simulation (shown on the $x$-axis) via halo abundance matching. Subsequently, the empirical cumulative (left panels) and differential (right panels) HMF from the simulation are compared with the reference HMF (HMFcalc; \citealt{Murray2013}). Without re-weighting (blue lines), the HMF of FIREbox can overestimate the reference HMF by up to $\sim{}0.2$ dex.  In contrast, the re-weighted cumulative and differential HMFs (orange lines and symbols) match their reference HMF nearly within statistical errors (shaded regions and error bars are the 16th to 84th percentiles calculated via bootstrapping).
}
\label{fig:Reweighting}
\end{figure}

We show a test of the re-weighting approach in Fig.~\ref{fig:Reweighting}. Without re-weighting, the cumulative and differential HMFs in \pf{} can exceed the expectations from HMFcalc by up to $\sim{}0.2$ dex, especially at $z\leq{}2$. After re-weighting, the HMFs typically match the reference HMFs close to statistical errors. 

\section{Comparison with FIRE-2 zoom-in simulations}
\label{app:CompFIRE2}

\begin{figure*}
\begin{tabular}{cc}
\includegraphics[width=80mm]{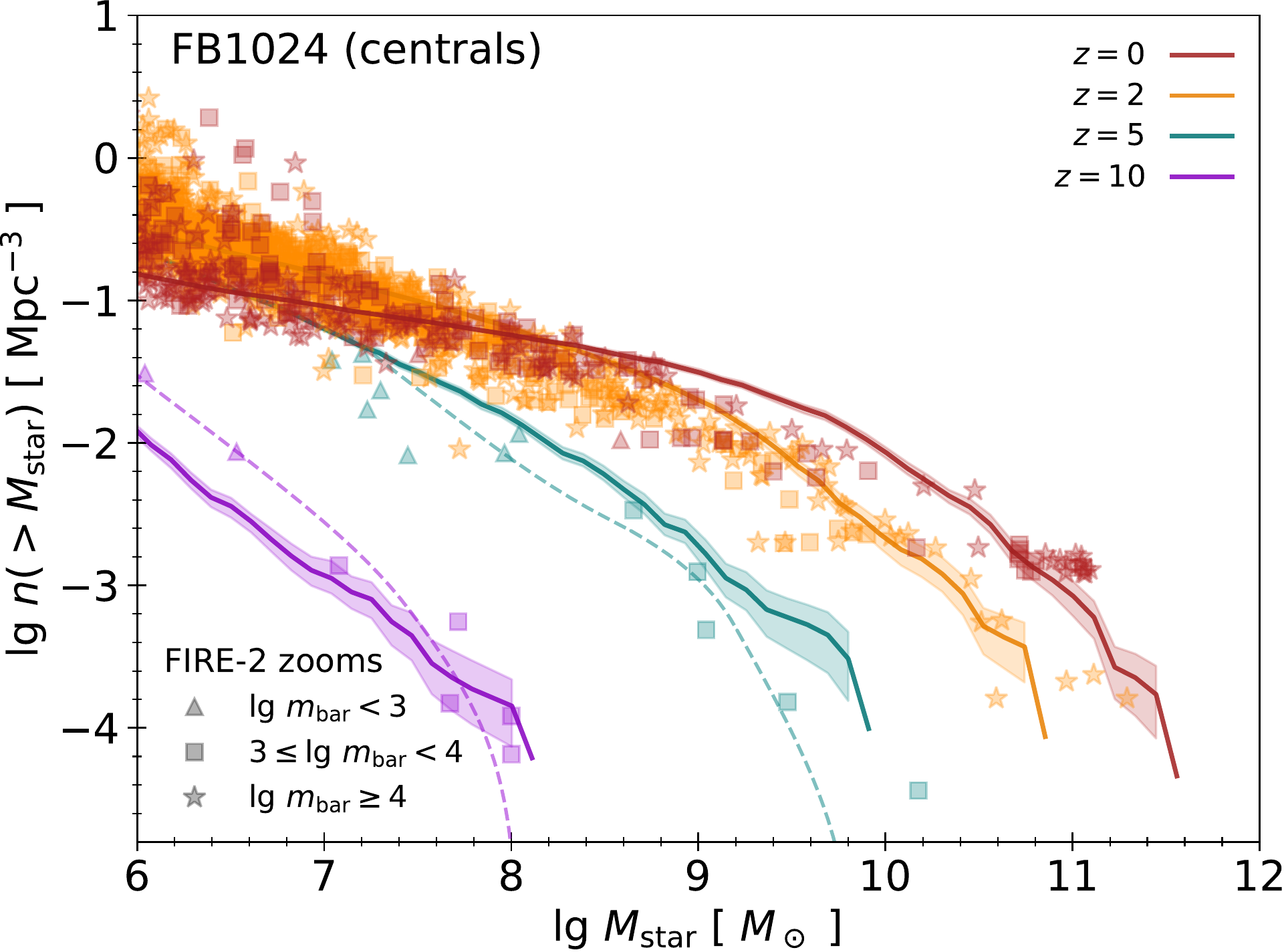} &
\includegraphics[width=80mm]{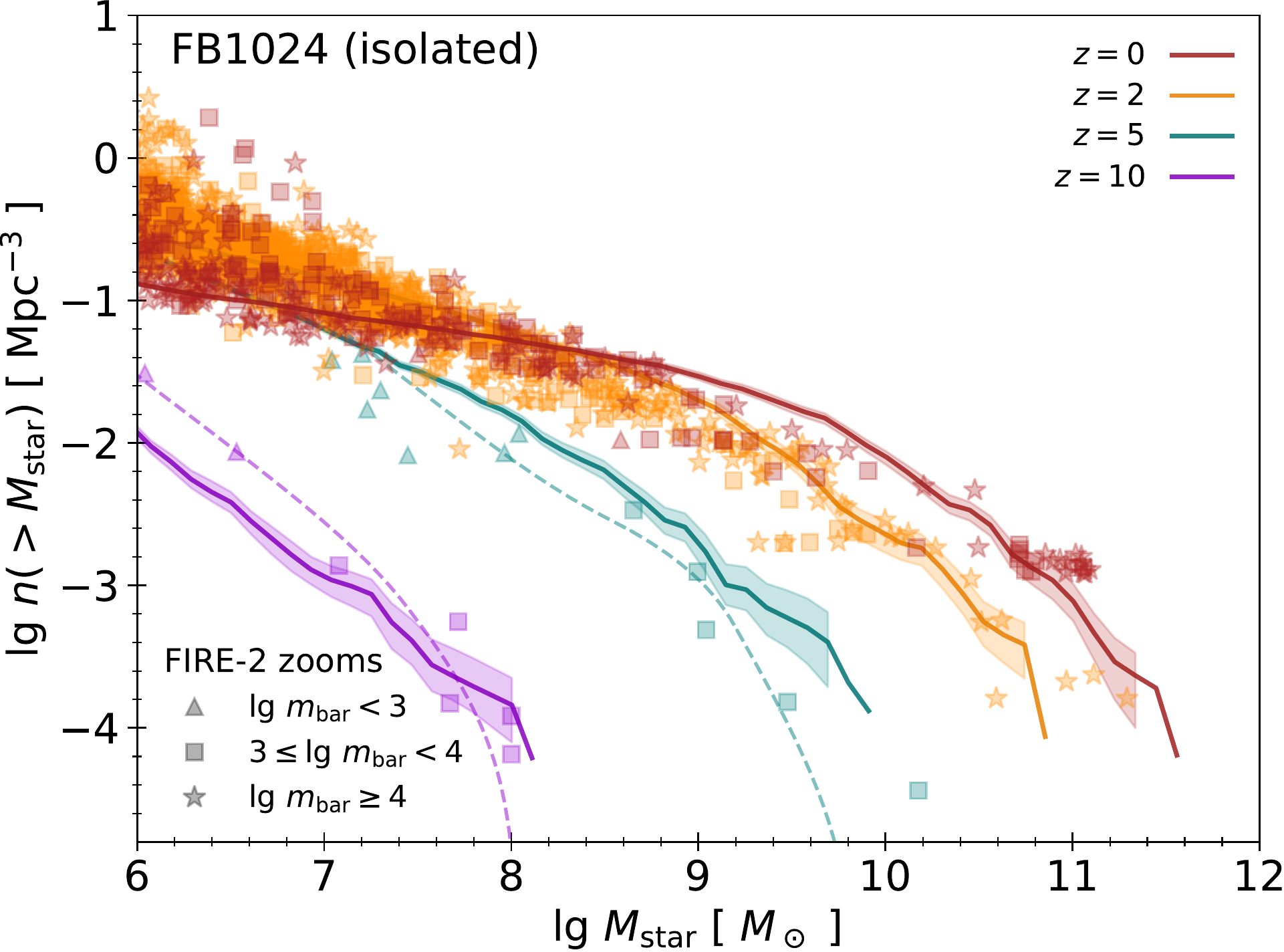}
\end{tabular}
\caption{Comparison between FIRE-2 cosmological zoom-in simulations and \pf{}. Solid lines in the left (right) panel show the cumulative galaxy stellar mass function (SMF) of central (isolated) galaxies with $M_{\rm star}>10^6$ $M_\odot$ in \pf{}. A galaxy is isolated if it is the central galaxy of a main halo and it does not lie within 3 times the virial radius of another main halo. The abundance of galaxies is re-weighted, see Appendix \ref{app:reweighting}.
The various symbols correspond to SMF estimates for central galaxies in FIRE-2 cosmological zoom-in simulations at $z=10$ and $z=5$ \citep{Ma2018b}, $z=2$ \citep{Angles-Alcazar2017a}, and $z=0$ \citep{Hopkins2018}, see Table \ref{tab:FIRE2zooms}. Dashed lines are fits provided in \protect\cite{Ma2018b}. For individual zoom-in simulations, the abundance of galaxies is derived from the expected abundance of their halos. Circles, squares, and stars indicate zoom-in simulations with a baryonic mass resolution of $<10^3$ $M_\odot$, $10^{3-4}$ $M_\odot$, and  $>10^4$ $M_\odot$. 
\pf{} agrees with the results of previous zoom-in simulations for massive galaxies but appears to predict a larger abundance of intermediate mass galaxies ($M_{\rm star}\sim{}10^9-10^{10} M_\odot$). This difference appears to be primarily related to the higher numerical resolution of zoom-in simulations of low mass galaxies. Restricting the analysis to isolated galaxies, as opposed to central galaxies, does not significantly lower the abundance of intermediate mass galaxies.
}
\label{fig:SMFcomp}
\end{figure*}

\begin{figure}
\begin{tabular}{cc}
\includegraphics[width=80mm]{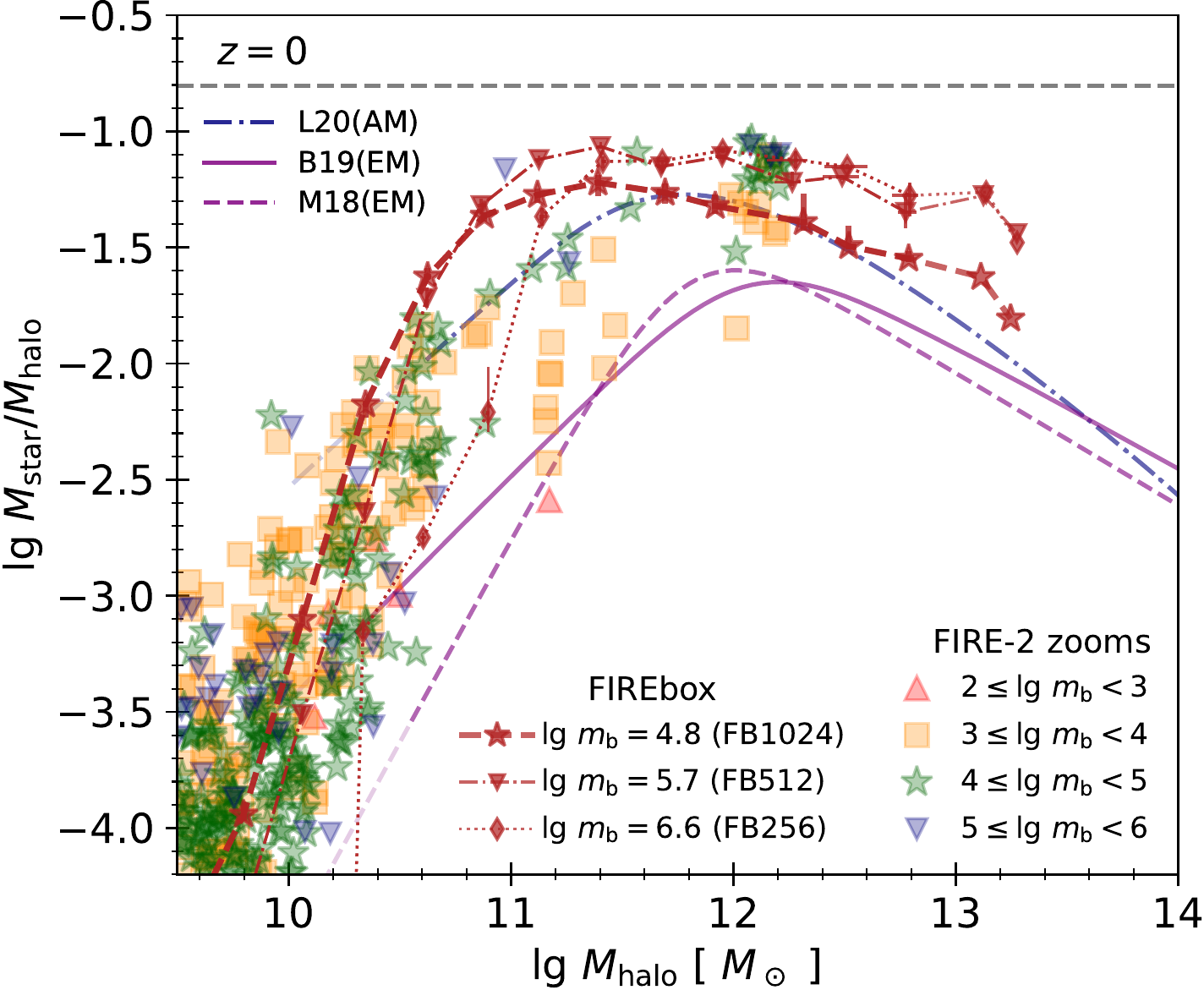}
\end{tabular}
\caption{Ratio between the stellar mass of central galaxies and their parent halos in \pf{}, its lower resolution re-runs, and in FIRE-2 zoom-in simulations at $z=0$. The red dashed line (red dot-dashed and dotted lines) show the median stellar-mass-to-halo-mass ratio in bins of halo mass for \pf{} galaxies (for galaxies from the lower resolution re-runs), while all other symbols report the mass ratios of individual galaxies in FIRE-2 zoom-ins of different mass resolution (see legend). Stellar masses are measured within $R_{\rm g}=3\,R_{\rm half}$, see section \ref{sect:halos}.
The figure includes estimates of the galaxy stellar mass -- halo mass relation via abundance matching (AM, \citealt{Leja2020}, dot-dashed line) and empirical modeling (EM, \citealt{Moster2018, Behroozi2019}, dashed and solid lines), see also Fig.~\ref{fig:SHMR}. 
Various resolution trends are apparent. At the mass resolution of \pf{} ($m_{\rm b}\sim{}10^5\,M_\odot$), galaxy stellar mass depends only weakly on resolution in halos of low ($<10^{10.5}$ $M_\odot$) and high ($\gtrsim{}10^{12}$ $M_\odot$) mass. However, the stellar mass appears quite resolution dependent for halos of intermediate mass ($M_{\rm halo}\sim{}10^{11}\,M_\odot$). At low resolution ($m_{\rm b}\gg{}10^5\,M_\odot$), galaxy stellar masses are generally overestimated (underestimated) in halos above (below) $\sim{}10^{11}\,M_\odot$ relative to the primary \fb{} run.}
\label{fig:SHMRcomp}
\end{figure}

\begin{table}
\begin{tabular}{cccccc}
Label & $m_{\rm b}$ [$10^3$ $M_\odot$] & $z_{\rm final}$ & cosmo & source \\
\hline \hline
\multicolumn{5}{c}{w/o Metal Diffusion}\\
\hline
m11a & 2.1 & 0 & $a$ &  A \\ 
m11b & 2.1 & 0 & $a$ & A \\ 
m11c & 2.1 & 0 & $a$ & A \\
m11q & 0.9, 7.1 & 0 & $a$ &B \\
m11v & 7.1 & 0 & $a$ &B \\
m12b & 57 & 0 & $a$ &B \\
m12c & 57 & 0 & $a$ &B \\
m12f & 7.1, 57 & 0 & $a$ &C \\
m12i & 7.1, 57 & 0 & $a$ &B \\
m12m & 7.1, 57 & 0 & $a$ &B \\
m12q & 57 & 0 & $a$ & B \\
A1 & 33 & 1 & $c$ & D \\
A2 & 33 & 1 & $c$ & D \\
A4 & 33 & 1 & $c$ & D \\
A8 & 33 & 1 & $c$ & D \\
\hline
\multicolumn{5}{c}{w/ Metal Diffusion}\\
\hline
m11d & 7.1 & 0 & $b$ & E \\
m11e & 7.1 & 0 & $b$ &E \\
m11h & 7.1 & 0 & $b$ &E \\
m11i & 7.1 & 0 & $b$ & E \\
m11q & 7.1 & 0 & $a$ &  B \\
m12b & 7.1, 57 & 0 & $a$ & F\\
m12c & 7.1, 57 & 0 & $a$ & F \\
m12f & 7.1, 57, 450 & 0 & $a$ &C \\
m12i & 7.1, 57, 450 & 0 & $a$ & G \\
m12m & 7.1, 57, 450 & 0 & $a$ &B \\
m12r & 7.1, 57 & 0 & $b$ & H\\
m12w & 7.1, 57 & 0 & $b$ &H \\
\hline \hline
\end{tabular}
\caption{FIRE-2 zoom-in simulations used as a point of reference for \pf{}. The first three columns list the simulation identifier, baryonic mass resolution, and the final redshift reached by each simulation. The fourth column states the adopted cosmology of each run. All runs adopt a standard, flat $\Lambda$CDM cosmology with $h\sim{}0.7$, $\Omega_{\rm m}=0.27-0.31$, and $\Omega_{\rm b}\sim{}0.0455-0.048$ broadly consistent with current observational constraints \citep{PlanckCollaboration2015a}. Specifically, cosmology $a$ corresponds to $h=0.702$, $\Omega_{\rm m}=0.272$, $\Omega_{\rm b}=0.0455$, cosmology $b$ to $h=0.68$, $\Omega_{\rm m}=0.31$, $\Omega_{\rm b}=0.048$, and cosmology $c$ to $h=0.697$, $\Omega_{\rm m}=0.2821$, $\Omega_{\rm b}=0.0461$. The final column lists the work that first describes the respective simulation with A: \protect\cite{Chan2018}, B: \protect\cite{Hopkins2018}, C: \protect\cite{Garrison-Kimmel2017a}, D: \protect\cite{Angles-Alcazar2017a},
E: \protect\cite{El-badry2018}, F: \protect\cite{Garrison-Kimmel2019}, G: \protect\cite{Wetzel2016}, and H: \protect\cite{Samuel2020}. The first 15 lines (the last 12 lines) list runs without (with) metal diffusion due to sub-grid turbulence \protect\citep{Hopkins2018}.}
\label{tab:FIRE2zooms}
\end{table}

The SMF in \pf{} shows a higher abundance of moderately low mass galaxies ($M_{\rm star}\sim{}10^9-10^{10}$ $M_\odot$) than is seen in galaxy surveys. Here, we compare \pf{} to other FIRE-2 zoom-in simulations to explore whether this difference is caused by the numerical resolution or the different set-up of \fb{} as a cosmological volume simulation. Overall, we include 41 separate FIRE-2 zoom-in simulations which target halos over a broad range of halo masses $M_{\rm halo}\sim{}10^{11}-10^{13}$ $M_\odot$ and are run to $z=1$ or $z=0$, see Table~\ref{tab:FIRE2zooms}.

We derive cumulative SMFs for galaxies in zoom-in simulations in an approximate fashion via abundance matching of the stellar masses of galaxies ($M_{\rm star}$) and the masses of their host halos ($M_{\rm halo}$). Ignoring scatter, the cumulative SMF $\Phi_{\rm star}(\lg{} M_{\rm star})$ equals the cumulative HMF $\Phi(\lg{} M_{\rm halo})$ and we can thus plot $M_{\rm star}$ vs $\Phi_{\rm star}(\lg{} M_{\rm star})$ for each galaxy from a zoom-in simulations. Cumulative HMFs are obtained from HMFcalc as described in Appendix \ref{app:reweighting}.

The left panel of Fig.~\ref{fig:SMFcomp} shows the SMF of central galaxies in \pf{}. We exclude satellite galaxies since the primary galaxies in zoom-in simulations are usually selected to be centrals or isolated galaxies. In each case, the abundances of the main (or isolated) halos are re-weighted to match the expected HMF of all halos (see Appendix \ref{app:reweighting}) to allow a more direct comparison with the SMF of zoom-in runs.

At high $z$, the SMF in \pf{} is in good agreement with the SMF predicted via abundance matching from the zoom-in runs. At low $z$, however, \pf{} predicts a higher abundances for $M_{\rm star}\sim{}10^9-10^{10}$ $M_\odot$ galaxies compared both with the available FIRE-2 zooms (Fig.~\ref{fig:SMFcomp}) and observations (Fig.~\ref{fig:SMFz}). We now investigate the origin of difference in some detail.

First, we would like to test whether a selection bias toward more isolated galaxies in zoom-in simulations could be responsible. To this end, we plot in the right hand panel of Fig.~\ref{fig:SMFcomp} the SMF of isolated galaxies in \pf{}. A galaxy is isolated if it does not lie within 3 times the virial radius of another main halo. A comparison with the left hand panel of Fig.~\ref{fig:SMFcomp} and with Fig.~\ref{fig:SMFz} reveals that the SMF in \pf{} in the stellar mass regime of interest does not strongly depend on the isolation criterion (all vs central vs isolated galaxies). Hence, differences in galaxy isolation do not appear to be responsible for the excess in moderately low mass galaxies in \pf{} at low $z$.

Also, we can largely exclude a statistical effect related to the scatter in the SHMR relation. This scatter is empirically constrained to about 0.2 dex in massive halos (see e.g., \citealt{Reddick2013, Zu2015a}), while numerical simulations (e.g., \citealt{Schaye2015, Pillepich2018a, Feldmann2019}) as well as semi-analytic (e.g., \citealt{Somerville2012}) and empirical models (e.g., \citealt{Hearin2013e}) suggest that the scatter increases with decreasing halo mass to potentially $\sim{}0.3$ dex at $M_{\rm halo}\sim{}10^{11}$ $M_\odot$ \citep{Wechsler2018}. However, Fig.~\ref{fig:SMFcomp} highlights that all FIRE-2 zoom-in simulations (out of a dozen) with $M_{\rm star}\sim{}10^9-10^{10}$ $M_\odot$ at $z=0$ have lower abundances, i.e., lower stellar masses for a given halo mass.

In Fig.~\ref{fig:SHMRcomp} we show the SHMR for central galaxies in both \pf{} and in the FIRE-2 zoom-ins. The figure highlights that stellar masses of galaxies in halos of intermediate mass ($M_{\rm halo}\sim{}10^{11}\,M_\odot$) are noticeably resolution dependent, varying by an order of magnitude (with large scatter) when increasing the mass resolution by 3 orders of magnitude. In addition, it appears that central galaxies residing in such halos are more massive (by $\sim{}0.2$ dex) in \pf{} compared with zoom-ins of a similar resolution. The latter result may indicate that the Lagrangian patches of the zoom-ins (or perhaps the box-size of \pf{}) are too small to adequately capture the cosmological environment at $z=0$. Given the resolution dependence, we caution that our predictions for central galaxies residing in $M_{\rm halo}\sim{}10^{11}\,M_\odot$ halos (and thus the SMF of galaxies with $M_{\rm star}\sim{}10^9-10^{10}\,M_\odot$) are uncertain. In contrast, stellar masses in low mass ($<10^{10.5}$ $M_\odot$) and massive ($\sim{}10^{12}$ $M_\odot$) halos appear close to converged.

%%%%%%%%%%%%%%%%%%%%%%%%%%%%%%%%%%%%%%%%%%%%%%%%%%

% Don't change these lines
\bsp	% typesetting comment
\label{lastpage}
\end{document}